\documentclass[longauth]{aaEC}

\usepackage{graphicx}
\usepackage{natbib}
\usepackage{scalerel}
\usepackage{lastpage}
\usepackage[table]{xcolor}
\usepackage[normalem]{ulem}

\def\Eq{\mbox{Eq.~}}

\def\Fig{\mbox{Fig.~}}

\def\Tab{\mbox{Table~}}

\def\Sec{\mbox{Sect.~}}
\def\Secs{\mbox{Sects.~}}

\usepackage{txfonts}
\usepackage[pdfencoding=auto,psdextra]{hyperref}
\hypersetup{
    colorlinks=true,
    linkcolor=blue,
    filecolor=magenta,      
    urlcolor=blue,
    citecolor=blue
}
\makeatletter
\renewcommand*\aa@pageof{, page \thepage{} of \pageref*{LastPage}}
\makeatother

\usepackage[utf8]{inputenc}

\usepackage{euclid}

\usepackage{amsmath}	
\usepackage{multicol}        
\usepackage{bm}		
\usepackage{pdflscape}	

\newcommand{\HS}{\texttt{HST2EUCLID}}

\usepackage[switch, modulo]{lineno}

\begin{document}

\nolinenumbers
%
%

\title{Euclid Quick Data Release (Q1)}
\subtitle{LEMON -- LEns MOdelling with Neural networks. Automated and fast modelling of \Euclid gravitational lenses with a singular isothermal ellipsoid mass profile}
   
\newcommand{\orcid}[1]{} 


%
%
\author{Euclid Collaboration: V.~Busillo\orcid{0009-0000-6049-1073}\thanks{\email{valerio.busillo@inaf.it}}\inst{\ref{aff1},\ref{aff2},\ref{aff3}}
\and C.~Tortora\orcid{0000-0001-7958-6531}\inst{\ref{aff1}}
\and R.~B.~Metcalf\orcid{0000-0003-3167-2574}\inst{\ref{aff4},\ref{aff5}}
\and J.~W.~Nightingale\orcid{0000-0002-8987-7401}\inst{\ref{aff6}}
\and M.~Meneghetti\orcid{0000-0003-1225-7084}\inst{\ref{aff5},\ref{aff7}}
\and F.~Gentile\orcid{0000-0002-8008-9871}\inst{\ref{aff8},\ref{aff5},\ref{aff9}}
\and R.~Gavazzi\orcid{0000-0002-5540-6935}\inst{\ref{aff10},\ref{aff11}}
\and F.~Zhong\orcid{0009-0001-8607-4945}\inst{\ref{aff12}}
\and R.~Li\inst{\ref{aff13}}
\and B.~Cl\'ement\orcid{0000-0002-7966-3661}\inst{\ref{aff14},\ref{aff15}}
\and G.~Covone\orcid{0000-0002-2553-096X}\inst{\ref{aff2},\ref{aff1},\ref{aff3}}
\and N.~R.~Napolitano\orcid{0000-0003-0911-8884}\inst{\ref{aff16},\ref{aff1},\ref{aff3}}
\and F.~Courbin\orcid{0000-0003-0758-6510}\inst{\ref{aff17},\ref{aff18}}
\and M.~Walmsley\orcid{0000-0002-6408-4181}\inst{\ref{aff19},\ref{aff20}}
\and E.~Jullo\orcid{0000-0002-9253-053X}\inst{\ref{aff10}}
\and J.~Pearson\orcid{0000-0001-8555-8561}\inst{\ref{aff21}}
\and D.~Scott\orcid{0000-0002-6878-9840}\inst{\ref{aff22}}
\and A.~M.~C.~Le~Brun\orcid{0000-0002-0936-4594}\inst{\ref{aff23}}
\and L.~Leuzzi\orcid{0009-0006-4479-7017}\inst{\ref{aff4},\ref{aff5}}
\and N.~Aghanim\orcid{0000-0002-6688-8992}\inst{\ref{aff24}}
\and B.~Altieri\orcid{0000-0003-3936-0284}\inst{\ref{aff25}}
\and A.~Amara\inst{\ref{aff26}}
\and S.~Andreon\orcid{0000-0002-2041-8784}\inst{\ref{aff27}}
\and H.~Aussel\orcid{0000-0002-1371-5705}\inst{\ref{aff28}}
\and C.~Baccigalupi\orcid{0000-0002-8211-1630}\inst{\ref{aff29},\ref{aff30},\ref{aff31},\ref{aff32}}
\and M.~Baldi\orcid{0000-0003-4145-1943}\inst{\ref{aff9},\ref{aff5},\ref{aff7}}
\and S.~Bardelli\orcid{0000-0002-8900-0298}\inst{\ref{aff5}}
\and P.~Battaglia\orcid{0000-0002-7337-5909}\inst{\ref{aff5}}
\and A.~Biviano\orcid{0000-0002-0857-0732}\inst{\ref{aff30},\ref{aff29}}
\and E.~Branchini\orcid{0000-0002-0808-6908}\inst{\ref{aff33},\ref{aff34},\ref{aff27}}
\and M.~Brescia\orcid{0000-0001-9506-5680}\inst{\ref{aff2},\ref{aff1}}
\and J.~Brinchmann\orcid{0000-0003-4359-8797}\inst{\ref{aff35},\ref{aff36}}
\and S.~Camera\orcid{0000-0003-3399-3574}\inst{\ref{aff37},\ref{aff38},\ref{aff39}}
\and G.~Ca\~nas-Herrera\orcid{0000-0003-2796-2149}\inst{\ref{aff40},\ref{aff41},\ref{aff42}}
\and V.~Capobianco\orcid{0000-0002-3309-7692}\inst{\ref{aff39}}
\and C.~Carbone\orcid{0000-0003-0125-3563}\inst{\ref{aff43}}
\and V.~F.~Cardone\inst{\ref{aff44},\ref{aff45}}
\and J.~Carretero\orcid{0000-0002-3130-0204}\inst{\ref{aff46},\ref{aff47}}
\and S.~Casas\orcid{0000-0002-4751-5138}\inst{\ref{aff48}}
\and M.~Castellano\orcid{0000-0001-9875-8263}\inst{\ref{aff44}}
\and G.~Castignani\orcid{0000-0001-6831-0687}\inst{\ref{aff5}}
\and S.~Cavuoti\orcid{0000-0002-3787-4196}\inst{\ref{aff1},\ref{aff3}}
\and K.~C.~Chambers\orcid{0000-0001-6965-7789}\inst{\ref{aff49}}
\and A.~Cimatti\inst{\ref{aff50}}
\and C.~Colodro-Conde\inst{\ref{aff51}}
\and G.~Congedo\orcid{0000-0003-2508-0046}\inst{\ref{aff52}}
\and C.~J.~Conselice\orcid{0000-0003-1949-7638}\inst{\ref{aff20}}
\and L.~Conversi\orcid{0000-0002-6710-8476}\inst{\ref{aff53},\ref{aff25}}
\and Y.~Copin\orcid{0000-0002-5317-7518}\inst{\ref{aff54}}
\and H.~M.~Courtois\orcid{0000-0003-0509-1776}\inst{\ref{aff55}}
\and M.~Cropper\orcid{0000-0003-4571-9468}\inst{\ref{aff56}}
\and A.~Da~Silva\orcid{0000-0002-6385-1609}\inst{\ref{aff57},\ref{aff58}}
\and H.~Degaudenzi\orcid{0000-0002-5887-6799}\inst{\ref{aff59}}
\and S.~de~la~Torre\inst{\ref{aff10}}
\and G.~De~Lucia\orcid{0000-0002-6220-9104}\inst{\ref{aff30}}
\and A.~M.~Di~Giorgio\orcid{0000-0002-4767-2360}\inst{\ref{aff60}}
\and J.~Dinis\orcid{0000-0001-5075-1601}\inst{\ref{aff57},\ref{aff58}}
\and H.~Dole\orcid{0000-0002-9767-3839}\inst{\ref{aff24}}
\and F.~Dubath\orcid{0000-0002-6533-2810}\inst{\ref{aff59}}
\and X.~Dupac\inst{\ref{aff25}}
\and S.~Dusini\orcid{0000-0002-1128-0664}\inst{\ref{aff61}}
\and S.~Escoffier\orcid{0000-0002-2847-7498}\inst{\ref{aff62}}
\and M.~Farina\orcid{0000-0002-3089-7846}\inst{\ref{aff60}}
\and R.~Farinelli\inst{\ref{aff5}}
\and F.~Faustini\orcid{0000-0001-6274-5145}\inst{\ref{aff63},\ref{aff44}}
\and S.~Ferriol\inst{\ref{aff54}}
\and F.~Finelli\orcid{0000-0002-6694-3269}\inst{\ref{aff5},\ref{aff64}}
\and S.~Fotopoulou\orcid{0000-0002-9686-254X}\inst{\ref{aff65}}
\and M.~Frailis\orcid{0000-0002-7400-2135}\inst{\ref{aff30}}
\and E.~Franceschi\orcid{0000-0002-0585-6591}\inst{\ref{aff5}}
\and S.~Galeotta\orcid{0000-0002-3748-5115}\inst{\ref{aff30}}
\and K.~George\orcid{0000-0002-1734-8455}\inst{\ref{aff66}}
\and W.~Gillard\orcid{0000-0003-4744-9748}\inst{\ref{aff62}}
\and B.~Gillis\orcid{0000-0002-4478-1270}\inst{\ref{aff52}}
\and C.~Giocoli\orcid{0000-0002-9590-7961}\inst{\ref{aff5},\ref{aff7}}
\and J.~Gracia-Carpio\inst{\ref{aff67}}
\and B.~R.~Granett\orcid{0000-0003-2694-9284}\inst{\ref{aff27}}
\and A.~Grazian\orcid{0000-0002-5688-0663}\inst{\ref{aff68}}
\and F.~Grupp\inst{\ref{aff67},\ref{aff66}}
\and S.~V.~H.~Haugan\orcid{0000-0001-9648-7260}\inst{\ref{aff69}}
\and W.~Holmes\inst{\ref{aff70}}
\and I.~Hook\orcid{0000-0002-2960-978X}\inst{\ref{aff71}}
\and F.~Hormuth\inst{\ref{aff72}}
\and A.~Hornstrup\orcid{0000-0002-3363-0936}\inst{\ref{aff73},\ref{aff74}}
\and P.~Hudelot\inst{\ref{aff11}}
\and K.~Jahnke\orcid{0000-0003-3804-2137}\inst{\ref{aff75}}
\and M.~Jhabvala\inst{\ref{aff76}}
\and B.~Joachimi\orcid{0000-0001-7494-1303}\inst{\ref{aff77}}
\and E.~Keih\"anen\orcid{0000-0003-1804-7715}\inst{\ref{aff78}}
\and S.~Kermiche\orcid{0000-0002-0302-5735}\inst{\ref{aff62}}
\and A.~Kiessling\orcid{0000-0002-2590-1273}\inst{\ref{aff70}}
\and B.~Kubik\orcid{0009-0006-5823-4880}\inst{\ref{aff54}}
\and M.~K\"ummel\orcid{0000-0003-2791-2117}\inst{\ref{aff66}}
\and M.~Kunz\orcid{0000-0002-3052-7394}\inst{\ref{aff79}}
\and H.~Kurki-Suonio\orcid{0000-0002-4618-3063}\inst{\ref{aff80},\ref{aff81}}
\and Q.~Le~Boulc'h\inst{\ref{aff82}}
\and S.~Ligori\orcid{0000-0003-4172-4606}\inst{\ref{aff39}}
\and P.~B.~Lilje\orcid{0000-0003-4324-7794}\inst{\ref{aff69}}
\and V.~Lindholm\orcid{0000-0003-2317-5471}\inst{\ref{aff80},\ref{aff81}}
\and I.~Lloro\orcid{0000-0001-5966-1434}\inst{\ref{aff83}}
\and G.~Mainetti\orcid{0000-0003-2384-2377}\inst{\ref{aff82}}
\and D.~Maino\inst{\ref{aff84},\ref{aff43},\ref{aff85}}
\and E.~Maiorano\orcid{0000-0003-2593-4355}\inst{\ref{aff5}}
\and O.~Mansutti\orcid{0000-0001-5758-4658}\inst{\ref{aff30}}
\and O.~Marggraf\orcid{0000-0001-7242-3852}\inst{\ref{aff86}}
\and K.~Markovic\orcid{0000-0001-6764-073X}\inst{\ref{aff70}}
\and M.~Martinelli\orcid{0000-0002-6943-7732}\inst{\ref{aff44},\ref{aff45}}
\and N.~Martinet\orcid{0000-0003-2786-7790}\inst{\ref{aff10}}
\and F.~Marulli\orcid{0000-0002-8850-0303}\inst{\ref{aff4},\ref{aff5},\ref{aff7}}
\and R.~Massey\orcid{0000-0002-6085-3780}\inst{\ref{aff87}}
\and S.~Maurogordato\inst{\ref{aff88}}
\and E.~Medinaceli\orcid{0000-0002-4040-7783}\inst{\ref{aff5}}
\and S.~Mei\orcid{0000-0002-2849-559X}\inst{\ref{aff89},\ref{aff90}}
\and Y.~Mellier\inst{\ref{aff91},\ref{aff11}}
\and E.~Merlin\orcid{0000-0001-6870-8900}\inst{\ref{aff44}}
\and G.~Meylan\inst{\ref{aff14}}
\and A.~Mora\orcid{0000-0002-1922-8529}\inst{\ref{aff92}}
\and M.~Moresco\orcid{0000-0002-7616-7136}\inst{\ref{aff4},\ref{aff5}}
\and L.~Moscardini\orcid{0000-0002-3473-6716}\inst{\ref{aff4},\ref{aff5},\ref{aff7}}
\and R.~Nakajima\orcid{0009-0009-1213-7040}\inst{\ref{aff86}}
\and C.~Neissner\orcid{0000-0001-8524-4968}\inst{\ref{aff93},\ref{aff47}}
\and S.-M.~Niemi\inst{\ref{aff40}}
\and C.~Padilla\orcid{0000-0001-7951-0166}\inst{\ref{aff93}}
\and S.~Paltani\orcid{0000-0002-8108-9179}\inst{\ref{aff59}}
\and F.~Pasian\orcid{0000-0002-4869-3227}\inst{\ref{aff30}}
\and K.~Pedersen\inst{\ref{aff94}}
\and V.~Pettorino\inst{\ref{aff40}}
\and S.~Pires\orcid{0000-0002-0249-2104}\inst{\ref{aff28}}
\and G.~Polenta\orcid{0000-0003-4067-9196}\inst{\ref{aff63}}
\and M.~Poncet\inst{\ref{aff95}}
\and L.~A.~Popa\inst{\ref{aff96}}
\and L.~Pozzetti\orcid{0000-0001-7085-0412}\inst{\ref{aff5}}
\and F.~Raison\orcid{0000-0002-7819-6918}\inst{\ref{aff67}}
\and R.~Rebolo\inst{\ref{aff51},\ref{aff97},\ref{aff98}}
\and A.~Renzi\orcid{0000-0001-9856-1970}\inst{\ref{aff99},\ref{aff61}}
\and J.~Rhodes\orcid{0000-0002-4485-8549}\inst{\ref{aff70}}
\and G.~Riccio\inst{\ref{aff1}}
\and E.~Romelli\orcid{0000-0003-3069-9222}\inst{\ref{aff30}}
\and M.~Roncarelli\orcid{0000-0001-9587-7822}\inst{\ref{aff5}}
\and R.~Saglia\orcid{0000-0003-0378-7032}\inst{\ref{aff66},\ref{aff67}}
\and Z.~Sakr\orcid{0000-0002-4823-3757}\inst{\ref{aff100},\ref{aff101},\ref{aff102}}
\and A.~G.~S\'anchez\orcid{0000-0003-1198-831X}\inst{\ref{aff67}}
\and D.~Sapone\orcid{0000-0001-7089-4503}\inst{\ref{aff103}}
\and B.~Sartoris\orcid{0000-0003-1337-5269}\inst{\ref{aff66},\ref{aff30}}
\and J.~A.~Schewtschenko\orcid{0000-0002-4913-6393}\inst{\ref{aff52}}
\and M.~Schirmer\orcid{0000-0003-2568-9994}\inst{\ref{aff75}}
\and P.~Schneider\orcid{0000-0001-8561-2679}\inst{\ref{aff86}}
\and T.~Schrabback\orcid{0000-0002-6987-7834}\inst{\ref{aff104}}
\and A.~Secroun\orcid{0000-0003-0505-3710}\inst{\ref{aff62}}
\and E.~Sefusatti\orcid{0000-0003-0473-1567}\inst{\ref{aff30},\ref{aff29},\ref{aff31}}
\and G.~Seidel\orcid{0000-0003-2907-353X}\inst{\ref{aff75}}
\and M.~Seiffert\orcid{0000-0002-7536-9393}\inst{\ref{aff70}}
\and S.~Serrano\orcid{0000-0002-0211-2861}\inst{\ref{aff105},\ref{aff106},\ref{aff107}}
\and P.~Simon\inst{\ref{aff86}}
\and C.~Sirignano\orcid{0000-0002-0995-7146}\inst{\ref{aff99},\ref{aff61}}
\and G.~Sirri\orcid{0000-0003-2626-2853}\inst{\ref{aff7}}
\and G.~Smadja\inst{\ref{aff54}}
\and L.~Stanco\orcid{0000-0002-9706-5104}\inst{\ref{aff61}}
\and J.~Steinwagner\orcid{0000-0001-7443-1047}\inst{\ref{aff67}}
\and P.~Tallada-Cresp\'{i}\orcid{0000-0002-1336-8328}\inst{\ref{aff46},\ref{aff47}}
\and A.~N.~Taylor\inst{\ref{aff52}}
\and I.~Tereno\inst{\ref{aff57},\ref{aff108}}
\and S.~Toft\orcid{0000-0003-3631-7176}\inst{\ref{aff109},\ref{aff110}}
\and R.~Toledo-Moreo\orcid{0000-0002-2997-4859}\inst{\ref{aff111}}
\and F.~Torradeflot\orcid{0000-0003-1160-1517}\inst{\ref{aff47},\ref{aff46}}
\and I.~Tutusaus\orcid{0000-0002-3199-0399}\inst{\ref{aff101}}
\and L.~Valenziano\orcid{0000-0002-1170-0104}\inst{\ref{aff5},\ref{aff64}}
\and J.~Valiviita\orcid{0000-0001-6225-3693}\inst{\ref{aff80},\ref{aff81}}
\and T.~Vassallo\orcid{0000-0001-6512-6358}\inst{\ref{aff66},\ref{aff30}}
\and A.~Veropalumbo\orcid{0000-0003-2387-1194}\inst{\ref{aff27},\ref{aff34},\ref{aff33}}
\and Y.~Wang\orcid{0000-0002-4749-2984}\inst{\ref{aff112}}
\and J.~Weller\orcid{0000-0002-8282-2010}\inst{\ref{aff66},\ref{aff67}}
\and G.~Zamorani\orcid{0000-0002-2318-301X}\inst{\ref{aff5}}
\and E.~Zucca\orcid{0000-0002-5845-8132}\inst{\ref{aff5}}
\and V.~Allevato\orcid{0000-0001-7232-5152}\inst{\ref{aff1}}
\and M.~Ballardini\orcid{0000-0003-4481-3559}\inst{\ref{aff113},\ref{aff114},\ref{aff5}}
\and M.~Bolzonella\orcid{0000-0003-3278-4607}\inst{\ref{aff5}}
\and E.~Bozzo\orcid{0000-0002-8201-1525}\inst{\ref{aff59}}
\and C.~Burigana\orcid{0000-0002-3005-5796}\inst{\ref{aff115},\ref{aff64}}
\and R.~Cabanac\orcid{0000-0001-6679-2600}\inst{\ref{aff101}}
\and M.~Calabrese\orcid{0000-0002-2637-2422}\inst{\ref{aff116},\ref{aff43}}
\and D.~Di~Ferdinando\inst{\ref{aff7}}
\and J.~A.~Escartin~Vigo\inst{\ref{aff67}}
\and L.~Gabarra\orcid{0000-0002-8486-8856}\inst{\ref{aff117}}
\and M.~Huertas-Company\orcid{0000-0002-1416-8483}\inst{\ref{aff51},\ref{aff118},\ref{aff119},\ref{aff120}}
\and S.~Matthew\orcid{0000-0001-8448-1697}\inst{\ref{aff52}}
\and N.~Mauri\orcid{0000-0001-8196-1548}\inst{\ref{aff50},\ref{aff7}}
\and A.~A.~Nucita\inst{\ref{aff121},\ref{aff122},\ref{aff123}}
\and A.~Pezzotta\orcid{0000-0003-0726-2268}\inst{\ref{aff67}}
\and M.~P\"ontinen\orcid{0000-0001-5442-2530}\inst{\ref{aff80}}
\and C.~Porciani\orcid{0000-0002-7797-2508}\inst{\ref{aff86}}
\and V.~Scottez\inst{\ref{aff91},\ref{aff124}}
\and M.~Tenti\orcid{0000-0002-4254-5901}\inst{\ref{aff7}}
\and M.~Viel\orcid{0000-0002-2642-5707}\inst{\ref{aff29},\ref{aff30},\ref{aff32},\ref{aff31},\ref{aff125}}
\and M.~Wiesmann\orcid{0009-0000-8199-5860}\inst{\ref{aff69}}
\and Y.~Akrami\orcid{0000-0002-2407-7956}\inst{\ref{aff126},\ref{aff127}}
\and S.~Alvi\orcid{0000-0001-5779-8568}\inst{\ref{aff113}}
\and I.~T.~Andika\orcid{0000-0001-6102-9526}\inst{\ref{aff128},\ref{aff129}}
\and S.~Anselmi\orcid{0000-0002-3579-9583}\inst{\ref{aff61},\ref{aff99},\ref{aff130}}
\and M.~Archidiacono\orcid{0000-0003-4952-9012}\inst{\ref{aff84},\ref{aff85}}
\and F.~Atrio-Barandela\orcid{0000-0002-2130-2513}\inst{\ref{aff131}}
\and D.~Bertacca\orcid{0000-0002-2490-7139}\inst{\ref{aff99},\ref{aff68},\ref{aff61}}
\and M.~Bethermin\orcid{0000-0002-3915-2015}\inst{\ref{aff132}}
\and A.~Blanchard\orcid{0000-0001-8555-9003}\inst{\ref{aff101}}
\and L.~Blot\orcid{0000-0002-9622-7167}\inst{\ref{aff133},\ref{aff130}}
\and S.~Borgani\orcid{0000-0001-6151-6439}\inst{\ref{aff134},\ref{aff29},\ref{aff30},\ref{aff31},\ref{aff125}}
\and M.~L.~Brown\orcid{0000-0002-0370-8077}\inst{\ref{aff20}}
\and S.~Bruton\orcid{0000-0002-6503-5218}\inst{\ref{aff135}}
\and A.~Calabro\orcid{0000-0003-2536-1614}\inst{\ref{aff44}}
\and B.~Camacho~Quevedo\orcid{0000-0002-8789-4232}\inst{\ref{aff105},\ref{aff107}}
\and A.~Cappi\inst{\ref{aff5},\ref{aff88}}
\and F.~Caro\inst{\ref{aff44}}
\and C.~S.~Carvalho\inst{\ref{aff108}}
\and T.~Castro\orcid{0000-0002-6292-3228}\inst{\ref{aff30},\ref{aff31},\ref{aff29},\ref{aff125}}
\and F.~Cogato\orcid{0000-0003-4632-6113}\inst{\ref{aff4},\ref{aff5}}
\and S.~Conseil\orcid{0000-0002-3657-4191}\inst{\ref{aff54}}
\and S.~Contarini\orcid{0000-0002-9843-723X}\inst{\ref{aff67}}
\and A.~R.~Cooray\orcid{0000-0002-3892-0190}\inst{\ref{aff136}}
\and O.~Cucciati\orcid{0000-0002-9336-7551}\inst{\ref{aff5}}
\and F.~De~Paolis\orcid{0000-0001-6460-7563}\inst{\ref{aff121},\ref{aff122},\ref{aff123}}
\and G.~Desprez\orcid{0000-0001-8325-1742}\inst{\ref{aff137}}
\and A.~D\'iaz-S\'anchez\orcid{0000-0003-0748-4768}\inst{\ref{aff138}}
\and S.~Di~Domizio\orcid{0000-0003-2863-5895}\inst{\ref{aff33},\ref{aff34}}
\and J.~M.~Diego\orcid{0000-0001-9065-3926}\inst{\ref{aff139}}
\and P.~Dimauro\orcid{0000-0001-7399-2854}\inst{\ref{aff44},\ref{aff140}}
\and A.~Enia\orcid{0000-0002-0200-2857}\inst{\ref{aff9},\ref{aff5}}
\and Y.~Fang\inst{\ref{aff66}}
\and A.~G.~Ferrari\orcid{0009-0005-5266-4110}\inst{\ref{aff7}}
\and P.~G.~Ferreira\orcid{0000-0002-3021-2851}\inst{\ref{aff117}}
\and A.~Finoguenov\orcid{0000-0002-4606-5403}\inst{\ref{aff80}}
\and A.~Franco\orcid{0000-0002-4761-366X}\inst{\ref{aff122},\ref{aff121},\ref{aff123}}
\and K.~Ganga\orcid{0000-0001-8159-8208}\inst{\ref{aff89}}
\and J.~Garc\'ia-Bellido\orcid{0000-0002-9370-8360}\inst{\ref{aff126}}
\and T.~Gasparetto\orcid{0000-0002-7913-4866}\inst{\ref{aff30}}
\and V.~Gautard\inst{\ref{aff8}}
\and E.~Gaztanaga\orcid{0000-0001-9632-0815}\inst{\ref{aff107},\ref{aff105},\ref{aff141}}
\and F.~Giacomini\orcid{0000-0002-3129-2814}\inst{\ref{aff7}}
\and F.~Gianotti\orcid{0000-0003-4666-119X}\inst{\ref{aff5}}
\and G.~Gozaliasl\orcid{0000-0002-0236-919X}\inst{\ref{aff142},\ref{aff80}}
\and M.~Guidi\orcid{0000-0001-9408-1101}\inst{\ref{aff9},\ref{aff5}}
\and C.~M.~Gutierrez\orcid{0000-0001-7854-783X}\inst{\ref{aff143}}
\and A.~Hall\orcid{0000-0002-3139-8651}\inst{\ref{aff52}}
\and W.~G.~Hartley\inst{\ref{aff59}}
\and S.~Hemmati\orcid{0000-0003-2226-5395}\inst{\ref{aff144}}
\and C.~Hern\'andez-Monteagudo\orcid{0000-0001-5471-9166}\inst{\ref{aff98},\ref{aff51}}
\and H.~Hildebrandt\orcid{0000-0002-9814-3338}\inst{\ref{aff145}}
\and J.~Hjorth\orcid{0000-0002-4571-2306}\inst{\ref{aff94}}
\and J.~J.~E.~Kajava\orcid{0000-0002-3010-8333}\inst{\ref{aff146},\ref{aff147}}
\and Y.~Kang\orcid{0009-0000-8588-7250}\inst{\ref{aff59}}
\and V.~Kansal\orcid{0000-0002-4008-6078}\inst{\ref{aff148},\ref{aff149}}
\and D.~Karagiannis\orcid{0000-0002-4927-0816}\inst{\ref{aff113},\ref{aff150}}
\and K.~Kiiveri\inst{\ref{aff78}}
\and C.~C.~Kirkpatrick\inst{\ref{aff78}}
\and S.~Kruk\orcid{0000-0001-8010-8879}\inst{\ref{aff25}}
\and M.~Lattanzi\orcid{0000-0003-1059-2532}\inst{\ref{aff114}}
\and J.~Le~Graet\orcid{0000-0001-6523-7971}\inst{\ref{aff62}}
\and L.~Legrand\orcid{0000-0003-0610-5252}\inst{\ref{aff151},\ref{aff152}}
\and M.~Lembo\orcid{0000-0002-5271-5070}\inst{\ref{aff113},\ref{aff114}}
\and F.~Lepori\orcid{0009-0000-5061-7138}\inst{\ref{aff153}}
\and G.~Leroy\orcid{0009-0004-2523-4425}\inst{\ref{aff154},\ref{aff87}}
\and J.~Lesgourgues\orcid{0000-0001-7627-353X}\inst{\ref{aff48}}
\and T.~I.~Liaudat\orcid{0000-0002-9104-314X}\inst{\ref{aff155}}
\and S.~J.~Liu\orcid{0000-0001-7680-2139}\inst{\ref{aff60}}
\and A.~Loureiro\orcid{0000-0002-4371-0876}\inst{\ref{aff156},\ref{aff157}}
\and J.~Macias-Perez\orcid{0000-0002-5385-2763}\inst{\ref{aff158}}
\and G.~Maggio\orcid{0000-0003-4020-4836}\inst{\ref{aff30}}
\and M.~Magliocchetti\orcid{0000-0001-9158-4838}\inst{\ref{aff60}}
\and F.~Mannucci\orcid{0000-0002-4803-2381}\inst{\ref{aff159}}
\and R.~Maoli\orcid{0000-0002-6065-3025}\inst{\ref{aff160},\ref{aff44}}
\and J.~Mart\'{i}n-Fleitas\orcid{0000-0002-8594-569X}\inst{\ref{aff92}}
\and C.~J.~A.~P.~Martins\orcid{0000-0002-4886-9261}\inst{\ref{aff161},\ref{aff35}}
\and L.~Maurin\orcid{0000-0002-8406-0857}\inst{\ref{aff24}}
\and M.~Miluzio\inst{\ref{aff25},\ref{aff162}}
\and P.~Monaco\orcid{0000-0003-2083-7564}\inst{\ref{aff134},\ref{aff30},\ref{aff31},\ref{aff29}}
\and C.~Moretti\orcid{0000-0003-3314-8936}\inst{\ref{aff32},\ref{aff125},\ref{aff30},\ref{aff29},\ref{aff31}}
\and G.~Morgante\inst{\ref{aff5}}
\and S.~Nadathur\orcid{0000-0001-9070-3102}\inst{\ref{aff141}}
\and K.~Naidoo\orcid{0000-0002-9182-1802}\inst{\ref{aff141}}
\and P.~Natoli\orcid{0000-0003-0126-9100}\inst{\ref{aff113},\ref{aff114}}
\and A.~Navarro-Alsina\orcid{0000-0002-3173-2592}\inst{\ref{aff86}}
\and S.~Nesseris\orcid{0000-0002-0567-0324}\inst{\ref{aff126}}
\and F.~Passalacqua\orcid{0000-0002-8606-4093}\inst{\ref{aff99},\ref{aff61}}
\and K.~Paterson\orcid{0000-0001-8340-3486}\inst{\ref{aff75}}
\and L.~Patrizii\inst{\ref{aff7}}
\and A.~Pisani\orcid{0000-0002-6146-4437}\inst{\ref{aff62},\ref{aff163}}
\and D.~Potter\orcid{0000-0002-0757-5195}\inst{\ref{aff153}}
\and S.~Quai\orcid{0000-0002-0449-8163}\inst{\ref{aff4},\ref{aff5}}
\and M.~Radovich\orcid{0000-0002-3585-866X}\inst{\ref{aff68}}
\and I.~Risso\orcid{0000-0003-2525-7761}\inst{\ref{aff164}}
\and P.-F.~Rocci\inst{\ref{aff24}}
\and S.~Sacquegna\orcid{0000-0002-8433-6630}\inst{\ref{aff121},\ref{aff122},\ref{aff123}}
\and M.~Sahl\'en\orcid{0000-0003-0973-4804}\inst{\ref{aff165}}
\and E.~Sarpa\orcid{0000-0002-1256-655X}\inst{\ref{aff32},\ref{aff125},\ref{aff31}}
\and A.~Schneider\orcid{0000-0001-7055-8104}\inst{\ref{aff153}}
\and M.~Schultheis\inst{\ref{aff88}}
\and D.~Sciotti\orcid{0009-0008-4519-2620}\inst{\ref{aff44},\ref{aff45}}
\and E.~Sellentin\inst{\ref{aff166},\ref{aff42}}
\and M.~Sereno\orcid{0000-0003-0302-0325}\inst{\ref{aff5},\ref{aff7}}
\and L.~C.~Smith\orcid{0000-0002-3259-2771}\inst{\ref{aff167}}
\and J.~Stadel\orcid{0000-0001-7565-8622}\inst{\ref{aff153}}
\and K.~Tanidis\orcid{0000-0001-9843-5130}\inst{\ref{aff117}}
\and C.~Tao\orcid{0000-0001-7961-8177}\inst{\ref{aff62}}
\and G.~Testera\inst{\ref{aff34}}
\and R.~Teyssier\orcid{0000-0001-7689-0933}\inst{\ref{aff163}}
\and S.~Tosi\orcid{0000-0002-7275-9193}\inst{\ref{aff33},\ref{aff164}}
\and A.~Troja\orcid{0000-0003-0239-4595}\inst{\ref{aff99},\ref{aff61}}
\and M.~Tucci\inst{\ref{aff59}}
\and C.~Valieri\inst{\ref{aff7}}
\and A.~Venhola\orcid{0000-0001-6071-4564}\inst{\ref{aff168}}
\and D.~Vergani\orcid{0000-0003-0898-2216}\inst{\ref{aff5}}
\and G.~Vernardos\orcid{0000-0001-8554-7248}\inst{\ref{aff169},\ref{aff170}}
\and G.~Verza\orcid{0000-0002-1886-8348}\inst{\ref{aff171}}
\and P.~Vielzeuf\orcid{0000-0003-2035-9339}\inst{\ref{aff62}}
\and N.~A.~Walton\orcid{0000-0003-3983-8778}\inst{\ref{aff167}}}
                                                                                   
\institute{INAF-Osservatorio Astronomico di Capodimonte, Via Moiariello 16, 80131 Napoli, Italy\label{aff1}
\and
Department of Physics "E. Pancini", University Federico II, Via Cinthia 6, 80126, Napoli, Italy\label{aff2}
\and
INFN section of Naples, Via Cinthia 6, 80126, Napoli, Italy\label{aff3}
\and
Dipartimento di Fisica e Astronomia "Augusto Righi" - Alma Mater Studiorum Universit\`a di Bologna, via Piero Gobetti 93/2, 40129 Bologna, Italy\label{aff4}
\and
INAF-Osservatorio di Astrofisica e Scienza dello Spazio di Bologna, Via Piero Gobetti 93/3, 40129 Bologna, Italy\label{aff5}
\and
School of Mathematics, Statistics and Physics, Newcastle University, Herschel Building, Newcastle-upon-Tyne, NE1 7RU, UK\label{aff6}
\and
INFN-Sezione di Bologna, Viale Berti Pichat 6/2, 40127 Bologna, Italy\label{aff7}
\and
CEA Saclay, DFR/IRFU, Service d'Astrophysique, Bat. 709, 91191 Gif-sur-Yvette, France\label{aff8}
\and
Dipartimento di Fisica e Astronomia, Universit\`a di Bologna, Via Gobetti 93/2, 40129 Bologna, Italy\label{aff9}
\and
Aix-Marseille Universit\'e, CNRS, CNES, LAM, Marseille, France\label{aff10}
\and
Institut d'Astrophysique de Paris, UMR 7095, CNRS, and Sorbonne Universit\'e, 98 bis boulevard Arago, 75014 Paris, France\label{aff11}
\and
School of Physics and Astronomy, Sun Yat-sen University, Guangzhou 519082, Zhuhai Campus, China\label{aff12}
\and
Institude for Astrophysics, School of Physics, Zhengzhou University, Zhengzhou, 450001, China\label{aff13}
\and
Institute of Physics, Laboratory of Astrophysics, Ecole Polytechnique F\'ed\'erale de Lausanne (EPFL), Observatoire de Sauverny, 1290 Versoix, Switzerland\label{aff14}
\and
SCITAS, Ecole Polytechnique F\'ed\'erale de Lausanne (EPFL), 1015 Lausanne, Switzerland\label{aff15}
\and
Dipartimento di Fisica "E. Pancini", Universita degli Studi di Napoli Federico II, Via Cinthia 6, 80126, Napoli, Italy\label{aff16}
\and
Institut de Ci\`{e}ncies del Cosmos (ICCUB), Universitat de Barcelona (IEEC-UB), Mart\'{i} i Franqu\`{e}s 1, 08028 Barcelona, Spain\label{aff17}
\and
Instituci\'o Catalana de Recerca i Estudis Avan\c{c}ats (ICREA), Passeig de Llu\'{\i}s Companys 23, 08010 Barcelona, Spain\label{aff18}
\and
David A. Dunlap Department of Astronomy \& Astrophysics, University of Toronto, 50 St George Street, Toronto, Ontario M5S 3H4, Canada\label{aff19}
\and
Jodrell Bank Centre for Astrophysics, Department of Physics and Astronomy, University of Manchester, Oxford Road, Manchester M13 9PL, UK\label{aff20}
\and
School of Physical Sciences, The Open University, Milton Keynes, MK7 6AA, UK\label{aff21}
\and
Department of Physics and Astronomy, University of British Columbia, Vancouver, BC V6T 1Z1, Canada\label{aff22}
\and
Laboratoire d'etude de l'Univers et des phenomenes eXtremes, Observatoire de Paris, Universit\'e PSL, Sorbonne Universit\'e, CNRS, 92190 Meudon, France\label{aff23}
\and
Universit\'e Paris-Saclay, CNRS, Institut d'astrophysique spatiale, 91405, Orsay, France\label{aff24}
\and
ESAC/ESA, Camino Bajo del Castillo, s/n., Urb. Villafranca del Castillo, 28692 Villanueva de la Ca\~nada, Madrid, Spain\label{aff25}
\and
School of Mathematics and Physics, University of Surrey, Guildford, Surrey, GU2 7XH, UK\label{aff26}
\and
INAF-Osservatorio Astronomico di Brera, Via Brera 28, 20122 Milano, Italy\label{aff27}
\and
Universit\'e Paris-Saclay, Universit\'e Paris Cit\'e, CEA, CNRS, AIM, 91191, Gif-sur-Yvette, France\label{aff28}
\and
IFPU, Institute for Fundamental Physics of the Universe, via Beirut 2, 34151 Trieste, Italy\label{aff29}
\and
INAF-Osservatorio Astronomico di Trieste, Via G. B. Tiepolo 11, 34143 Trieste, Italy\label{aff30}
\and
INFN, Sezione di Trieste, Via Valerio 2, 34127 Trieste TS, Italy\label{aff31}
\and
SISSA, International School for Advanced Studies, Via Bonomea 265, 34136 Trieste TS, Italy\label{aff32}
\and
Dipartimento di Fisica, Universit\`a di Genova, Via Dodecaneso 33, 16146, Genova, Italy\label{aff33}
\and
INFN-Sezione di Genova, Via Dodecaneso 33, 16146, Genova, Italy\label{aff34}
\and
Instituto de Astrof\'isica e Ci\^encias do Espa\c{c}o, Universidade do Porto, CAUP, Rua das Estrelas, PT4150-762 Porto, Portugal\label{aff35}
\and
Faculdade de Ci\^encias da Universidade do Porto, Rua do Campo de Alegre, 4150-007 Porto, Portugal\label{aff36}
\and
Dipartimento di Fisica, Universit\`a degli Studi di Torino, Via P. Giuria 1, 10125 Torino, Italy\label{aff37}
\and
INFN-Sezione di Torino, Via P. Giuria 1, 10125 Torino, Italy\label{aff38}
\and
INAF-Osservatorio Astrofisico di Torino, Via Osservatorio 20, 10025 Pino Torinese (TO), Italy\label{aff39}
\and
European Space Agency/ESTEC, Keplerlaan 1, 2201 AZ Noordwijk, The Netherlands\label{aff40}
\and
Institute Lorentz, Leiden University, Niels Bohrweg 2, 2333 CA Leiden, The Netherlands\label{aff41}
\and
Leiden Observatory, Leiden University, Einsteinweg 55, 2333 CC Leiden, The Netherlands\label{aff42}
\and
INAF-IASF Milano, Via Alfonso Corti 12, 20133 Milano, Italy\label{aff43}
\and
INAF-Osservatorio Astronomico di Roma, Via Frascati 33, 00078 Monteporzio Catone, Italy\label{aff44}
\and
INFN-Sezione di Roma, Piazzale Aldo Moro, 2 - c/o Dipartimento di Fisica, Edificio G. Marconi, 00185 Roma, Italy\label{aff45}
\and
Centro de Investigaciones Energ\'eticas, Medioambientales y Tecnol\'ogicas (CIEMAT), Avenida Complutense 40, 28040 Madrid, Spain\label{aff46}
\and
Port d'Informaci\'{o} Cient\'{i}fica, Campus UAB, C. Albareda s/n, 08193 Bellaterra (Barcelona), Spain\label{aff47}
\and
Institute for Theoretical Particle Physics and Cosmology (TTK), RWTH Aachen University, 52056 Aachen, Germany\label{aff48}
\and
Institute for Astronomy, University of Hawaii, 2680 Woodlawn Drive, Honolulu, HI 96822, USA\label{aff49}
\and
Dipartimento di Fisica e Astronomia "Augusto Righi" - Alma Mater Studiorum Universit\`a di Bologna, Viale Berti Pichat 6/2, 40127 Bologna, Italy\label{aff50}
\and
Instituto de Astrof\'{\i}sica de Canarias, V\'{\i}a L\'actea, 38205 La Laguna, Tenerife, Spain\label{aff51}
\and
Institute for Astronomy, University of Edinburgh, Royal Observatory, Blackford Hill, Edinburgh EH9 3HJ, UK\label{aff52}
\and
European Space Agency/ESRIN, Largo Galileo Galilei 1, 00044 Frascati, Roma, Italy\label{aff53}
\and
Universit\'e Claude Bernard Lyon 1, CNRS/IN2P3, IP2I Lyon, UMR 5822, Villeurbanne, F-69100, France\label{aff54}
\and
UCB Lyon 1, CNRS/IN2P3, IUF, IP2I Lyon, 4 rue Enrico Fermi, 69622 Villeurbanne, France\label{aff55}
\and
Mullard Space Science Laboratory, University College London, Holmbury St Mary, Dorking, Surrey RH5 6NT, UK\label{aff56}
\and
Departamento de F\'isica, Faculdade de Ci\^encias, Universidade de Lisboa, Edif\'icio C8, Campo Grande, PT1749-016 Lisboa, Portugal\label{aff57}
\and
Instituto de Astrof\'isica e Ci\^encias do Espa\c{c}o, Faculdade de Ci\^encias, Universidade de Lisboa, Campo Grande, 1749-016 Lisboa, Portugal\label{aff58}
\and
Department of Astronomy, University of Geneva, ch. d'Ecogia 16, 1290 Versoix, Switzerland\label{aff59}
\and
INAF-Istituto di Astrofisica e Planetologia Spaziali, via del Fosso del Cavaliere, 100, 00100 Roma, Italy\label{aff60}
\and
INFN-Padova, Via Marzolo 8, 35131 Padova, Italy\label{aff61}
\and
Aix-Marseille Universit\'e, CNRS/IN2P3, CPPM, Marseille, France\label{aff62}
\and
Space Science Data Center, Italian Space Agency, via del Politecnico snc, 00133 Roma, Italy\label{aff63}
\and
INFN-Bologna, Via Irnerio 46, 40126 Bologna, Italy\label{aff64}
\and
School of Physics, HH Wills Physics Laboratory, University of Bristol, Tyndall Avenue, Bristol, BS8 1TL, UK\label{aff65}
\and
Universit\"ats-Sternwarte M\"unchen, Fakult\"at f\"ur Physik, Ludwig-Maximilians-Universit\"at M\"unchen, Scheinerstrasse 1, 81679 M\"unchen, Germany\label{aff66}
\and
Max Planck Institute for Extraterrestrial Physics, Giessenbachstr. 1, 85748 Garching, Germany\label{aff67}
\and
INAF-Osservatorio Astronomico di Padova, Via dell'Osservatorio 5, 35122 Padova, Italy\label{aff68}
\and
Institute of Theoretical Astrophysics, University of Oslo, P.O. Box 1029 Blindern, 0315 Oslo, Norway\label{aff69}
\and
Jet Propulsion Laboratory, California Institute of Technology, 4800 Oak Grove Drive, Pasadena, CA, 91109, USA\label{aff70}
\and
Department of Physics, Lancaster University, Lancaster, LA1 4YB, UK\label{aff71}
\and
Felix Hormuth Engineering, Goethestr. 17, 69181 Leimen, Germany\label{aff72}
\and
Technical University of Denmark, Elektrovej 327, 2800 Kgs. Lyngby, Denmark\label{aff73}
\and
Cosmic Dawn Center (DAWN), Denmark\label{aff74}
\and
Max-Planck-Institut f\"ur Astronomie, K\"onigstuhl 17, 69117 Heidelberg, Germany\label{aff75}
\and
NASA Goddard Space Flight Center, Greenbelt, MD 20771, USA\label{aff76}
\and
Department of Physics and Astronomy, University College London, Gower Street, London WC1E 6BT, UK\label{aff77}
\and
Department of Physics and Helsinki Institute of Physics, Gustaf H\"allstr\"omin katu 2, 00014 University of Helsinki, Finland\label{aff78}
\and
Universit\'e de Gen\`eve, D\'epartement de Physique Th\'eorique and Centre for Astroparticle Physics, 24 quai Ernest-Ansermet, CH-1211 Gen\`eve 4, Switzerland\label{aff79}
\and
Department of Physics, P.O. Box 64, 00014 University of Helsinki, Finland\label{aff80}
\and
Helsinki Institute of Physics, Gustaf H{\"a}llstr{\"o}min katu 2, University of Helsinki, Helsinki, Finland\label{aff81}
\and
Centre de Calcul de l'IN2P3/CNRS, 21 avenue Pierre de Coubertin 69627 Villeurbanne Cedex, France\label{aff82}
\and
SKA Observatory, Jodrell Bank, Lower Withington, Macclesfield, Cheshire SK11 9FT, UK\label{aff83}
\and
Dipartimento di Fisica "Aldo Pontremoli", Universit\`a degli Studi di Milano, Via Celoria 16, 20133 Milano, Italy\label{aff84}
\and
INFN-Sezione di Milano, Via Celoria 16, 20133 Milano, Italy\label{aff85}
\and
Universit\"at Bonn, Argelander-Institut f\"ur Astronomie, Auf dem H\"ugel 71, 53121 Bonn, Germany\label{aff86}
\and
Department of Physics, Institute for Computational Cosmology, Durham University, South Road, Durham, DH1 3LE, UK\label{aff87}
\and
Universit\'e C\^{o}te d'Azur, Observatoire de la C\^{o}te d'Azur, CNRS, Laboratoire Lagrange, Bd de l'Observatoire, CS 34229, 06304 Nice cedex 4, France\label{aff88}
\and
Universit\'e Paris Cit\'e, CNRS, Astroparticule et Cosmologie, 75013 Paris, France\label{aff89}
\and
CNRS-UCB International Research Laboratory, Centre Pierre Binetruy, IRL2007, CPB-IN2P3, Berkeley, USA\label{aff90}
\and
Institut d'Astrophysique de Paris, 98bis Boulevard Arago, 75014, Paris, France\label{aff91}
\and
Aurora Technology for European Space Agency (ESA), Camino bajo del Castillo, s/n, Urbanizacion Villafranca del Castillo, Villanueva de la Ca\~nada, 28692 Madrid, Spain\label{aff92}
\and
Institut de F\'{i}sica d'Altes Energies (IFAE), The Barcelona Institute of Science and Technology, Campus UAB, 08193 Bellaterra (Barcelona), Spain\label{aff93}
\and
DARK, Niels Bohr Institute, University of Copenhagen, Jagtvej 155, 2200 Copenhagen, Denmark\label{aff94}
\and
Centre National d'Etudes Spatiales -- Centre spatial de Toulouse, 18 avenue Edouard Belin, 31401 Toulouse Cedex 9, France\label{aff95}
\and
Institute of Space Science, Str. Atomistilor, nr. 409 M\u{a}gurele, Ilfov, 077125, Romania\label{aff96}
\and
Consejo Superior de Investigaciones Cientificas, Calle Serrano 117, 28006 Madrid, Spain\label{aff97}
\and
Universidad de La Laguna, Departamento de Astrof\'{\i}sica, 38206 La Laguna, Tenerife, Spain\label{aff98}
\and
Dipartimento di Fisica e Astronomia "G. Galilei", Universit\`a di Padova, Via Marzolo 8, 35131 Padova, Italy\label{aff99}
\and
Institut f\"ur Theoretische Physik, University of Heidelberg, Philosophenweg 16, 69120 Heidelberg, Germany\label{aff100}
\and
Institut de Recherche en Astrophysique et Plan\'etologie (IRAP), Universit\'e de Toulouse, CNRS, UPS, CNES, 14 Av. Edouard Belin, 31400 Toulouse, France\label{aff101}
\and
Universit\'e St Joseph; Faculty of Sciences, Beirut, Lebanon\label{aff102}
\and
Departamento de F\'isica, FCFM, Universidad de Chile, Blanco Encalada 2008, Santiago, Chile\label{aff103}
\and
Universit\"at Innsbruck, Institut f\"ur Astro- und Teilchenphysik, Technikerstr. 25/8, 6020 Innsbruck, Austria\label{aff104}
\and
Institut d'Estudis Espacials de Catalunya (IEEC),  Edifici RDIT, Campus UPC, 08860 Castelldefels, Barcelona, Spain\label{aff105}
\and
Satlantis, University Science Park, Sede Bld 48940, Leioa-Bilbao, Spain\label{aff106}
\and
Institute of Space Sciences (ICE, CSIC), Campus UAB, Carrer de Can Magrans, s/n, 08193 Barcelona, Spain\label{aff107}
\and
Instituto de Astrof\'isica e Ci\^encias do Espa\c{c}o, Faculdade de Ci\^encias, Universidade de Lisboa, Tapada da Ajuda, 1349-018 Lisboa, Portugal\label{aff108}
\and
Cosmic Dawn Center (DAWN)\label{aff109}
\and
Niels Bohr Institute, University of Copenhagen, Jagtvej 128, 2200 Copenhagen, Denmark\label{aff110}
\and
Universidad Polit\'ecnica de Cartagena, Departamento de Electr\'onica y Tecnolog\'ia de Computadoras,  Plaza del Hospital 1, 30202 Cartagena, Spain\label{aff111}
\and
Infrared Processing and Analysis Center, California Institute of Technology, Pasadena, CA 91125, USA\label{aff112}
\and
Dipartimento di Fisica e Scienze della Terra, Universit\`a degli Studi di Ferrara, Via Giuseppe Saragat 1, 44122 Ferrara, Italy\label{aff113}
\and
Istituto Nazionale di Fisica Nucleare, Sezione di Ferrara, Via Giuseppe Saragat 1, 44122 Ferrara, Italy\label{aff114}
\and
INAF, Istituto di Radioastronomia, Via Piero Gobetti 101, 40129 Bologna, Italy\label{aff115}
\and
Astronomical Observatory of the Autonomous Region of the Aosta Valley (OAVdA), Loc. Lignan 39, I-11020, Nus (Aosta Valley), Italy\label{aff116}
\and
Department of Physics, Oxford University, Keble Road, Oxford OX1 3RH, UK\label{aff117}
\and
Instituto de Astrof\'isica de Canarias (IAC); Departamento de Astrof\'isica, Universidad de La Laguna (ULL), 38200, La Laguna, Tenerife, Spain\label{aff118}
\and
Universit\'e PSL, Observatoire de Paris, Sorbonne Universit\'e, CNRS, LERMA, 75014, Paris, France\label{aff119}
\and
Universit\'e Paris-Cit\'e, 5 Rue Thomas Mann, 75013, Paris, France\label{aff120}
\and
Department of Mathematics and Physics E. De Giorgi, University of Salento, Via per Arnesano, CP-I93, 73100, Lecce, Italy\label{aff121}
\and
INFN, Sezione di Lecce, Via per Arnesano, CP-193, 73100, Lecce, Italy\label{aff122}
\and
INAF-Sezione di Lecce, c/o Dipartimento Matematica e Fisica, Via per Arnesano, 73100, Lecce, Italy\label{aff123}
\and
ICL, Junia, Universit\'e Catholique de Lille, LITL, 59000 Lille, France\label{aff124}
\and
ICSC - Centro Nazionale di Ricerca in High Performance Computing, Big Data e Quantum Computing, Via Magnanelli 2, Bologna, Italy\label{aff125}
\and
Instituto de F\'isica Te\'orica UAM-CSIC, Campus de Cantoblanco, 28049 Madrid, Spain\label{aff126}
\and
CERCA/ISO, Department of Physics, Case Western Reserve University, 10900 Euclid Avenue, Cleveland, OH 44106, USA\label{aff127}
\and
Technical University of Munich, TUM School of Natural Sciences, Physics Department, James-Franck-Str.~1, 85748 Garching, Germany\label{aff128}
\and
Max-Planck-Institut f\"ur Astrophysik, Karl-Schwarzschild-Str.~1, 85748 Garching, Germany\label{aff129}
\and
Laboratoire Univers et Th\'eorie, Observatoire de Paris, Universit\'e PSL, Universit\'e Paris Cit\'e, CNRS, 92190 Meudon, France\label{aff130}
\and
Departamento de F{\'\i}sica Fundamental. Universidad de Salamanca. Plaza de la Merced s/n. 37008 Salamanca, Spain\label{aff131}
\and
Universit\'e de Strasbourg, CNRS, Observatoire astronomique de Strasbourg, UMR 7550, 67000 Strasbourg, France\label{aff132}
\and
Center for Data-Driven Discovery, Kavli IPMU (WPI), UTIAS, The University of Tokyo, Kashiwa, Chiba 277-8583, Japan\label{aff133}
\and
Dipartimento di Fisica - Sezione di Astronomia, Universit\`a di Trieste, Via Tiepolo 11, 34131 Trieste, Italy\label{aff134}
\and
California Institute of Technology, 1200 E California Blvd, Pasadena, CA 91125, USA\label{aff135}
\and
Department of Physics \& Astronomy, University of California Irvine, Irvine CA 92697, USA\label{aff136}
\and
Kapteyn Astronomical Institute, University of Groningen, PO Box 800, 9700 AV Groningen, The Netherlands\label{aff137}
\and
Departamento F\'isica Aplicada, Universidad Polit\'ecnica de Cartagena, Campus Muralla del Mar, 30202 Cartagena, Murcia, Spain\label{aff138}
\and
Instituto de F\'isica de Cantabria, Edificio Juan Jord\'a, Avenida de los Castros, 39005 Santander, Spain\label{aff139}
\and
Observatorio Nacional, Rua General Jose Cristino, 77-Bairro Imperial de Sao Cristovao, Rio de Janeiro, 20921-400, Brazil\label{aff140}
\and
Institute of Cosmology and Gravitation, University of Portsmouth, Portsmouth PO1 3FX, UK\label{aff141}
\and
Department of Computer Science, Aalto University, PO Box 15400, Espoo, FI-00 076, Finland\label{aff142}
\and
Instituto de Astrof\'\i sica de Canarias, c/ Via Lactea s/n, La Laguna 38200, Spain. Departamento de Astrof\'\i sica de la Universidad de La Laguna, Avda. Francisco Sanchez, La Laguna, 38200, Spain\label{aff143}
\and
Caltech/IPAC, 1200 E. California Blvd., Pasadena, CA 91125, USA\label{aff144}
\and
Ruhr University Bochum, Faculty of Physics and Astronomy, Astronomical Institute (AIRUB), German Centre for Cosmological Lensing (GCCL), 44780 Bochum, Germany\label{aff145}
\and
Department of Physics and Astronomy, Vesilinnantie 5, 20014 University of Turku, Finland\label{aff146}
\and
Serco for European Space Agency (ESA), Camino bajo del Castillo, s/n, Urbanizacion Villafranca del Castillo, Villanueva de la Ca\~nada, 28692 Madrid, Spain\label{aff147}
\and
ARC Centre of Excellence for Dark Matter Particle Physics, Melbourne, Australia\label{aff148}
\and
Centre for Astrophysics \& Supercomputing, Swinburne University of Technology,  Hawthorn, Victoria 3122, Australia\label{aff149}
\and
Department of Physics and Astronomy, University of the Western Cape, Bellville, Cape Town, 7535, South Africa\label{aff150}
\and
DAMTP, Centre for Mathematical Sciences, Wilberforce Road, Cambridge CB3 0WA, UK\label{aff151}
\and
Kavli Institute for Cosmology Cambridge, Madingley Road, Cambridge, CB3 0HA, UK\label{aff152}
\and
Department of Astrophysics, University of Zurich, Winterthurerstrasse 190, 8057 Zurich, Switzerland\label{aff153}
\and
Department of Physics, Centre for Extragalactic Astronomy, Durham University, South Road, Durham, DH1 3LE, UK\label{aff154}
\and
IRFU, CEA, Universit\'e Paris-Saclay 91191 Gif-sur-Yvette Cedex, France\label{aff155}
\and
Oskar Klein Centre for Cosmoparticle Physics, Department of Physics, Stockholm University, Stockholm, SE-106 91, Sweden\label{aff156}
\and
Astrophysics Group, Blackett Laboratory, Imperial College London, London SW7 2AZ, UK\label{aff157}
\and
Univ. Grenoble Alpes, CNRS, Grenoble INP, LPSC-IN2P3, 53, Avenue des Martyrs, 38000, Grenoble, France\label{aff158}
\and
INAF-Osservatorio Astrofisico di Arcetri, Largo E. Fermi 5, 50125, Firenze, Italy\label{aff159}
\and
Dipartimento di Fisica, Sapienza Universit\`a di Roma, Piazzale Aldo Moro 2, 00185 Roma, Italy\label{aff160}
\and
Centro de Astrof\'{\i}sica da Universidade do Porto, Rua das Estrelas, 4150-762 Porto, Portugal\label{aff161}
\and
HE Space for European Space Agency (ESA), Camino bajo del Castillo, s/n, Urbanizacion Villafranca del Castillo, Villanueva de la Ca\~nada, 28692 Madrid, Spain\label{aff162}
\and
Department of Astrophysical Sciences, Peyton Hall, Princeton University, Princeton, NJ 08544, USA\label{aff163}
\and
INAF-Osservatorio Astronomico di Brera, Via Brera 28, 20122 Milano, Italy, and INFN-Sezione di Genova, Via Dodecaneso 33, 16146, Genova, Italy\label{aff164}
\and
Theoretical astrophysics, Department of Physics and Astronomy, Uppsala University, Box 515, 751 20 Uppsala, Sweden\label{aff165}
\and
Mathematical Institute, University of Leiden, Einsteinweg 55, 2333 CA Leiden, The Netherlands\label{aff166}
\and
Institute of Astronomy, University of Cambridge, Madingley Road, Cambridge CB3 0HA, UK\label{aff167}
\and
Space physics and astronomy research unit, University of Oulu, Pentti Kaiteran katu 1, FI-90014 Oulu, Finland\label{aff168}
\and
Department of Physics and Astronomy, Lehman College of the CUNY, Bronx, NY 10468, USA\label{aff169}
\and
American Museum of Natural History, Department of Astrophysics, New York, NY 10024, USA\label{aff170}
\and
Center for Computational Astrophysics, Flatiron Institute, 162 5th Avenue, 10010, New York, NY, USA\label{aff171}}
%
%
\abstract{The \Euclid mission aims to survey around \num{14000} $\deg^{2}$ of extragalactic sky, providing around $10^{5}$ gravitational lens images. Modelling of gravitational lenses is fundamental to estimate the total mass of the lens galaxy, along with its dark matter content. Traditional modelling of gravitational lenses is computationally intensive and requires manual input. In this paper, we use a Bayesian neural network, LEns MOdelling with Neural networks (LEMON), to model \Euclid gravitational lenses with a singular isothermal ellipsoid mass profile. Our method estimates key lens mass profile parameters, such as the Einstein radius, while also predicting the light parameters of foreground galaxies and their uncertainties. We validate LEMON's performance on both mock \Euclid datasets, real lenses observed with \HST (HST) that have been degraded to match observations with the same depth of the \Euclid Wide Survey, and real \Euclid lenses, demonstrating the ability of LEMON to predict parameters of both simulated and real lenses. Results show promising accuracy and reliability in predicting the Einstein radius, mass and light ellipticities, effective radius, Sérsic index, lens magnitude, and unlensed source position for simulated lens galaxies. The application to real data, including the latest Quick Release 1 strong lens candidates, provides encouraging results in the recovery of the parameters for real lenses. We also verified that LEMON has the potential to accelerate traditional modelling methods, by giving to the classical optimiser the LEMON predictions as starting points, resulting in a speed-up of up to 26 times the original time needed to model a sample of gravitational lenses, a result that would be impossible with randomly initialised guesses. Moreover, LEMON can be used to cross-validate results from the traditional modelling methods, and thus has the potential to reduce the failure rate of the \Euclid modelling pipeline. This work represents a significant step towards efficient, automated gravitational lens modelling, which is crucial for handling the large data volumes expected from \Euclid.}
%
%
%
%
   \titlerunning{\Euclid Q1: LEMON on \Euclid data}
   \authorrunning{Euclid Consortium: V. Busillo et al.}
   
   \maketitle

\nolinenumbers

\section{Introduction}
Gravitational strong lensing is a rare astrophysical phenomenon that occurs when a massive object, such as a galaxy, bends the light coming from a background distant object. This effect is a prediction of Einstein's general relativity, and is an important tool in the study of galaxy evolution, given that a gravitational lens can magnify distant galaxies, allowing for the detailed study of structures in galaxies at high redshift \citep{Coe2013}. Moreover, the distortion of the background object gives information on the properties of the deflector. In particular, when the redshifts of source and lens are known, galaxy-galaxy strong lensing allows us to determine with extreme precision the projected total mass within the Einstein radius of the foreground galaxy, without requiring any modelling a priori. By assuming a model for the deflector mass profile, it is possible to infer more quantities related to the dark matter content of galaxies, such as the total dark matter mass and the dark matter fraction \citep{Koopmans2006, Tortora2010, Auger2010, Sengul2022}, along with other properties, such as the mass density slope and the initial mass function \citep{Gavazzi2007, Koopmans2009, Treu2010, Sonnenfeld2013, Shajib2021}; finally, through gravitational lensing, it is also possible to detect substructures of the lens galaxy \citep{Vegetti2010}. Galaxy-galaxy strong lensing is also a powerful probe of the small-scale structure of galaxy clusters \citep[e.g.][]{Meneghetti_2020,Meneghetti_2022,Meneghetti_2023}.

The \Euclid mission will observe around \num{14000} $\deg^{2}$ of extragalactic sky from the Sun-Earth Lagrange point L2, surveying of the order of $10^{9}$ galaxies (for a description of the Euclid Wide Survey, we refer the reader to \citealt{Scaramella-EP1, Mellier2024}, while an overall description of the mission can be found in \citealt{Laureijs11}). Following \Euclid's launch in 2023, ESA and collaborators initiated the Early Release Observations programme (ERO, \citealt{EROcite, Cuillandre+24_EROprogramme}), targeting 17 astronomical objects, including galaxy clusters, nearby galaxies, globular clusters, and star-forming regions, with the objective of demonstrating the telescope's capabilities and providing early scientific insights \citep{Atek2024, Cuillandre2024, ERONearbyGals,  Kluge2024, Marleau2024, Massari2024, Saifollahi2024}. Among these targets, the Perseus cluster field is especially of interest for strong lensing science, with two main projects underway: the ERO Lens Search Experiment (ELSE), in which a blind visual search of galaxy-scale strong lensing systems is carried out \citep{AcevedoBarroso24}, finding 16 lens candidates (of which five have a convincing lens model) and a parallel search of strong lenses via the use of neural networks \citep{Pearce-Casey24} aimed at testing their performance on a real \Euclid field. A similar search with neural networks in the other ERO fields is carried out in \cite{Nagam25}. The first statistically relevant sample of around $\num{500}$ strong lenses in \Euclid has finally been collected with the data from the Euclid Quick Data Release 1 (Q1, \citealt{Q1cite, Q1-TP001}), which also represents the first publicly available dataset of Euclid Wide Survey (EWS)-like images, processed with the same pipeline and reaching the same depth as the final survey. Despite covering only \num{63.1} $\deg^{2}$, Q1 provides a powerful testbed for \Euclid's lens searching capabilities, with the strong lensing discovery engine at the forefront of these efforts \citep{Q1-SP048, Q1-SP052, Q1-SP053, Q1-SP054, Q1-SP059}.

High-resolution imaging, together with a large field of view, is crucial for detecting a large number of gravitational lenses. However, the substantial amount of data coming from the mission will necessarily require automated methods for both lens finding and lens modelling. Convolutional neural networks (CNNs) are already used extensively in lens-finding tasks, and generally perform well \citep{Petrillo+17_CNN,Petrillo+19_CNN,Petrillo+19_LinKS,Metcalf2019, canameras2020holismokes, Li2020, Rezaei2022, Rojas+22, nagam2}. However, modelling of gravitational lenses has traditionally relied on complex and time-consuming techniques that require manual inputs, such as maximum likelihood and Markov chain Monte Carlo methods \citep{Keeton2016, Nightingale2021_PyAutoFit, Nightingale2021, Harvey2024}. The advent of machine learning has allowed for the automation and acceleration of this process. In particular, CNNs and Bayesian neural networks (BNNs) have shown remarkable results in recovering lens parameters \citep{Perreault2017, Pearson2019, Pearson2021, Schuldt2021, Schuldt2023, Gentile2023}. A recent work by \citet{Park2021} demonstrated that BNNs can successfully recover the posterior probability density functions of lens model parameters with sufficient accuracy for time delay cosmography, achieving sub-percent precision on the Hubble constant when applied to ensembles of lensed quasars. Complementary approaches using simulation-based inference, such as neural posterior estimation, have shown even better uncertainty calibration and accuracy compared to BNNs for strong lens modelling on ground-based data \citep{Poh2022}. Furthermore, \citet{Legin2022} introduced an efficient hierarchical inference framework that corrects for selection biases in strong lensing surveys using neural networks, allowing for fast and unbiased inference of population-level lens parameters.

The main algorithm in this work is LEns MOdelling with Neural networks (LEMON; \citealt{Gentile2023}), a BNN able to perform fast automated analysis of strong gravitational lenses. In the original work, \cite{Gentile2023} used LEMON to estimate three parameters of a strong gravitational lens modelled as a singular isothermal ellipsoid (SIE), the Einstein radius and the two components of the ellipticity, for mock \textit{Hubble} Space Telescope (HST) and \Euclid gravitational lenses. Along with the expected values, LEMON is also able to output uncertainties for each parameter.

In this work, we expand the results of \cite{Gentile2023}, by predicting both mass and light parameters of the foreground galaxies in isolated galaxy-galaxy gravitational lenses, considering a single Sérsic model for the light profile and using mock \Euclid lenses with contaminants in the image for both training and testing. We also verify the behaviour of both uncertainty components and verify the capability of LEMON to generalise to real Euclidised HST lens images.

The paper is structured as follows. In Sect. \ref{sec:Data} we give details on the various datasets used for the analysis. In Sect. \ref{sec:LEMON_introduction} we briefly introduce the LEMON algorithm. In  Sect. \ref{sec:Methods} we describe the training procedure, the metrics used to quantify the performance of LEMON, and the calibration procedure for the total uncertainty. Section \ref{sec:test_set_results} is dedicated to the results of the analysis on the simulated test set, while Sect. \ref{sec:real_lenses_results} shows the results associated with real Euclidised lenses and real \Euclid lenses found in the Perseus ERO and Q1 fields. Section \ref{sec:classical_methods_speed_up} shows how the joint use of LEMON and classical modelling methods could provide both a speed up of the \Euclid modelling pipeline and a reduction of its failure late. We finally give our conclusions in Sect. \ref{sec:Conclusions}.

\section{Data}\label{sec:Data}
In the following, we give details on the various datasets used for the analysis. In Sect. \ref{sec:Metcalf_lenses}, the \Euclid mock lenses used for training and testing are described. We then describe the real Euclidised lenses used for testing LEMON performance on real lenses in Sect. \ref{sec:real_lenses}.

\subsection{\Euclid mock lenses}\label{sec:Metcalf_lenses}

\begin{table}
    \centering
    \caption{Parameter ranges for the \Euclid mock lenses.}
    \label{tab:Metcalf_parameters}
    \begin{tabular}{cc}
    \hline
    \noalign{\vskip 2pt}
    \hline
         Parameters&Range\\
    \hline
         Einstein radius $R_{\textrm{Ein}}/\mathrm{arcsec}$ &$[0.51, 4.44]$\\
         $x$-ellipticity $\epsilon_{x}$   &$[-0.80,0.80]$\\ 
         $y$-ellipticity  $\epsilon_{y}$      &$[-0.80,0.80]$\\
         Lens effective radius $R_{\textrm{e, lens}}/\mathrm{arcsec}$&$[0.05,2.00]$\\
         Lens Sérsic index $n_{\mathrm{lens}}$&$[2.00, 8.00]$\\
         Lens magnitude $m_{\mathrm{lens}}$&$[13.95, 23.56]$\\
         Source $x$ position $x_{\textrm{src}}/\textrm{arcsec}$&$[-5.00,4.00]$\\
         Source $y$ position $y_{\textrm{src}}/\textrm{arcsec}$&$[-4.00,8.00]$\\
    \hline
    \end{tabular}
    \tablefoot{Only the Sérsic index was extracted from a uniform distribution. This choice was made to avoid introducing an informative prior for this parameter. All the other parameters were extracted from non-analytic, empirical distributions derived from the Flagship simulation \protect\citep{EuclidSkyFlagship}. The coverage of the parameters in the reported intervals is thus non-uniform. Notice that the intervals for the two ellipticity components are the same for both mass and light distributions, and as such have been reported together.}
\end{table}

\begin{figure*}
    \centering
    \includegraphics[width=1\linewidth]{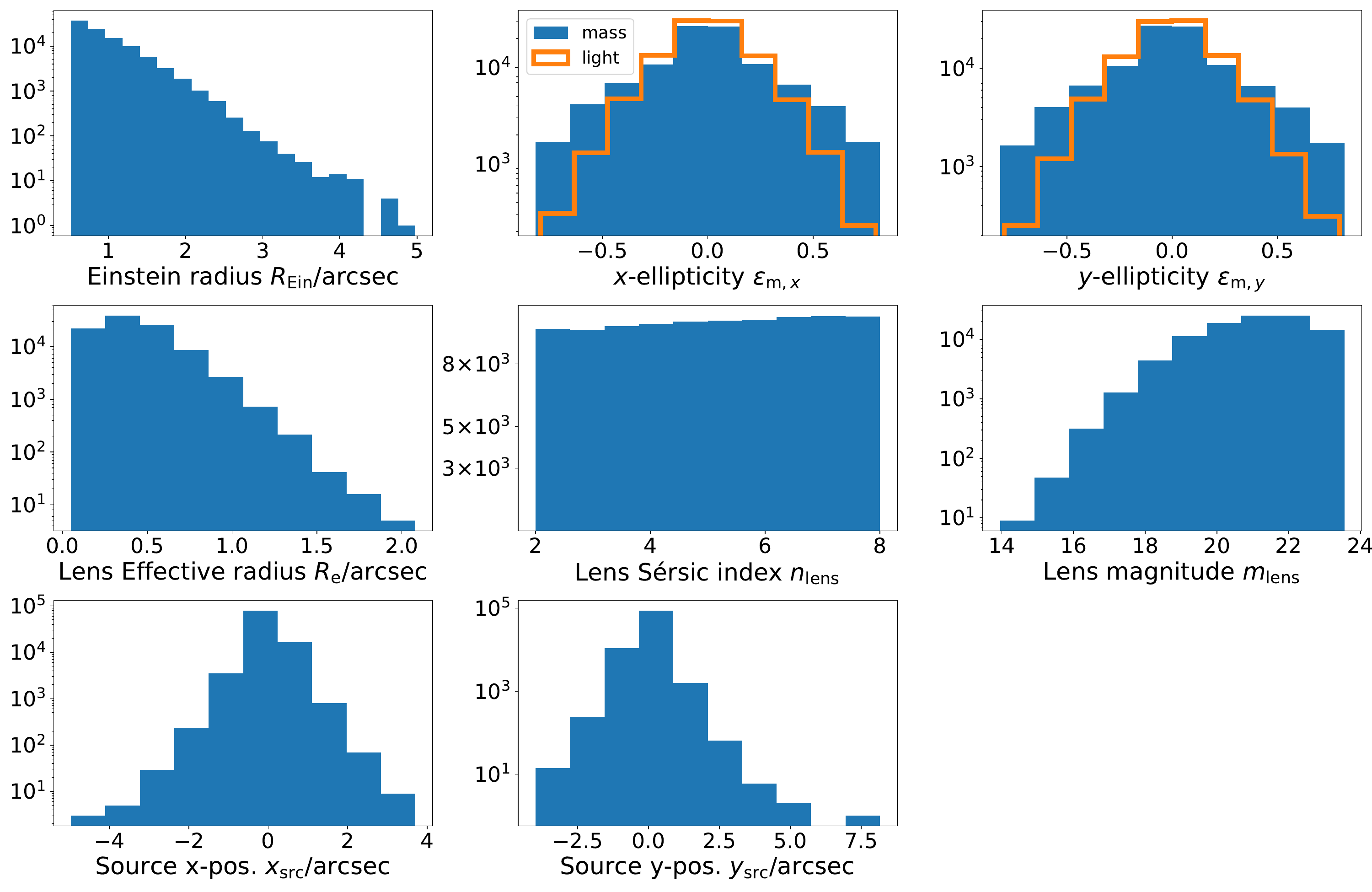}
    \caption{Distribution of mass and light parameters for the \Euclid mock lenses.}
    \label{fig:parameters_distribution}
\end{figure*}

For training and performance testing of LEMON, we used two simulation sets.
\begin{itemize}
    \item The first, or main dataset, consists of \num{100000} gravitational lenses, modelled with a SIE mass profile plus external shear, with $\gamma_{x,y}\in[-0.03, 0.03]$. In this set, the mass and light distributions are completely unaligned.
    \item The second dataset is identical to the main dataset, but it allows for a wider range of external shear components, $\gamma_{x,y}\in[-0.4, 0.4]$. This dataset was used to test LEMON's ability to recover the shear components. The network trained with this dataset shows performances comparable to the main network we used in the main text, while also managing to recover the shear parameters. More information are available in Appendix \ref{sec:shear_recovery}.
\end{itemize}

All the datasets are simulated by the Strong Lensing Science Working Group of the Euclid Consortium (Metcalf et al., in prep.). The lens images were made by simulating the \Euclid visible instrument (VIS, \citealt{Cropper2024}) camera output, at the depth of the EWS. The samples were selected to capture the full range of environmental scenarios that may occur when observing a gravitational lens, including objects along the line of sight as well as companion or background objects surrounding the lens system.

The lenses were constructed by first considering galaxies in the \Euclid Flagship simulation \citep{EuclidSkyFlagship}, taken from a sky area of the order of $100\,\textrm{deg}^{2}$, under the requirements that the galaxy chosen is brighter than the lens galaxy magnitude limit of 24 in the \IE band, which is the VIS imaging band, and categorised as a central galaxy. Central galaxies are galaxies that are positioned at the centres of the dark matter halos.  Large halos also contain satellite galaxies,  which should not produce isolated lenses like the ones we wish to model here. For each selected candidate, the apparent magnitude, effective radius, position angle, axis ratio and redshift were taken from the Flagship and converted into a surface brightness model. In general, the light model is composed of a disc and a spheroid component, but in our catalogue, only galaxies dominated by the spheroidal component are selected. This spheroid component is represented by a single Sérsic profile. The total flux and effective radius is taken from the Flagship simulation. The Sérsic index is uniformly distributed between 2 and 8. A mass model is then chosen by considering analytical profiles centred on each galaxy.

The mass in the lens was normalised with the ratio of dark matter to stellar mass within one effective radius. The stellar mass and effective radius were taken from the Flagship simulation.  The dark matter fraction within one effective radius was taken to be normally distributed with a mean of 0.6 and a standard deviation of 0.1, consistent with observations \citep{Mukherjee2022}.

Once the lens galaxies had been selected, a source galaxy was placed in the relative background. This increases the sample size of lenses in the dataset, in order to offset the inherent rarity of `natural lenses' (i.e. lenses in the Flagship simulation that naturally have a source behind them when the image is traced onto the observer plane). The redshift of the source was sampled randomly from the following distribution:
\begin{equation}
    p(z)\propto z^{\gamma} \textrm{e}^{-z^{2}/z_{0}^{2}}\,,
\end{equation}
where $\gamma = -0.23$ and $z_{0} = 3.06$ are values found by fitting to the COSMOS 2020 \texttt{Farmer} catalogue $I$-band number counts \citep{2022ApJS..258...11W}. The surface brightness is represented by between one and four Sérsic profiles.  
One profile has a brightness and size based on randomly selected sources from the Hubble Ultra Deep Field (HUDF) with similar redshifts (see \citealt{2008AandA…482..403M,2010AandA…514A..93M} for a discussion of this sample).  Sixty percent of the time additional smaller components are added.  They are positioned according to a Gaussian distribution centred on the main source with a scale size equal to the scale size of the ‘host’ source. Each component is given a random orientation, and axis ratio.  Their fluxes are drawn from a power-law distribution with index 4. This is a simplistic attempt to emulate some of the complex morphologies of high redshift sources.

Ray-shooting through the light cones was performed, to construct an image of the lens.  The deflection caused by other objects along the line of sight is not included, so that the mass distribution remains a SIE in all cases.  The lens was then rejected or accepted based on an observability criterion.  
This criterion seeks to filter out the many lenses candidates that would not be classified as lenses.  The procedure is as follows. 
The lens light was subtracted from the lensed source light.  Pixels that were greater than 4 times the background noise were considered to be in a source image.  This region was divided into disconnected images.  Images with a signal-to-noise (S/N) of less than 10 were discarded.  If there was more than one image remaining, the object was classified as a lens.  If there was one image, it could be an arc or a ring.  If there was a hole in the single image it was considered a lens.  Finally, if the image was crossed by a critical curve it was classified as a lens.  From the visual inspection of many simulated lenses, it was found that this criterion accepts almost all the lenses that an inspector would identify as a lens.

\begin{figure}
    \centering
    \includegraphics[width=1\linewidth]{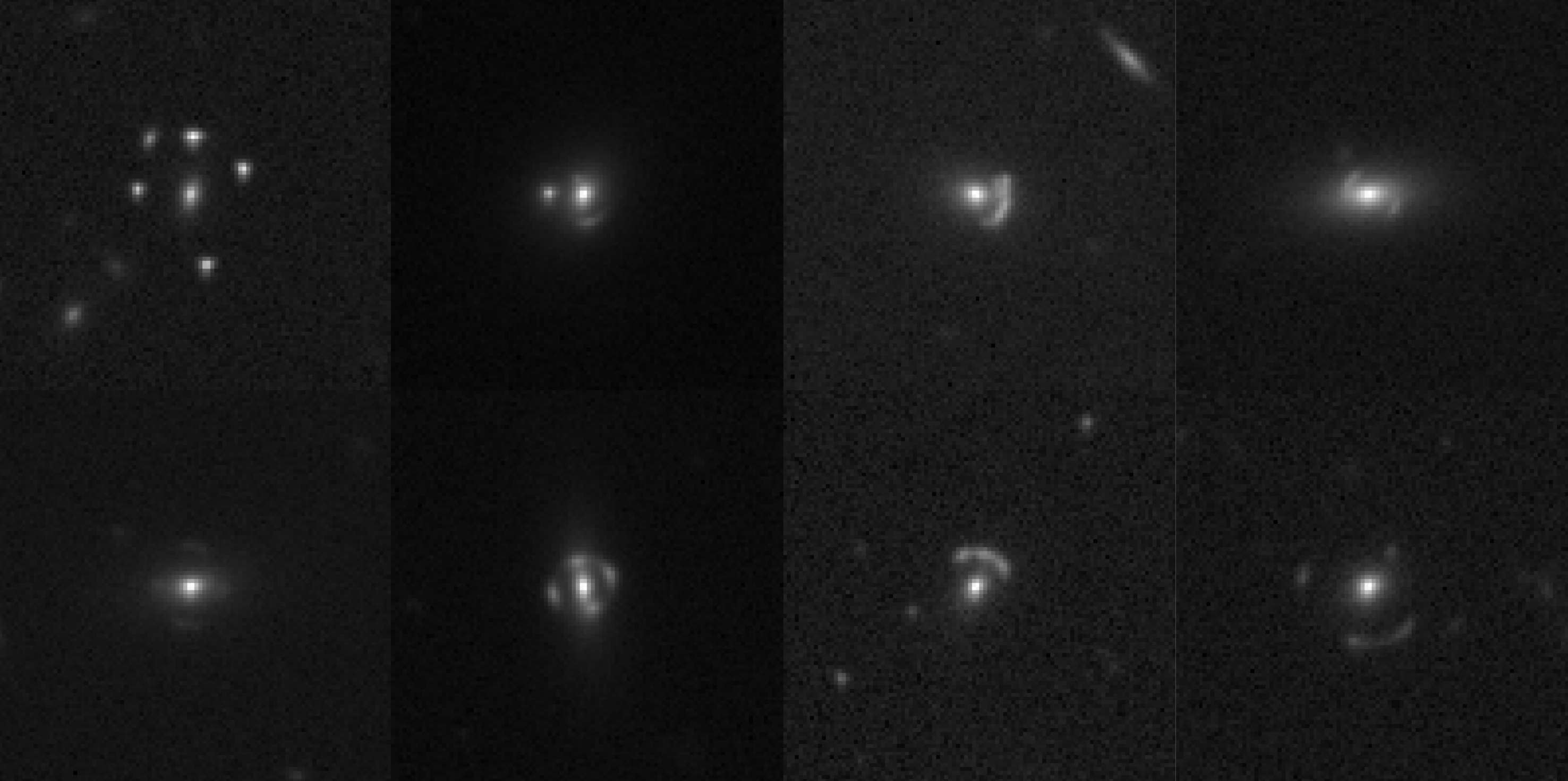}
    \caption{$20\arcsec \times 20\arcsec$ original VIS images of some examples of \Euclid mock lenses used for training LEMON.}
    \label{fig:train_lenses_example}
\end{figure}

A summary of the ranges of parameters for the mock lenses is listed in \Tab\ref{tab:Metcalf_parameters}, with the respective distributions shown in Fig. \ref{fig:parameters_distribution}. We underline the fact that, except for the Sérsic index, all the parameters are drawn from non-analytic, empirical distributions derived from the Flagship simulation, so the coverage for these parameters is not uniform. An example of some mock \Euclid lens images used for training is shown in Fig. \ref{fig:train_lenses_example}.

The eight parameters listed in \Tab\ref{tab:Metcalf_parameters} are the main ones that we want to predict with LEMON. The recovery of a ninth parameter, the Einstein mass, is discussed in Appendix \ref{sec:Einstein_mass}. They are listed as follows.

\begin{enumerate}
\item Einstein radius, in units of arcseconds, obtained via the formula
\begin{equation}
    R_{\textrm{Ein}} = \frac{\num{648000}}{\pi}\,\sqrt{\frac{A_{\textrm{Ein}}}{\pi}}\,,
\end{equation}
where $A_{\textrm{Ein}}$ is the Einstein area in square radians, defined as the area of the largest critical curve, and the factor \num{648000} corresponds to converting $\pi$ radians into arcseconds. Because $A_{\textrm{Ein}}$ is calculated numerically on a fixed resolution grid, it takes discrete values separated by 0.054 square arcseconds.

\item Ellipticity $x$ component, defined as
\begin{equation}
    \epsilon_{x} = \frac{1-q}{1+q}\cos(2\phi)\,,
\end{equation}
where $q = b/a$ is the axis ratio of the mass or light distribution, while $b$ and $a$ are, respectively, the semi-minor and semi-major axes of the respective ellipsoidal distribution. The angle $\phi$ is, instead, the position angle of the mass or light distribution, which indicates the angle between the semi-major axis of the ellipsoidal distribution and the positive $x$ axis of the standard Cartesian reference frame.
\item Ellipticity $y$ component, defined as
\begin{equation}
    \epsilon_{y} = \frac{1-q}{1+q}\sin(2\phi)\,,
\end{equation}
where $q$ and $\phi$ are defined as above.
\item Circularised effective radius of the lens, $R_{\textrm{e}}$, i.e. the radius enclosing one-half of the lens galaxy light, obtained from the major axis effective radius, $R_{\textrm{e, maj}}$, via the formula $R_{\textrm{e}} = \sqrt{q}\, R_{\textrm{e, maj}}$. We refer to it as $R_{\textrm{e, lens}}$ in the following.
\item Lens Sérsic index, which regulates the slope of the surface brightness profile. It corresponds to the parameter $n$ that appears in the Sérsic profile definition
\begin{equation}
    I(r)=I_{\textrm{e}}\exp\left\{ -b_{n}\left[\left(\frac{r}{R_{\textrm{e}}}\right)^{1/n}-1\right]\right\}\,,
\end{equation}
where $I_{\textrm{e}} = I(R_{\textrm{e}})$ and $b_{n}$ is a parameter, solution to the equation $ \Gamma(2n) = 2\,\gamma(2n,b_{n})$, where $\Gamma$ and $\gamma$ are the complete and incomplete gamma functions, respectively. We refer to it as $n_{\textrm{lens}}$ in the following.
\item Lens magnitude, defined as the apparent magnitude in the \IE band of the main lens. It should be noted that, due to the fact that input images are based on counts instead of fluxes, we need to take into account zero-point shifts when predicting magnitudes of non-simulated lenses.
\item The $x$ and $y$ position of the source (in arcseconds), defined as the unlensed $x$ and $y$ co-ordinates of the source’s brightest component, measured relative to the image centre.
\end{enumerate}

The original simulated images have dimensions of $20\arcsec \times 20\arcsec$, as is shown in Fig. \ref{fig:train_lenses_example}, but during the training of LEMON we create $10\arcsec \times 10\arcsec$ cut-outs of these images, centred on the lenses, in order to better see the lensing features.

\subsection{Euclidised lenses}\label{sec:real_lenses}
In addition to the set of simulated images, we tested the performance of LEMON using a sample of known strong gravitational lenses. These lenses were observed during several strong lensing surveys with HST or serendipitously discovered in the HST archive. Section \ref{sec:res_euclidezed} gives more details on the dataset. We used the code \HS\  \citep{EP-Bergamini} to Euclidise the HST images, mimicking \Euclid observations of these sources at the same depth and resolution of the EWS in the \IE band.

We summarise here the procedure implemented in \HS. For more details, we refer the reader to \cite{EP-Bergamini}. First, we converted the pixel values of the HST images from units of electrons per second (${\rm e}\,{\rm s}^{-1}$) to physical flux densities (${\rm erg}\,{\rm s}^{-1}\,{\rm cm}^{-2}\,{\rm Hz}^{-1}$). We performed this operation for all available HST images in photometric bands overlapping with the \IE passband. For the lenses in the dataset used in this paper, we used only HST observations with the Advanced Camera for Surveys (ACS) in the F814W band. Notice that the F814W band does not perfectly overlap with the \IE band. The Euclidisation procedure, however, assumes a constant spectral energy density (SED) for the lens galaxy, such that the magnitude in the \IE band is equal to the one in the F814W band. This means that there is no photon loss, even if the F814W band is narrower than the \IE band.

In the second step, we accounted for the \Euclid PSF. The \Euclid PSF used by \HS\ in this procedure coincides with the one used to produce the \Euclid mock lenses described in Sect. \ref{sec:Metcalf_lenses}, and was derived during the pre-launch VIS and NISP detector characterisation. This model matches well the true \Euclid PSFs. For more details on the PSF, we refer the reader to \cite{EP-Bergamini}. Since the input images were already convolved with the HST PSF, we did not convolve them directly with the \Euclid PSF model. Instead, we created a convolution kernel that matches the HST to the \Euclid PSF. Since we could not measure the PSF for each HST lens, we used a model for the ACS observations in the F814W band computed with the software \texttt{TinyTim}.\footnote{\url{https://www.stsci.edu/hst/instrumentation/focus-and-pointing/focus/tiny-tim-hst-psf-modeling}} The matched kernel for each pair of HST and \Euclid PSFs was computed with the \texttt{Python} function \texttt{create\_matching\_kernel} of the \texttt{Photutils} package\footnote{\url{https://photutils.readthedocs.io/en/stable}} \citep{photutils}. We then re-binned the images from the HST pixel grid to the correct \Euclid pixel scale. The resulting images have pixel scales of $100\,{\rm mas}\,{\rm px}^{-1}$ in the \IE\ band. The re-binned images, expressed in units of physical flux densities, were then converted into units of electrons per second using the \Euclid zero point.

Finally, we added Poisson photon noise. To accomplish this task, we assumed that the noise in the input HST background subtracted images is negligible compared to the noise in the \Euclid observations. This approximation is supported by the significantly greater depth of the HST observations compared with the \Euclid images. We adjusted the noise level to achieve the nominal S/N of 10 for an extended source of magnitude 24.5 within an aperture of \ang{;;0.65} radius expected for the EWS \citep[][]{Scaramella-EP1}.

\section{Introduction to LEMON}\label{sec:LEMON_introduction}
LEMON \citep{Gentile2023} is a BNN \citep{Charnock2020}, a machine-learning algorithm that employs the feature-recognition capabilities of a CNN to predict model parameters of a gravitational lens, using Bayesian statistics to estimate the respective uncertainties. The key point of LEMON is its ability, by using BNNs, to estimate the two main sources of uncertainty in machine-learning inference applications, namely epistemic uncertainty and aleatoric uncertainty.

\begin{itemize}
    \item Aleatoric uncertainty: also known as statistical uncertainty, is associated with the intrinsic quality of the data analysed by the algorithm. The most common sources of aleatoric uncertainty are a low S/N, corruptions in the images (e.g. masked regions), and source blending. This kind of uncertainty cannot be reduced with training, because it depends on the quality of the image being analysed.
    \item Epistemic uncertainty: also known as systematic uncertainty, it is caused by a lack of knowledge about the best model. It refers to the ignorance of the machine-learning algorithm about a certain region of the parameter space, and hence to the state of the machine rather than to the state of the images. This kind of uncertainty can be reduced with a higher completeness of the training set.
\end{itemize}

The aleatoric uncertainty for a certain image is given as an output by LEMON, together with the predicted parameter value (for details on how the aleatoric uncertainty is predicted by LEMON, refer to \citealt{Gentile2023}). The epistemic uncertainty for an image can be obtained thanks to the dropout layers of LEMON: by randomly switching off some connections between the neurons of the neural network, predicting repeatedly the parameters associated with a single image is equivalent to sampling from the posterior probability of the parameters. The epistemic uncertainty can then be evaluated by measuring the standard deviation of this sampled distribution.

The total uncertainty for a generic parameter, $p$, can be obtained via the following procedure \citep{Gentile2023}. For each image, LEMON predicts the mean value, $\overline{p}$, and the aleatoric uncertainty, $\sigma_{\textrm{A}}$, for $p$. A value $p_{i}$ is then sampled from a Gaussian distribution centred on $\overline{p}$ with a standard deviation equal to $\sigma_{\textrm{A}}$. This is repeated $\mathcal{N}$ times (we used $\mathcal{N} = 50$ to optimise for speed), sampling the posterior distribution of the parameter $p$. Finally, we compute the standard deviation of the distribution associated with the extractions of $p_{i}$, obtaining the total uncertainty.

LEMON employs the \texttt{ResNet} architecture \citep{He2015}, a deep residual network known for its success on computer vision tasks \citep{Russakovsky2015}. The architecture of LEMON is composed of: convolutional layers, used for feature extraction; dropout layers, to facilitate Bayesian uncertainty estimation through Monte Carlo sampling; finally, a fully connected layer made by $2N_{\textrm{param}}$ nodes, with $N_{\textrm{param}}$ number of parameters estimated by the network. A corresponding number of aleatoric uncertainties is estimated through the minimisation of the following Gaussian log-likelihood loss function:
\begin{equation}
    \ln\mathcal{L}(\textbf{y},\hat{\textbf{y}}) = \sum_{j=1}^{N_{\textrm{P}}}\left[-\frac{1}{2\sigma^{2}_{i,j}}\lVert y_{i,j}-\hat{y}_{i,j}\rVert-\frac{1}{2}\ln(\sigma^{2}_{i,j})\right]\,,\label{eq:LEMON_loss}
\end{equation}
where $\hat{\mathbf{y}}$ and $\mathbf{y}$ are the arrays of the predicted and the true parameter values, respectively,\footnote{A single predicted value $\hat{y}_{i}$ is the mean of the distribution of the $\mathcal{N}$  parameter extractions $p_{j}$.} while $\sigma_{i,j}$ is the aleatoric uncertainty predicted for the parameter $j$ of image $i$. The second term in \Eq\eqref{eq:LEMON_loss} is an additional regularisation term, introduced to prevent overly large or small variance estimates. Additional details on the implementation can be found in \cite{Gentile2023}.

\section{Methods}\label{sec:Methods}
In this section we first describe the training procedure for LEMON in Sect. \ref{sec:LEMON_training}. Section \ref{sec:metrics} is dedicated to the description of the metrics that we used to evaluate the performance of LEMON. Finally, Sect. \ref{sec:LEMON_calibration} is dedicated to discussing the calibration procedure for the total uncertainty.

\subsection{Training the network}\label{sec:LEMON_training}
We prepared the training of LEMON by taking the full sample of \num{100000} mock \Euclid lenses and splitting it into a training set formed of \num{80000} lenses, a validation set, and a test set formed by \num{10000} lenses each. We shuffled the three sets to avoid possible biases during learning. Unlike \cite{Gentile2023}, we used the \texttt{ResNet-50} architecture \citep{He2015} when implementing LEMON. We verified that this choice does not influence the results, confirming the analysis of \cite{Wagner-Carena2024}, which shows that marginal improvements over the \texttt{ResNet-34} architecture happen only at very large training set sizes. Before feeding the images into LEMON, we applied a pre-processing step whereby pixel values were first rescaled to the range [0,1], and then transformed using a square root scaling to enhance the visibility of low-intensity features.

Training was performed via stochastic gradient descent, with a batch size of 64 images. For the optimiser, we used ADAM \citep{Kingma2017}, with a starting learning rate of $10^{-4}$. To prevent overfitting, we used both the \texttt{ReduceLROnPlateau} and \texttt{EarlyStopping} callbacks of the \texttt{keras} \texttt{Python} library. The former reduces the learning rate of the optimiser when no improvement is seen for a given number of epochs, while the latter stops the training when the validation loss stops decreasing. Dropout layers, needed for estimating the epistemic uncertainty, are employed at both training and testing time, with a drop rate of 0.05. We performed the training via the \texttt{tensorflow} \texttt{Python} library \citep{Tensorflow2015}, using a single Nvidia RTX 4080 graphics processing unit (GPU).

\subsection{Metrics}\label{sec:metrics}
To evaluate the performance of LEMON in recovering the lens parameters, we made use of various statistical estimators.

\begin{itemize}
    \item Bias, defined as\footnote{The writing $\textrm{median}\{x_{i}|i\}$ denotes the median of the set $\{x_{i}|i=1,\dots,N\}$ of vector components $x_{i}$.}
        \begin{equation}
            \mu:=\textrm{median}\{\hat{y}_i-y_i|i\}\,.
        \end{equation}
This is an indicator of the accuracy of the prediction: higher bias values indicate a systematic overestimation of the predictions, and vice versa for lower bias values. For comparisons with forward modelling, we implement the uncertainty on the estimates by using the weighted version of the median, with weights equal to the reciprocal of the variance associated with the predictions.
\item Root mean square error (RMSE), defined as
\begin{equation}
\textrm{RMSE}:=\sqrt{\frac{1}{N}\sum_{i=1}^{N}\left(\hat{y_{i}}-y_{i}\right)^{2}}\,,
\end{equation}
where $\hat{y}_{i}$ and $y_{i}$ are the generic components of the $\hat{\mathbf{y}}$ and $\mathbf{y}$ arrays, respectively, which have length $N$. This metric is a measure of predictive power: a value of zero indicates a perfect fit to the true parameter values. For comparisons with forward modelling, we implement the uncertainty on the estimates by using the function \texttt{root\_mean\_squared\_error} from the \texttt{scikit-learn} Python library (which natively supports weights for the inputs), with weights equal to the reciprocal of the variance associated with the predictions.
\item Normalised median absolute deviation (NMAD), defined as
\begin{equation}
\textrm{NMAD}:=1.48\,\textrm{median}\{\left|(\hat{y}_i-y_i)-\textrm{median}\{\hat{y}_i-y_i|i\}\right|\,|i\}\,.
\end{equation}

\begin{figure*}
    \centering
    \includegraphics[width=1\linewidth]{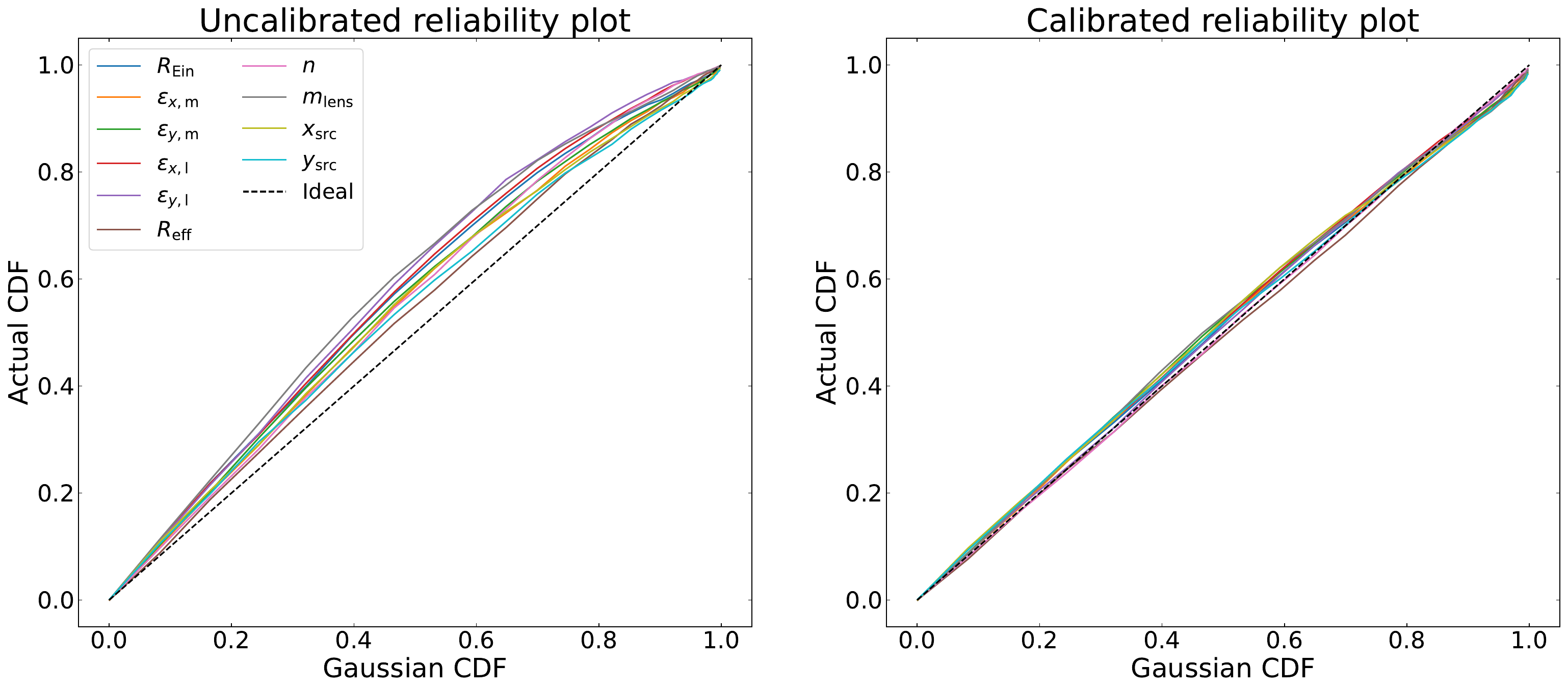}
    \caption{Reliability plots for the parameters predicted by LEMON. The lines represent, for each parameter, the discrepancy between the ideal cumulative statistical coverage from a Gaussian distribution ($x$ axis) and the corresponding empirical one ($y$ axis). The closer the curves are to the ideal relation (dashed black line), the more reliable the uncertainty estimates are.  \emph{Left}: Trends before the calibration procedure. \emph{Right}: Trends after the calibration procedure.}
    \label{fig:reliability_plot}
\end{figure*}

The NMAD is roughly analogous to the standard deviation ($\sigma \approx \kappa \, \textrm{MAD}$, where $\textrm{MAD}$ is the median absolute deviation and $\kappa$ is a constant scale factor, which is approximately equal to $1.48$ for normally distributed data). For comparisons with forward modelling, we substitute the median with the weighted median, with weights equal to the reciprocal of the variance associated with the predictions.

\item  Coefficient of determination, defined as
\begin{equation}
    R^{2}:=1-\frac{\sum_{i=1}^{N}\left(\hat{y}_{i}-y_{i}\right)^{2}}{\sum_{i=1}^{N}\left(y_{i}-\overline{y}\right)^{2}}\,,
\end{equation}
where
\begin{equation}
    \overline{y}=\frac{1}{N}\sum_{i=1}^{N}y_{i}
\end{equation}
is the mean value of the true parameter over all the lenses. This metric provides a measure of how well the parameters are predicted by LEMON. Given a fixed parameter, if it is perfectly predicted by the network, $\hat{y}_{i} = y_{i}$ for all the points of the dataset, and thus $R^{2} = 1$. A baseline network, which always outputs the mean value of the true parameter (i.e. $\hat{y}_{i} = \overline{y}$), will produce $R^{2} = 0$. If the network performs worse than the baseline model (i.e. the mean of the data provides a better fit than than the predicted values), the values of $R^{2}$ will be negative.\footnote{Notice that the coefficient of determination is not the square of a quantity $R$, and therefore can also assume negative values.} For comparisons with forward modelling, we implement the uncertainty on the estimates by using the function \texttt{r2\_score} from the \texttt{scikit-learn} Python library (which natively supports weights for the inputs), with weights equal to the reciprocal of the variance associated with the predictions.

\end{itemize}

\subsection{Calibration procedure}\label{sec:LEMON_calibration}
Uncertainties produced by BNNs are known to be affected by a systematic over- or under-estimation of the confidence intervals \citep{Guo2017}. Therefore, a calibration procedure for the uncertainties is required. Following \cite{Gentile2023}, we implement the Platt-scaling method \citep{Kull2017}, which consists of rescaling the predicted total uncertainty of a given parameter by a factor $s:\sigma \rightarrow s\sigma$, in order to match the empirical cumulative distribution function (CDF) with the CDF of a Gaussian distribution.

\begin{figure*}
    \centering
    \includegraphics[width=0.23\linewidth]{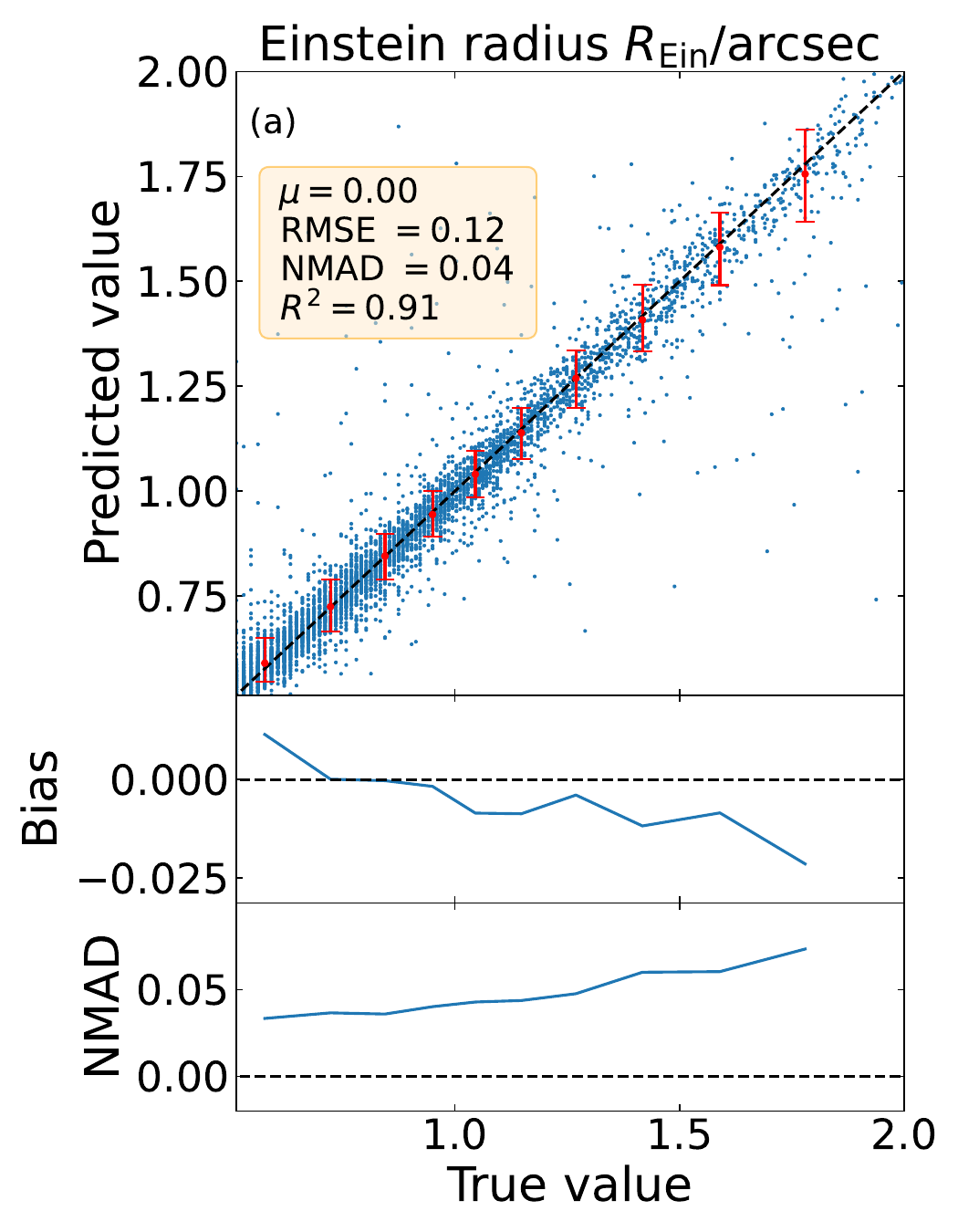}
    \includegraphics[width=0.23\linewidth]{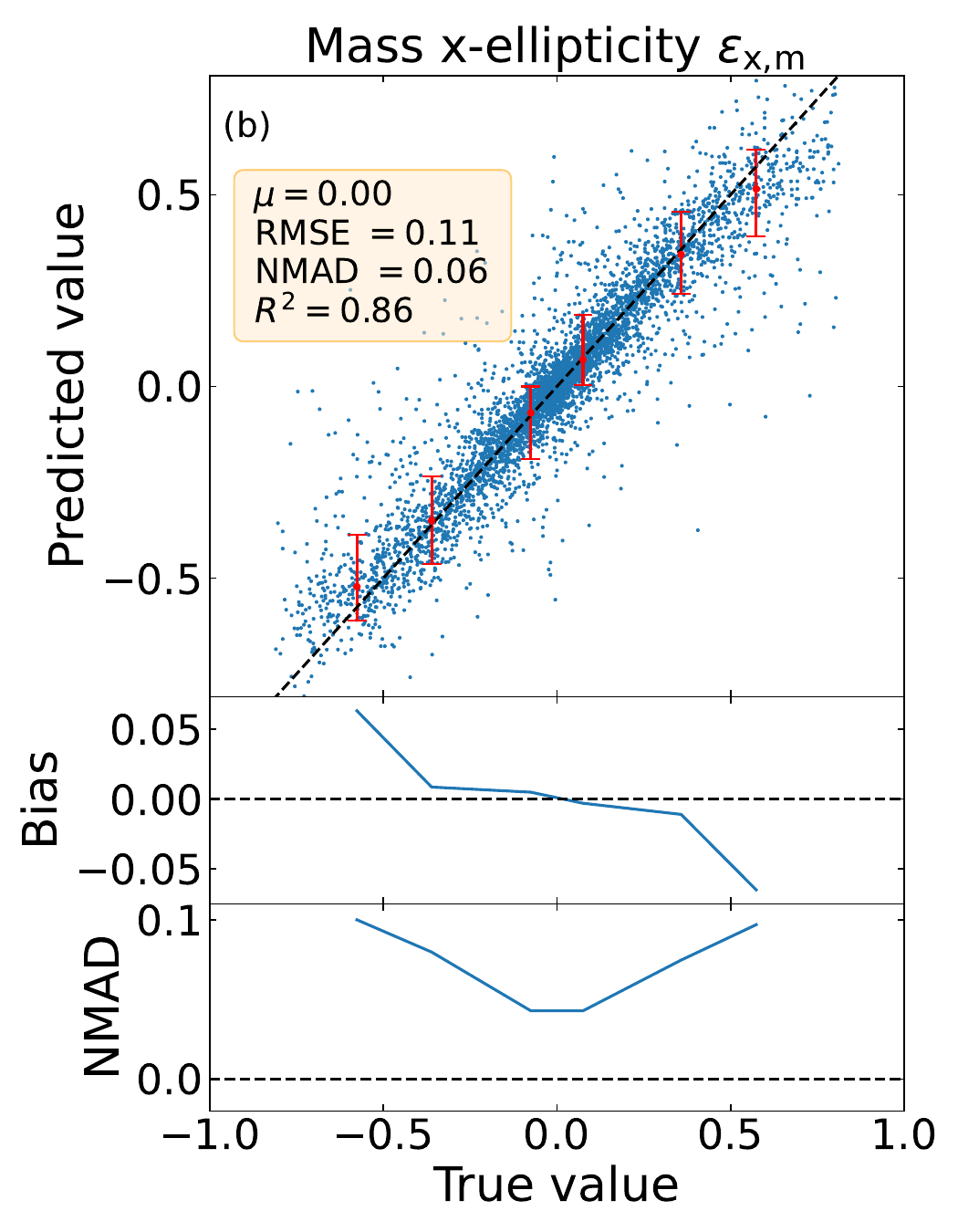}
    \includegraphics[width=0.23\linewidth]{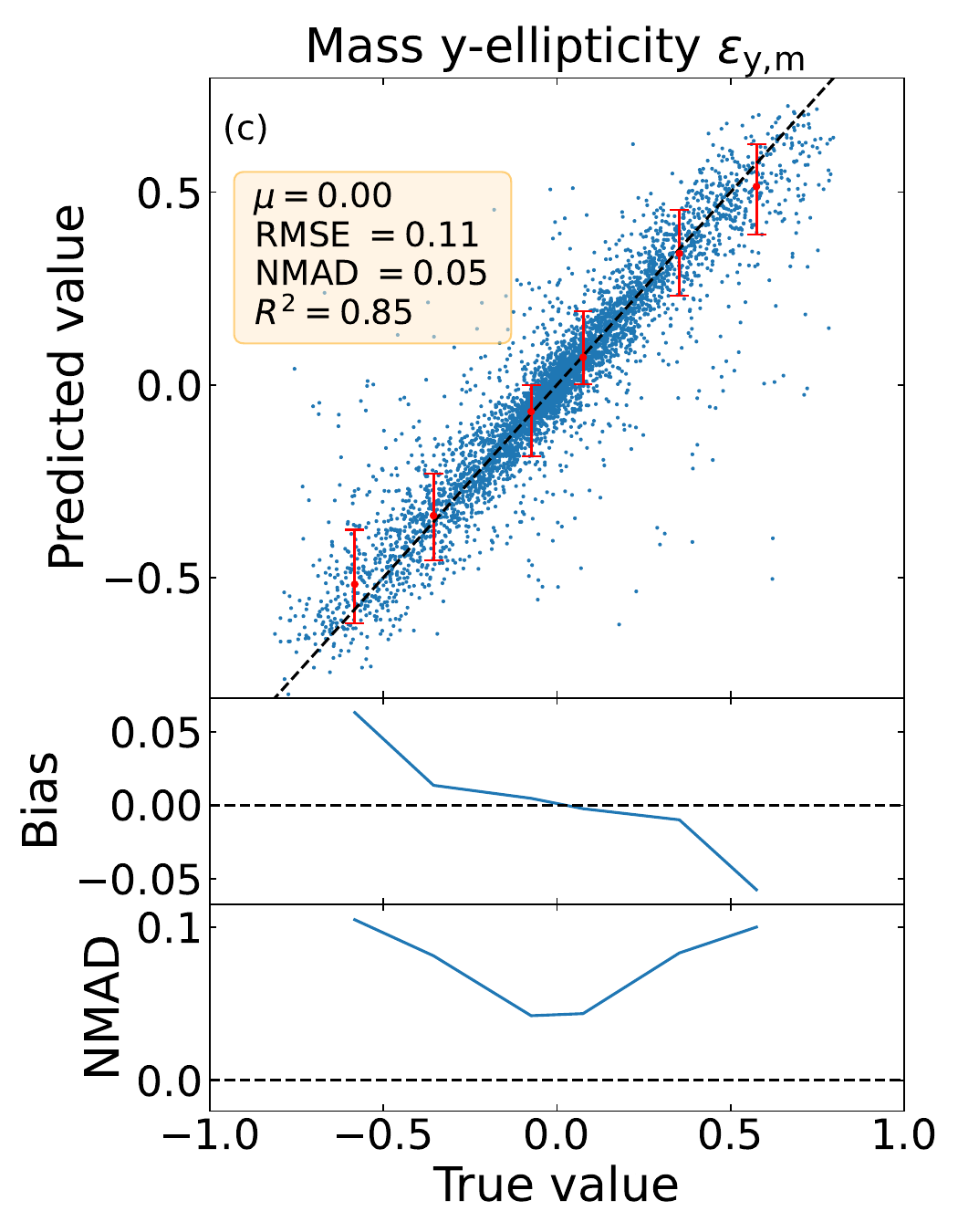}
    \includegraphics[width=0.23\linewidth]{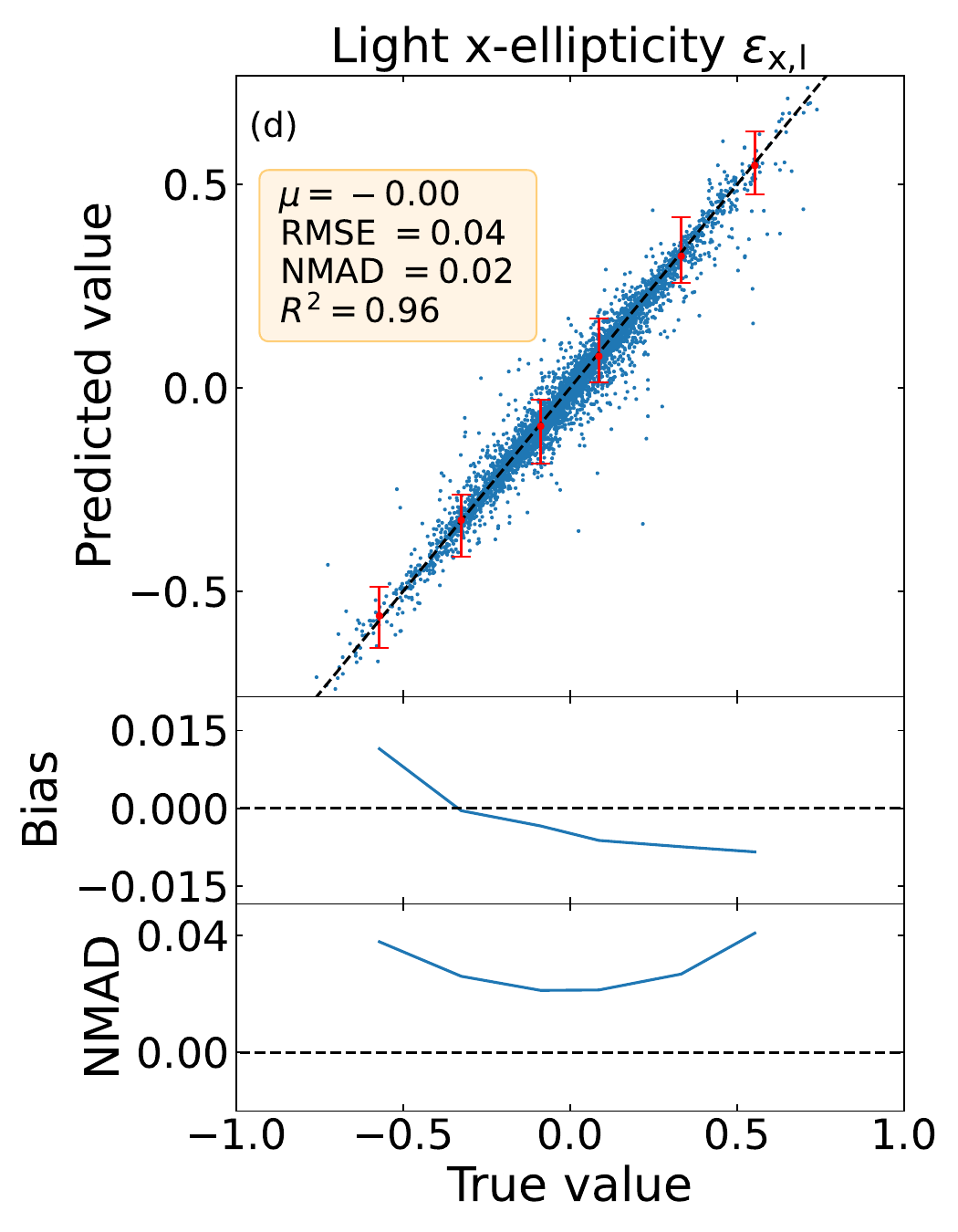}
\\
    \includegraphics[width=0.23\linewidth]{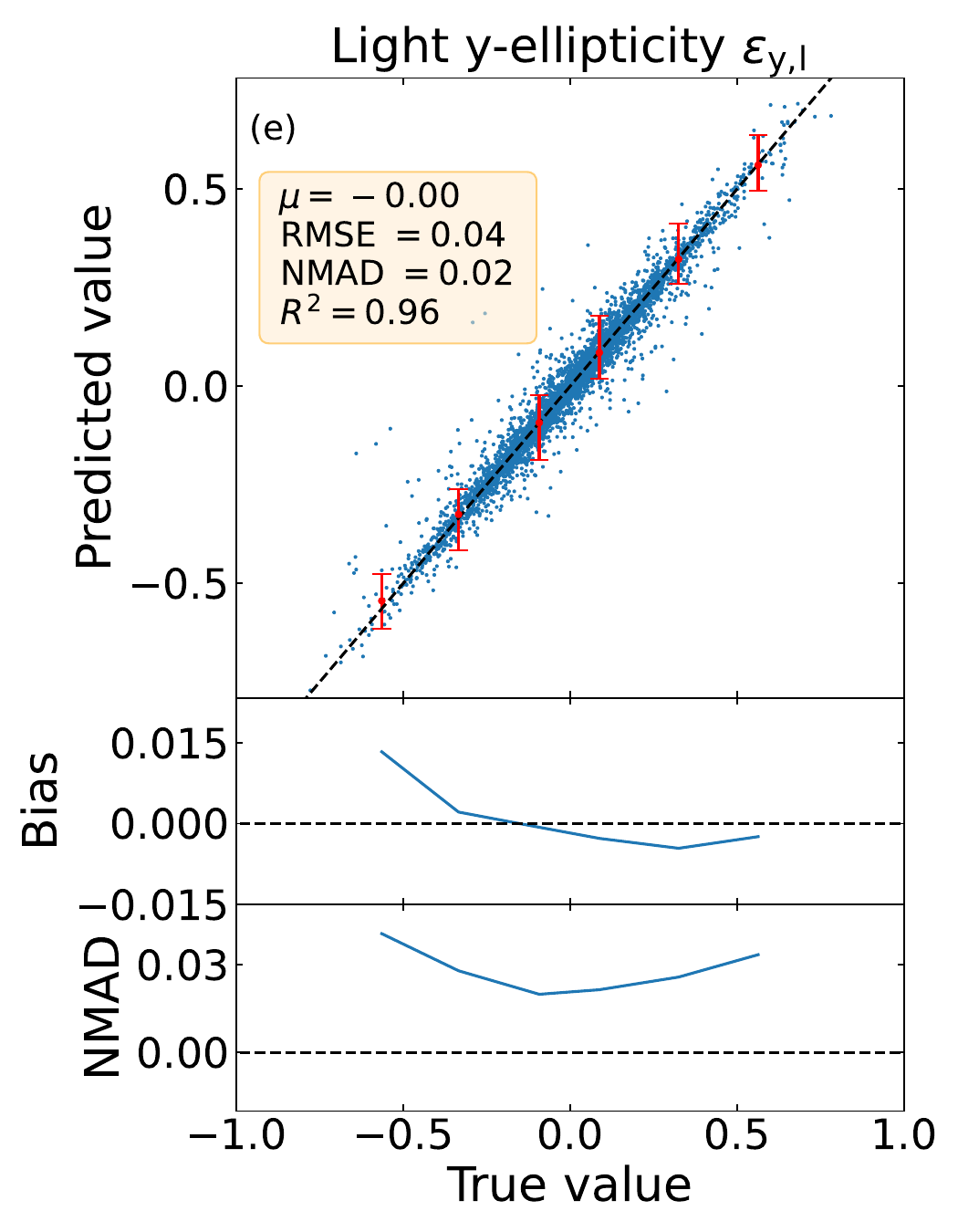}
    \includegraphics[width=0.23\linewidth]{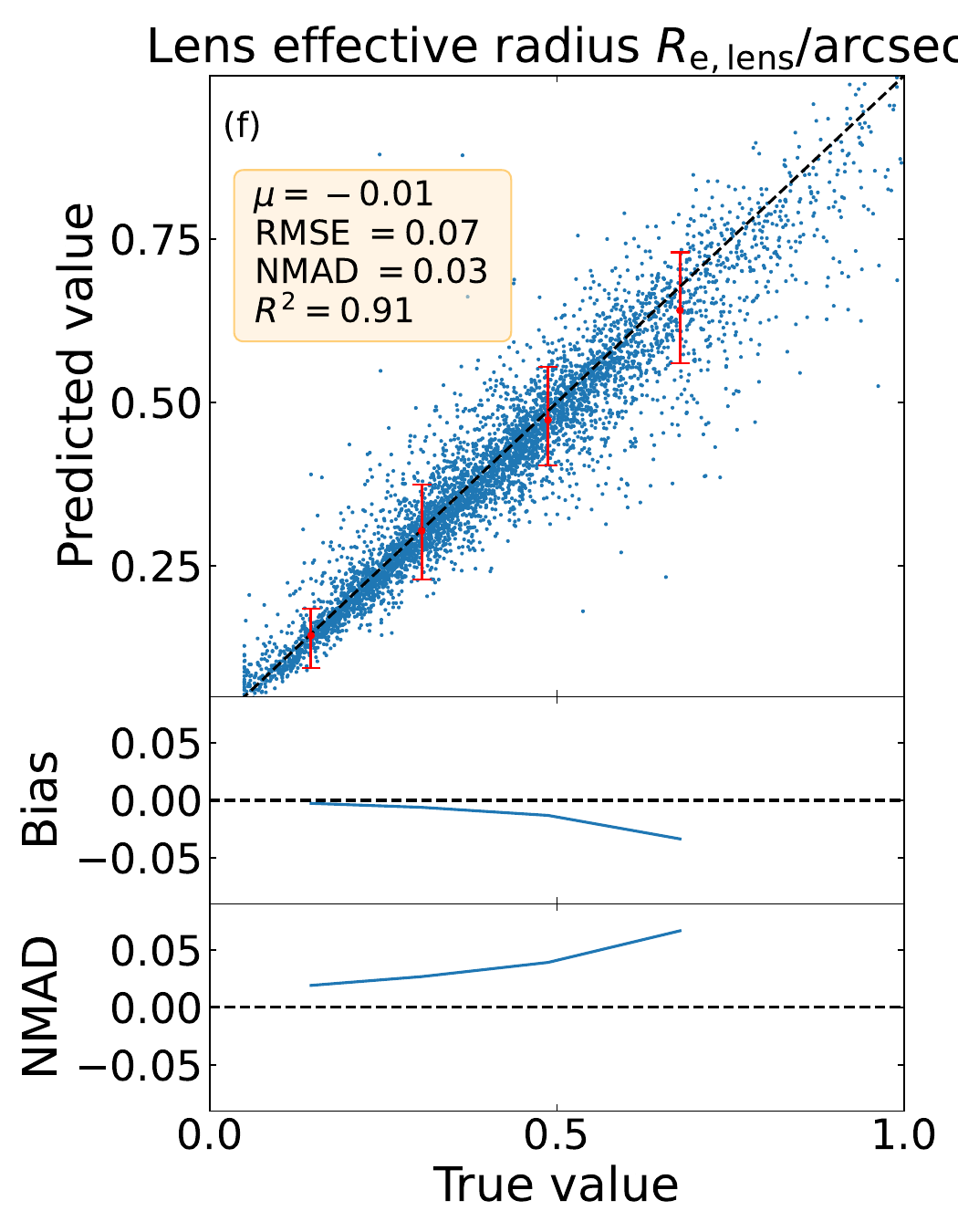}
    \includegraphics[width=0.23\linewidth]{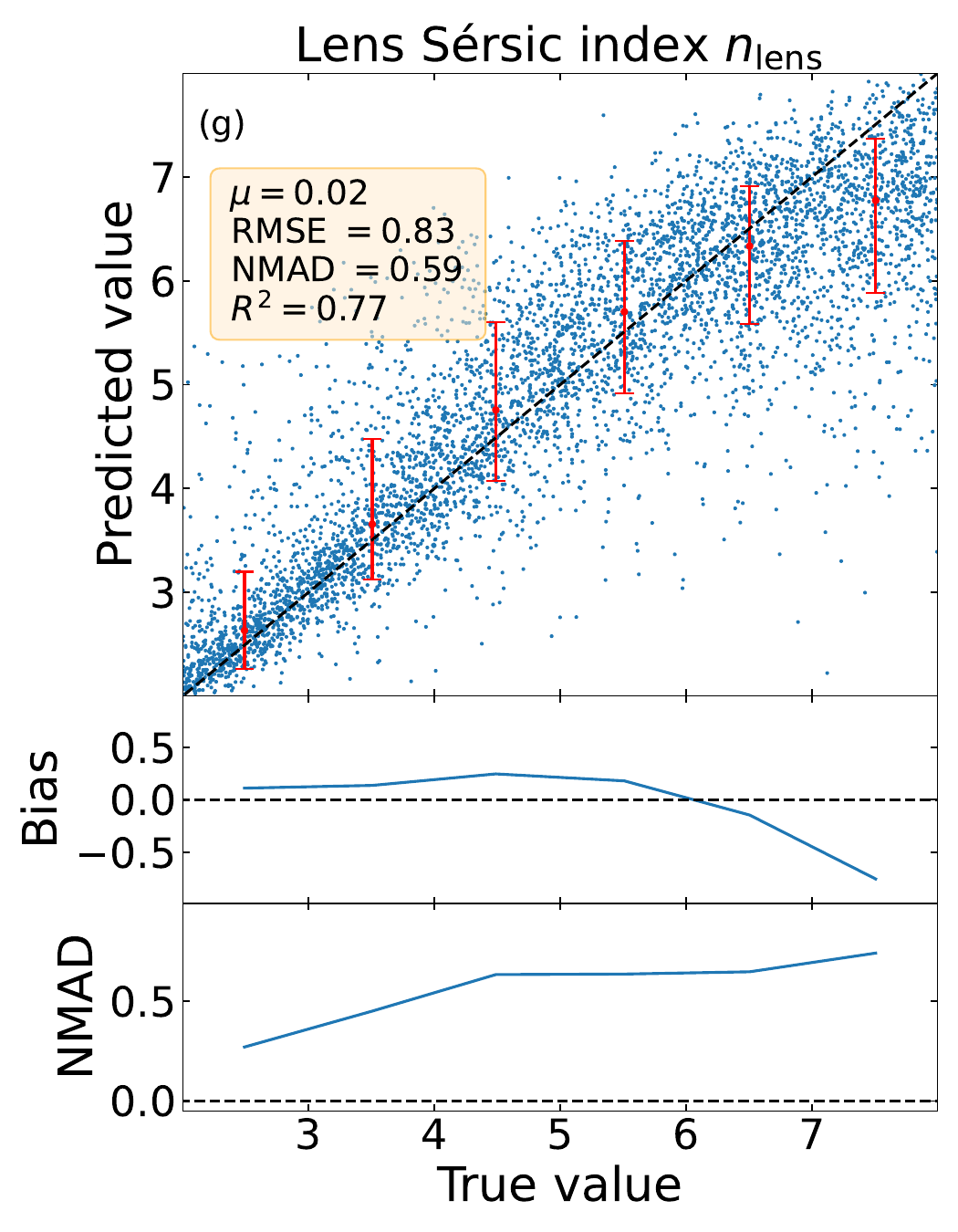}
    \includegraphics[width=0.23\linewidth]{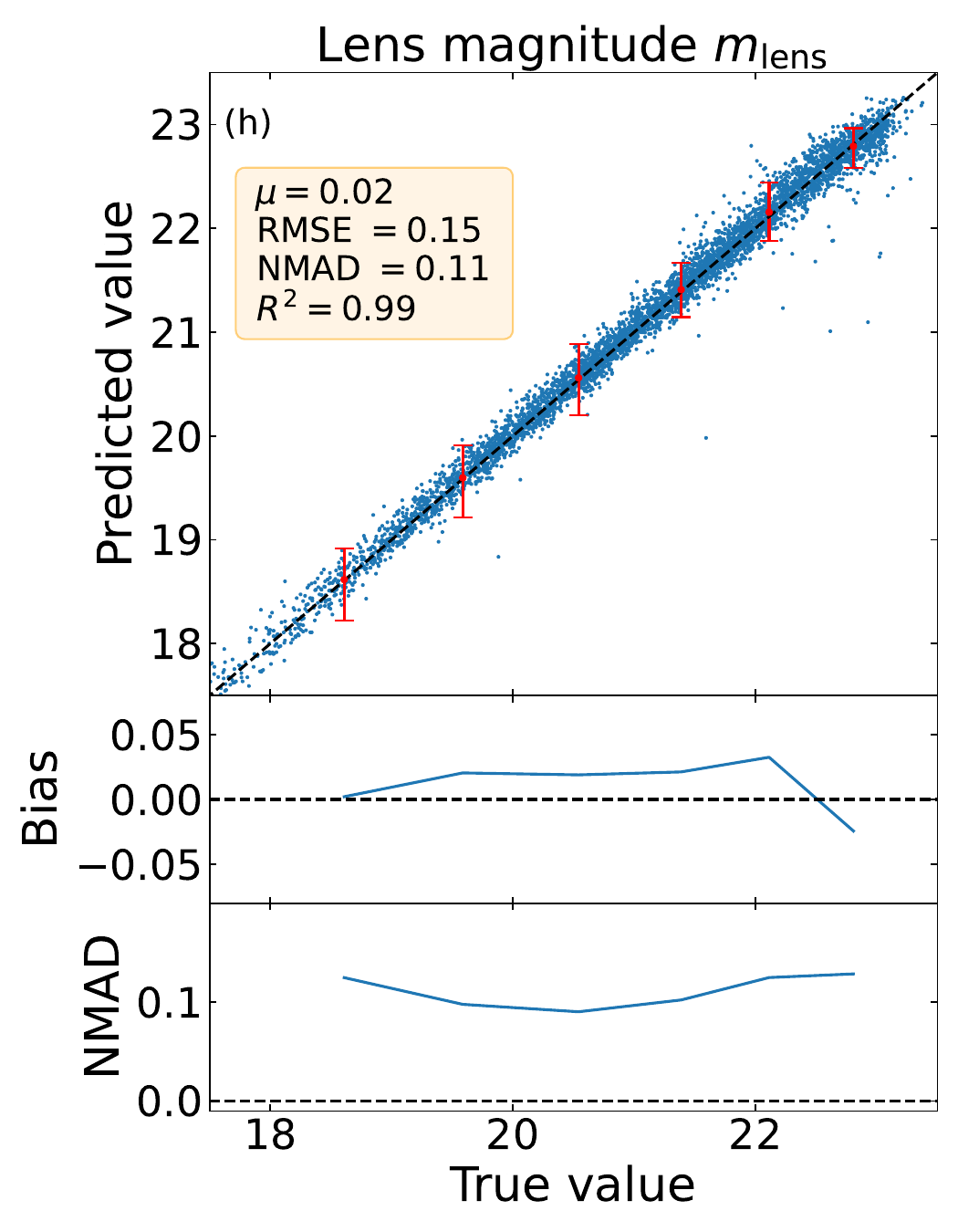}
\\
    \includegraphics[width=0.23\linewidth]{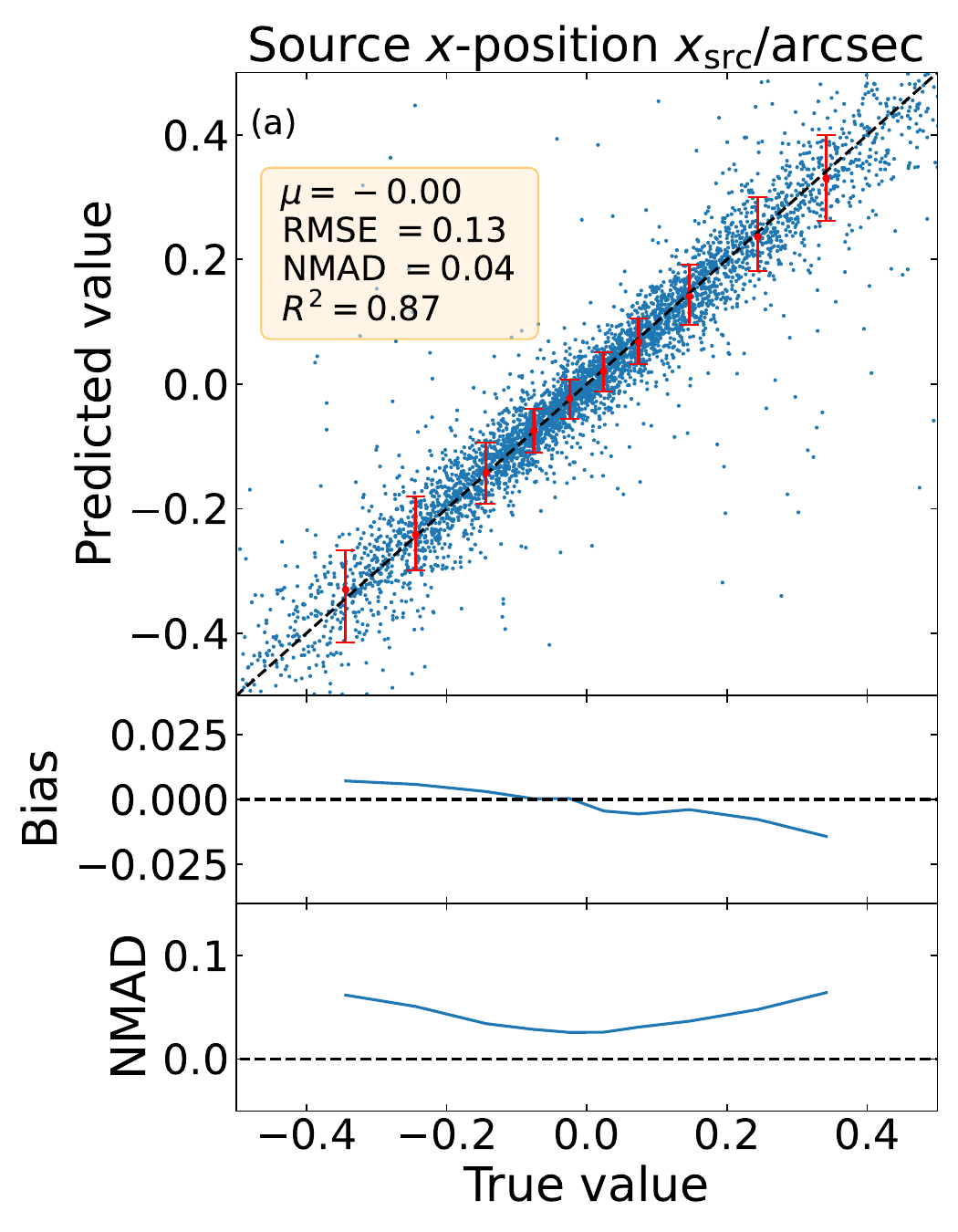}
    \includegraphics[width=0.23\linewidth]{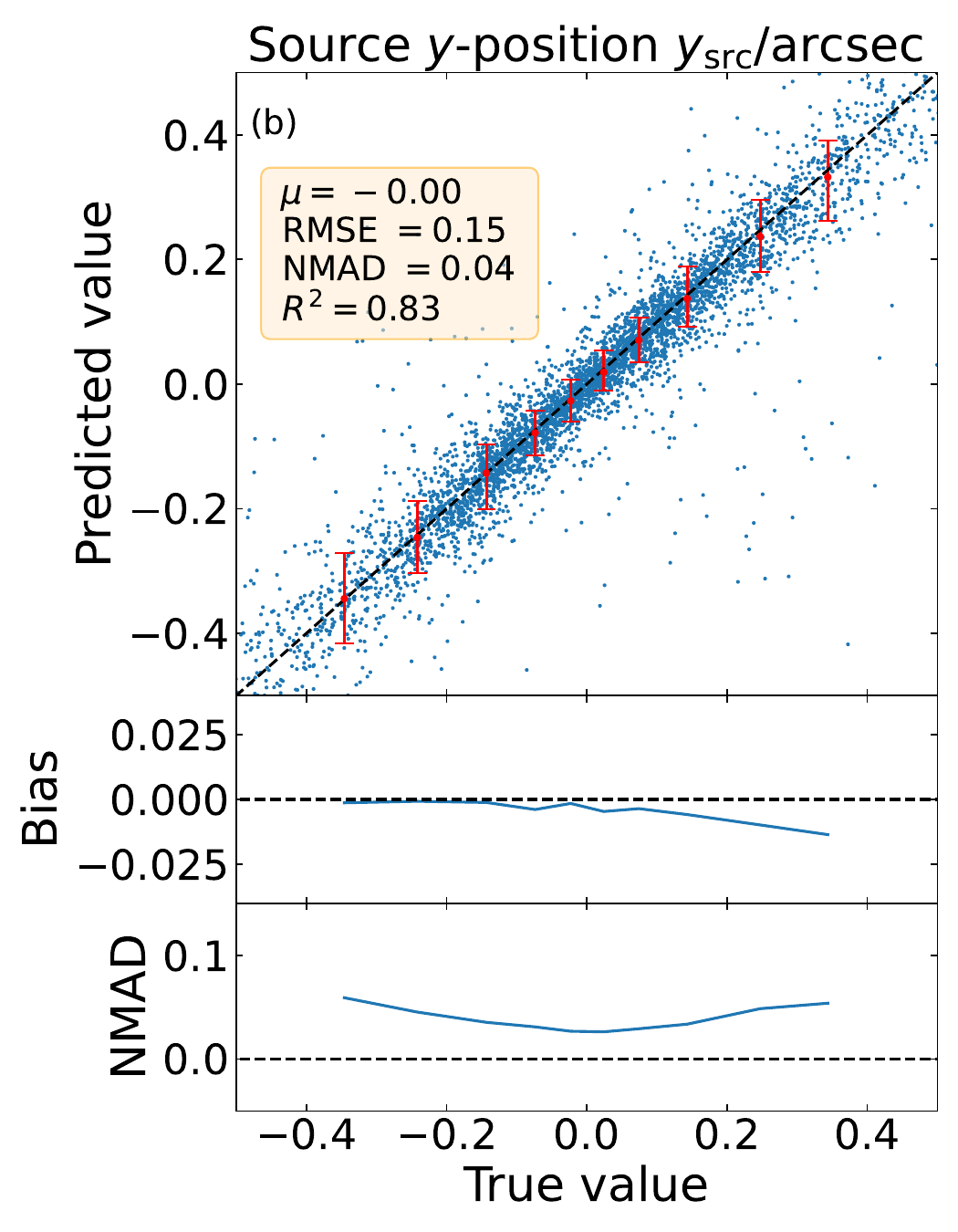}
    \includegraphics[width=0.23\linewidth]{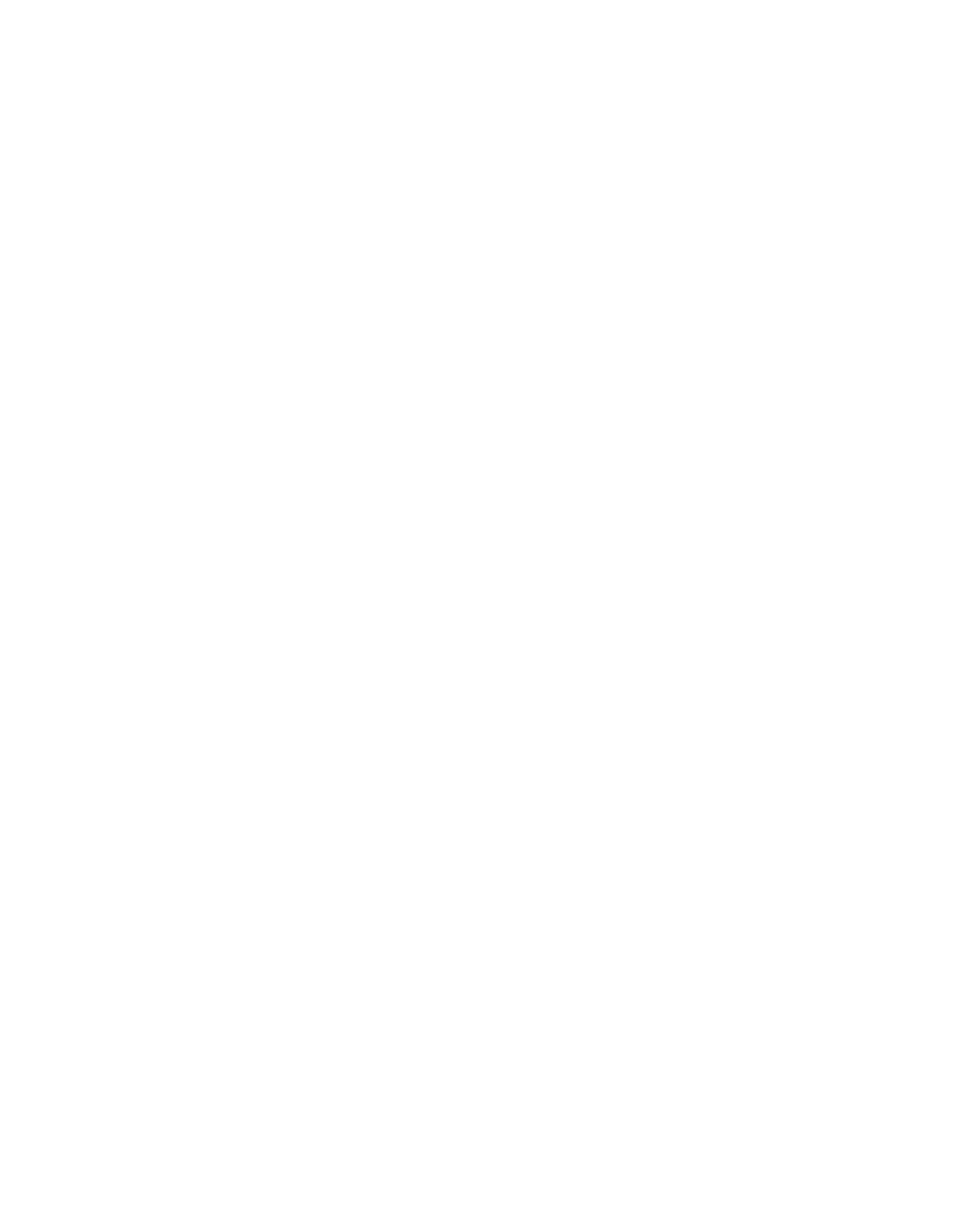}
    \includegraphics[width=0.23\linewidth]{Figures/blank.pdf}
    \caption{Recovery plot of the parameters of \Euclid mock lenses, showing how well the predicted model parameters from LEMON ($y$ co-ordinate) reproduce the corresponding true value ($x$ co-ordinate), for each lens of the test set (blue points). The ideal recovery line is shown as a dashed black line. The median trend of the scatter plots, along with the respective scatters associated with the 16th and 84th percentiles, are shown as red points and bars, respectively. For each panel, we also report the respective cumulative metrics. The bottom two rows of each parameter show the bias and the NMAD, respectively, as defined in \Sec\protect\ref{sec:metrics}, as a function of the true value of the parameters estimated by LEMON. For clarity, we have plotted only \num{5000} random points from the test set.}
    \label{fig:recovery_plot}
\end{figure*}

To obtain the scaling factors, we first proceeded by splitting the validation set into $70\%$ for the calibration training set and $30\%$ for the calibration test set. We then built the uncalibrated reliability plot, by putting on the $x$ axis the values extracted from the CDF of a Gaussian distribution and on the $y$ axis the respective values extracted from the actual CDF. We fitted the resulting trend with a $\beta$-function, and found the Platt-scaling factor, $s$, by minimising the difference between the $\beta$-function and the bisector of the reliability plot. This was done for each parameter, obtaining the values reported in \Tab\ref{tab:Scaling_factors_list}. Figure \ref{fig:reliability_plot} shows the uncalibrated reliability plot for each parameter in the left panel, and the calibrated reliability plot on the right panel.

It should be noted that the Platt-scaling calibration we use for the uncertainties has been performed on a dataset of simulated \Euclid-like lenses. As such, while this correction may hold for similar lenses, it could be not robust when switching to real data. An analysis of the robustness of this method on real lenses is discussed in Appendix \ref{sec:robustness_scaling_test}.
\begin{table}
    \centering
\caption{Rescaling factors used for the calibration procedure of each parameter predicted by LEMON.}
    \begin{tabular}{cc}
    \hline
    \noalign{\vskip 2pt}
    \hline
         Parameter& Scaling factor\\
    \hline
    \noalign{\vskip 2pt}
         Einstein radius $R_{\textrm{Ein}}/\textrm{arcsec}$& $0.84$\\
         Mass $x$-ellipticity $\epsilon_{x,\mathrm{m}}$& $0.92$\\
         Mass $y$-ellipticity $\epsilon_{y,\mathrm{m}}$& $0.91$\\
         Light $x$-ellipticity $\epsilon_{x,\mathrm{l}}$& $0.78$\\
         Light $y$-ellipticity $\epsilon_{y,\mathrm{l}}$& $0.79$\\
         Lens effective radius $R_{\textrm{e, lens}}/\textrm{arcsec}$& $0.89$\\
         Lens Sérsic index $n_{\textrm{lens}}$& $0.85$\\
         Lens magnitude $m_{\textrm{lens}}$& $0.79$\\
         Source $x$ position $x_{\textrm{src}}/\textrm{arcsec}$ & $0.88$\\
         Source $y$ position $y_{\textrm{src}}/\textrm{arcsec}$ & $0.88$\\
         
    \hline
    \end{tabular}
    \label{tab:Scaling_factors_list}
\end{table}

\section{Results for the test set}\label{sec:test_set_results}
In the following, we present the results of the analysis on the mock \Euclid test set. In Sect. \ref{sec:parameters_recovery_test_set}, we show the recovery of the parameters, along with the trends for the bias and the NMAD for each parameter. In Sect. \ref{sec:LEMON_uncertainties}, we analyse the behaviour of the uncertainties for each parameter.

\subsection{Recovery of parameters for \Euclid mock lenses}\label{sec:parameters_recovery_test_set}

\begin{figure*}
    \centering
    \includegraphics[width=1\linewidth]{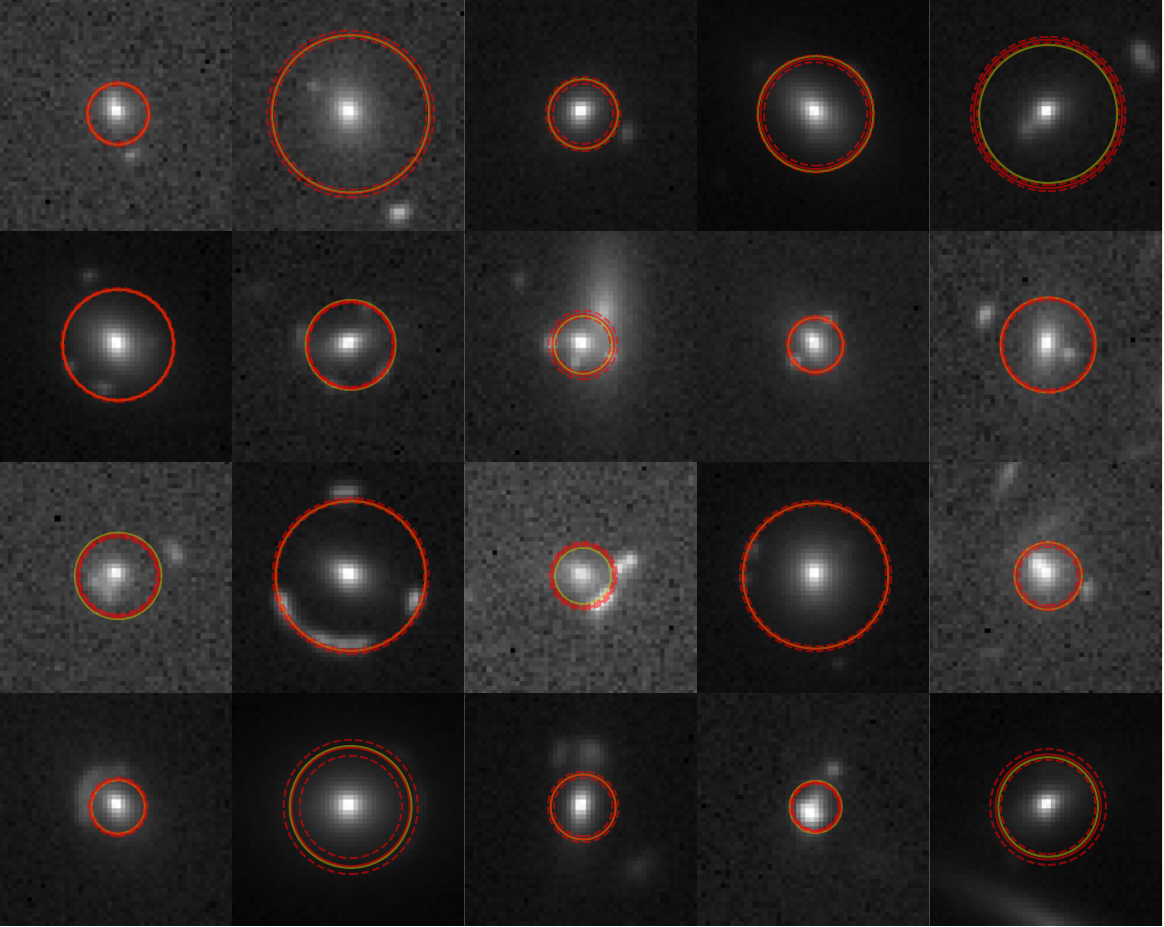}
    \caption{Random selection of 20 $10\arcsec \times 10\arcsec$ cut-outs of simulated lenses taken from the test set. Red circles show the best predictions for the Einstein radius from LEMON, with corresponding 16th and 84th percentiles shown with dashed red circles. Yellow circles show the true values for the Einstein radii. Predictions and true values are very similar, so in some cases the corresponding circles completely overlap.}
    \label{fig:panel_test_set_predictions}
\end{figure*}

\begin{figure*}
    \centering
    \includegraphics[width=1\linewidth]{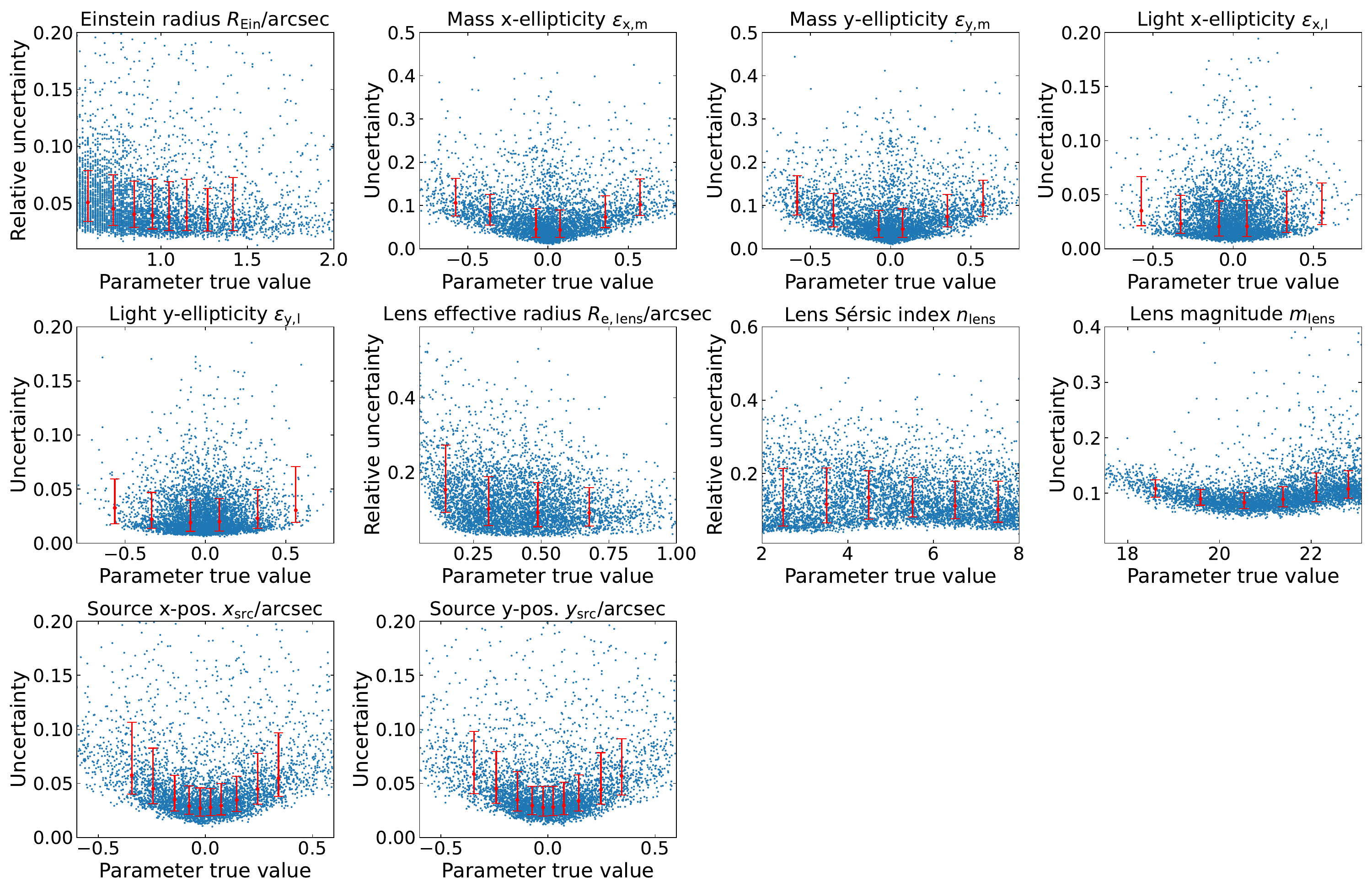}
    \caption{Calibrated relative uncertainty as a function of the true value of the lens parameters. For the ellipticity components, lens magnitude, and source co-ordinates, we consider the absolute uncertainty. For clarity, we plotted only \num{5000} random points. The median trends of the scatter plots, along with the respective scatters associated with the 16th and 84th percentiles, are shown as red points and bars, respectively.}
    \label{fig:calibrated_uncertainty_trend}
\end{figure*}

Figure \ref{fig:recovery_plot} shows the recovery of the parameters of the \Euclid mock lenses for the test set, for both the mass and light profiles, with the bottom two rows of each parameter showing the bias and NMAD metrics in sequential bins. Overall, the recovery of the parameters for the simulated \Euclid lenses of the test set is good, as is shown for example in Fig. \ref{fig:panel_test_set_predictions}, where we compare the Einstein radius predictions for a random selection of 20 test set lenses with the corresponding ground truth.

The Einstein radius is well recovered, with a very slight increase in NMAD and decrease in bias from $\ang{;;1.0}$ upwards, due to a skewed prior towards zero for the Einstein radius parameter, as we can see from Fig. \ref{fig:recovery_plot}a. The mass ellipticity components are well recovered, showing a slight increase in bias only for values outside the interval $[-0.5,0.5]$ and a NMAD trend that always stays below $0.1$. The recovery is even tighter for the light ellipticity, showing an $R^{2}$ of $0.96$ (compared against an $R^{2}$ of $0.85\textrm{--}0.86$ for the mass ellipticity components) and a NMAD trend that always lies below $0.04$. This slight discrepancy is to be expected, because the network cannot rely on light information to infer the shape of the mass distribution, and has instead to rely only on the configuration of the lensed source to derive the mass shape parameters, which is an intrinsically more uncertain procedure.

The effective radius shows the same behaviour as the Einstein radius, being slightly underestimated at values of $R_{\textrm{e}}\gtrsim \ang{;;0.50}$. This is also expected from the skewed prior for the effective radius parameter. It should be noted from Fig. \ref{fig:recovery_plot}d, however, that the lowest bias value is around $\ang{;;-0.05}$, which is negligible. The worst recovered parameter is the Sérsic index, which shows the lowest value of the coefficient of determination, $R^{2} = 0.77$, and a large scatter, which increases for an increasing value of $n_{\textrm{lens}}$. The median trend also shows a deviation from the linear trend for high values of the Sérsic index, as is shown from the bias trend in Fig. \ref{fig:recovery_plot}e. This is, however, in agreement with \texttt{GALNET}, a CNN that has shown remarkable abilities in recovering the light parameters of simulated and real KiDS galaxies\footnote{The algorithm makes use of the local PSF to enhance the predicting abilities of the network, but does not offer insights on the uncertainty associated with the predictions.} \citep{Li2022}.  For the magnitude, there is a very slight overestimate, of order $\mu\simeq 10^{-2}$, which is shown as a constant trend for the bias in Fig. \ref{fig:recovery_plot}f, up until around a magnitude of $22$, where the bias starts to decrease towards negative values. The scatter stays instead constant, with a value of the NMAD of around $0.1$. The unlensed source co-ordinates, finally, are well recovered, with no bias and $R^2$ of around $0.83\textrm{--}0.87$. This effect is most likely due to the very high difference between the flux of the lens and the flux of the source images, not allowing the network to find the correct magnitude.

\subsection{Behaviour of LEMON uncertainties}\label{sec:LEMON_uncertainties}

Figure \ref{fig:calibrated_uncertainty_trend} shows the trends for the calibrated total relative uncertainties\footnote{We define the total relative uncertainty as the predicted total uncertainty by LEMON divided by the corresponding predicted mean value.} for all predicted parameters (except for the ellipticity components and lens \IE magnitude, where we consider the absolute uncertainty). From the figure, we can see that the relative uncertainty on the Einstein radius is of order $\sim5\times10^{-2}$ on all the relevant range of the parameter, with a constant scatter throughout the interval range. Regarding the mass ellipticity components, we can see that the uncertainty increases for $\epsilon_{\textrm{m,l}}\rightarrow\pm0.8$, probably due to the fact that the distribution of the ellipticities is peaked around zero and thus there are fewer examples on the edges of the parameters. A similar trend is visible for the light ellipticity components, but with much smaller uncertainties. This reflects the fact that the network has an easier time predicting the shape of the light component directly from the lens galaxy image, compared to inferring the shape of the mass component through the lensed images' configurations. The trend for the lens effective radius is similar to the one for the Einstein radius, with an increase at low values due to the predicted values of the parameter approaching zero. The scatter of the Sérsic index is systematically high for all values of $n_{\textrm{lens}}$, probably due to an intrinsic difficulty in recovering this parameter through photometry alone. The relative uncertainty for this parameter, however, mainly remains below $40\%$. For both low and high values of lens magnitude, we find that the absolute uncertainty slightly increases.
Finally, we note that the uncertainties in the unlensed source co-ordinates exhibit trends similar to those of the ellipticity components.

We also compared the trends of aleatoric and epistemic uncertainties for the same parameters of Fig. \ref{fig:calibrated_uncertainty_trend}. We verified that the epistemic uncertainties are systematically lower than the corresponding aleatoric uncertainties for all parameters. The comparison, along with a brief discussion, is shown in Fig. \ref{fig:aleatoric_epistemic_uncertainties_comparison} of Appendix \ref{sec:epistemic_analysis}. Additionally, we analysed the behaviour of total uncertainty with the S/N of the source in Appendix \ref{sec:SNR_analysis}, which gave us information about how each parameter is recovered by the network (through either the light or the lensed source configuration).

\section{Results for real lenses}\label{sec:real_lenses_results}

\begin{figure*}
    \centering
    \includegraphics[width=1\linewidth]{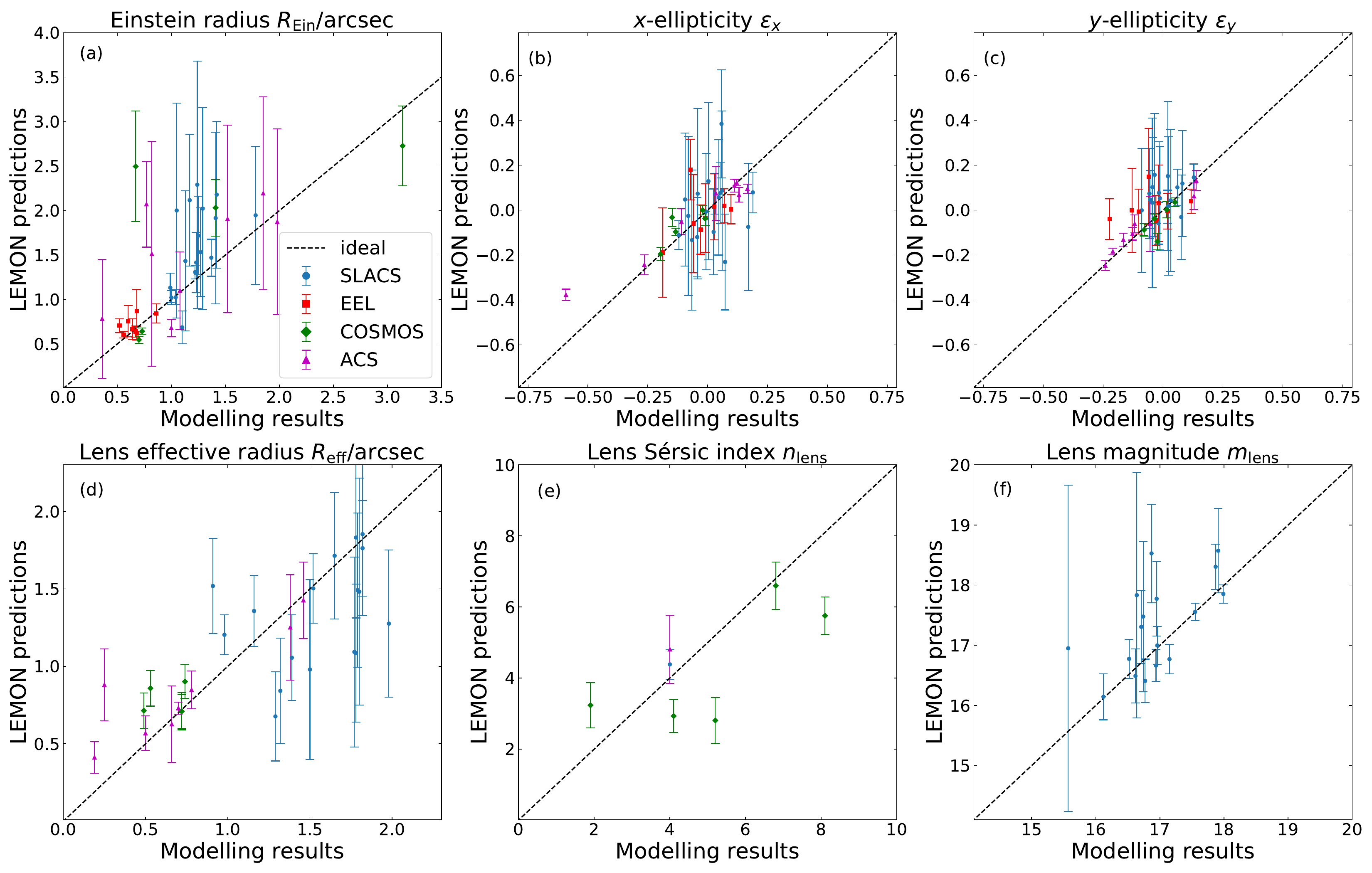}
    \caption{Predicted model parameters from LEMON ($y$ co-ordinate) against the corresponding model values from the literature ($x$ co-ordinate). The ideal recovery line is shown as a dashed black line. The SLACS and ACS best predicted values for the Sérsic indices are evaluated as the median of the predictions for all the lenses of these catalogues against a fixed value associated with the one for a De Vaucouleurs profile ($n_{\textrm{lens}}=4$), with uncertainty given by the NMAD of the predictions. Lenses with modelling values for the effective radius bigger than $2\arcsec$ and EEL lenses with modelling values for the Einstein radius lower than $\ang{;;0.5}$ have been removed due to being outside the training space of LEMON.}
    \label{fig:recovery_plot_Euclidized_lenses}
\end{figure*}

\subsection{LEMON applied to Euclidised lenses}\label{sec:res_euclidezed}
We now apply LEMON to a sub-sample of real Euclidised lenses described in Sect. \ref{sec:real_lenses}. This sub-sample is formed from four different main catalogues, from which the reference parameter values from standard modelling have been taken (where available).

\begin{itemize}
    \item A sample of 29 Sloan Lens Advanced Camera for Surveys (SLACS) catalogue systems, described in \cite{Bolton2008} and \cite{Auger2009}. Einstein radius, axis ratio, and position angle (estimated by assuming a SIE mass model for the lenses without external shear, and then used to derive the corresponding mass ellipticity components) are taken from the first three columns of table 5 in \cite{Bolton2008}, after correcting the position angle orientation from east of north to north of west; the effective radius and lens magnitude are instead taken from the columns named $r_{{\rm e},I}$ and $m_{I}$ of table 3 in \cite{Auger2009}.
    \item The sample of 13 early-type source with early-type lens systems (EELs) presented in \cite{Oldham2017}. Einstein radius, mass axis ratio, and mass position angle values are all taken from table 2, after the appropriate orientation conversions for the position angle. The mass model considered by the authors is an elliptical power law with external shear.
    \item A sample of five lenses found in the COSMOS survey \citep{Scoville2007}. All parameters are taken from table 2, except for the Einstein radius, which is taken from the errata by \cite{Faure2008} due to an error in the determination of the mass parameters in \cite{Scoville2007}. To model the systems, the authors use the code \texttt{Lenstool}, assuming a SIE model with external shear. Light ellipticity components have been derived from the parameters $1-q$ and $\textrm{PA}$ of table 2, and used for comparison.
    \item A sample of 13 lenses taken from those found by visually inspecting the whole imaging data taken with the ACS through the F814W filter up to the 31st of August 2011 \citep{Pawase2014}. All parameters are taken from table 3, and light axis ratio and position angle have been converted to light ellipticity components for comparison. Notice, however, that the Einstein radius for these lenses is not reported, and as such we compared with the radius of the arc (in arcseconds) as a substitute.
\end{itemize}

\begin{figure*}
    \centering
    \includegraphics[width=1\linewidth]{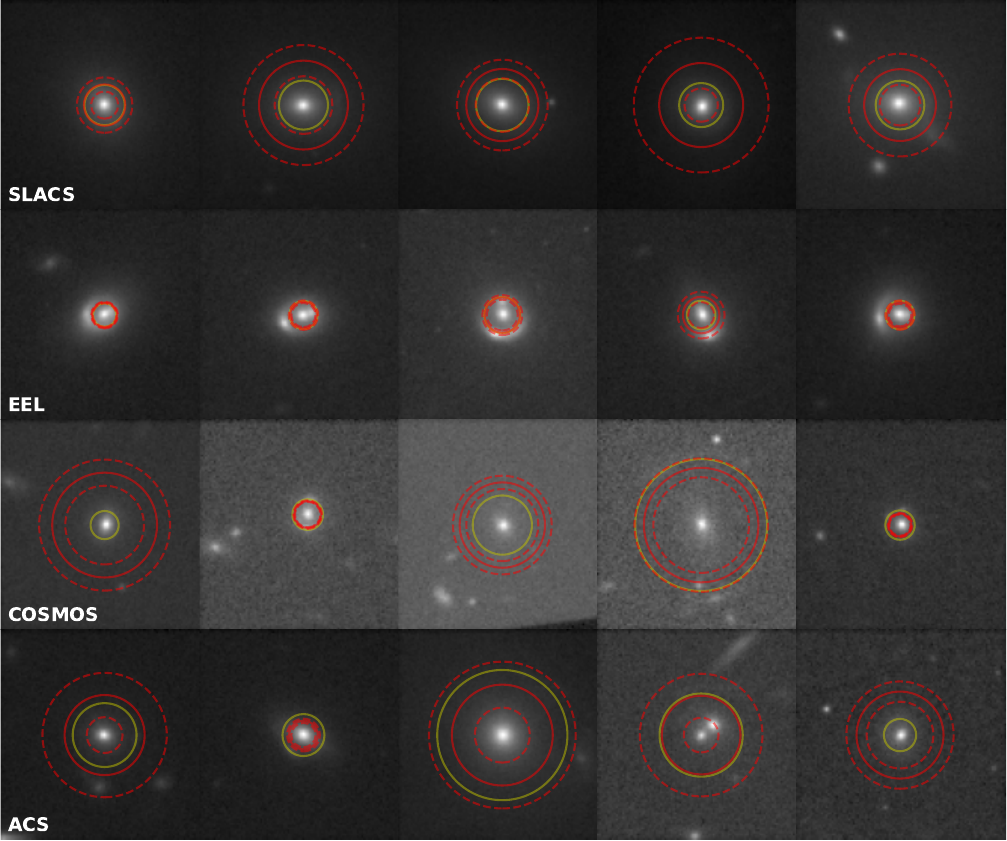}
    \caption{Random selection of 20 $10\arcsec \times 10\arcsec$ cut-outs of Euclidised lenses taken from the SLACS, EEL, COSMOS, and ACS subsamples. Red circles show the best predictions for the Einstein radius from LEMON, with corresponding 16th and 84th percentiles shown with dashed red circles. Yellow circles show the values for the Einstein radii reported in \protect\cite{Bolton2008}, \protect\cite{Oldham2017}, \protect\cite{Faure2008}, and \protect\cite{Pawase2014}, respectively. For ACS lenses, the radius of the arc is taken instead, due to lack of an Einstein radius measurement.}
    \label{fig:panel_Euclidized_lenses_predictions}
\end{figure*}

All of the Euclidised lenses used in this work have been spectroscopically confirmed. The comparison between forward modelling and LEMON predictions for mass and light parameters is shown\footnote{Notice that we aggregated mass and light ellipticity comparisons together. Mass and light ellipticity results from the literature have been compared with the matching type of parameter prediction from LEMON.} in Fig. \ref{fig:recovery_plot_Euclidized_lenses}, while the associated metrics for each parameter are reported in Table \ref{tab:metrics_Euclidized_lenses}. A panel with some of the lenses from each catalogue is shown in Fig. \ref{fig:panel_Euclidized_lenses_predictions}.

\begin{table*}
    \centering
    \caption{Cumulative metrics associated with the parameters of the Euclidised lenses shown in \Fig\protect\ref{fig:recovery_plot_Euclidized_lenses}.}
    \begin{tabular}{lcccccc}
    \hline
    \hline
              Metrics&$R_{\textrm{Ein}}$ ($\textrm{arcsec}$)&  $\epsilon_{x}$&  $\epsilon_{y}$&  $R_{\textrm{e, lens}}$ ($\textrm{arcsec}$)& $n_{\textrm{lens}}$&$m_{\textrm{lens}}$\\
              \hline
              Bias ($\mu$)&$-0.03\hphantom{\;\;\,}$&  0.00&  0.00&  0.03& $-0.35\hphantom{\;\;\,}$& 0.01\\
              RMSE&0.14&  0.07&  0.04&  0.18& 1.05& 0.30\\
              NMAD&0.11&  0.04&  0.03&  0.06& 1.10& 0.21\\
 Coefficient of determination ($R^{2}$)& 0.53& 0.87& 0.88& 0.64& $-0.47\hphantom{\;\;}$& 0.69\\
    \hline
    \end{tabular}
    \label{tab:metrics_Euclidized_lenses}
\end{table*}

\begin{figure*}
    \centering
    \includegraphics[width=0.45\linewidth]{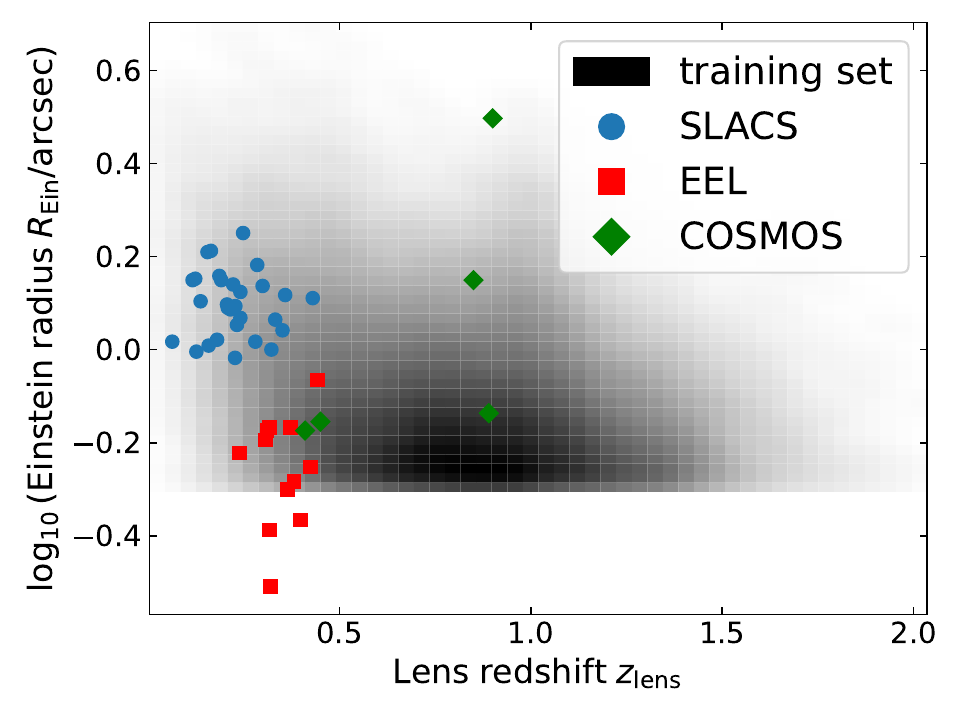}
        \includegraphics[width=0.45\linewidth]{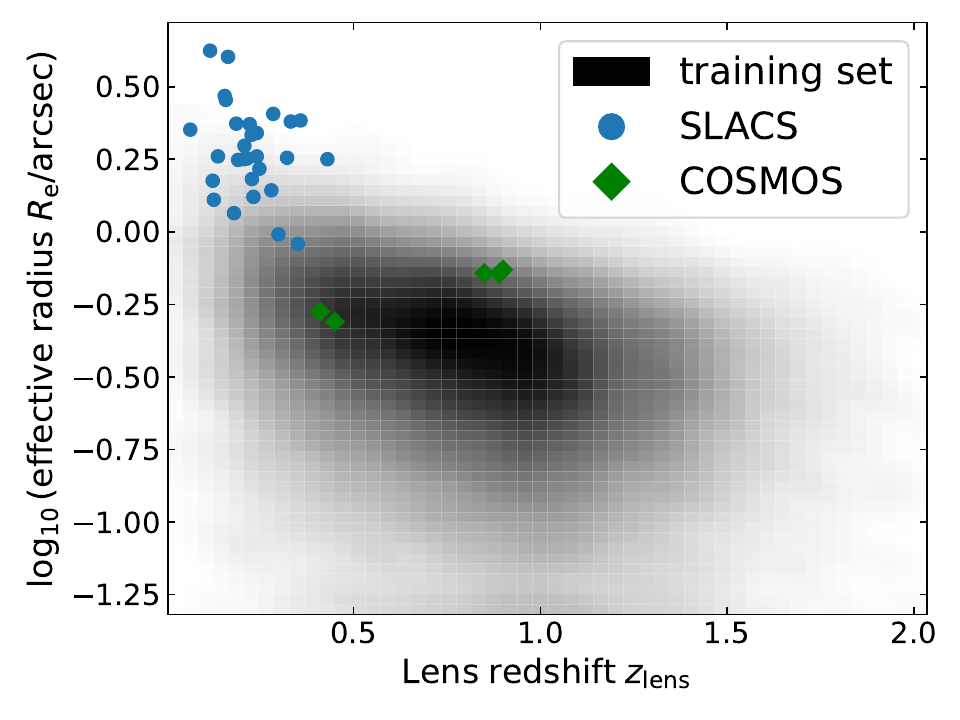}
    \caption{Distributions of the Einstein radius (\emph{left panel}) and the lens effective radius (\emph{right panel}) as a function of lens redshift. Black regions show the training set, while coloured points show the position of the Euclidised lenses. Notice that these points are predictions from forward modelling.}
    \label{fig:radii_vs_z}
\end{figure*}

From the table, we can see that the Einstein radius predictions match well the literature results, with a coefficient of determination of $R^2 = 0.53$. The mass and light ellipticities also match the values from the literature well, with $R^{2} = 0.87\textrm{--}0.88$. The effective radius is slightly underpredicted with respect to classical modelling above $\ang{;;1}$. This, however, is consistent with results from \cite{Li2022}, in which light parameters of galaxies were predicted with CNNs and compared with classical modelling predictions from \texttt{2DPHOT} \citep{LaBarbera2008}. In both cases, there is a systematic underprediction at high values of effective radii. Due to a low number of points available for the Sérsic index parameter, in addition to the fact that for SLACS and ACS lenses the classical modelling has been performed by assuming a fixed $n_{\textrm{lens}}=4$ value for this parameter, we cannot draw any particularly relevant conclusions. Finally, the lens magnitude for the SLACS lenses matches well the results from classical modelling, with $R^{2} = 0.69$.

\begin{figure*}
    \centering
    \includegraphics[width=1\linewidth]{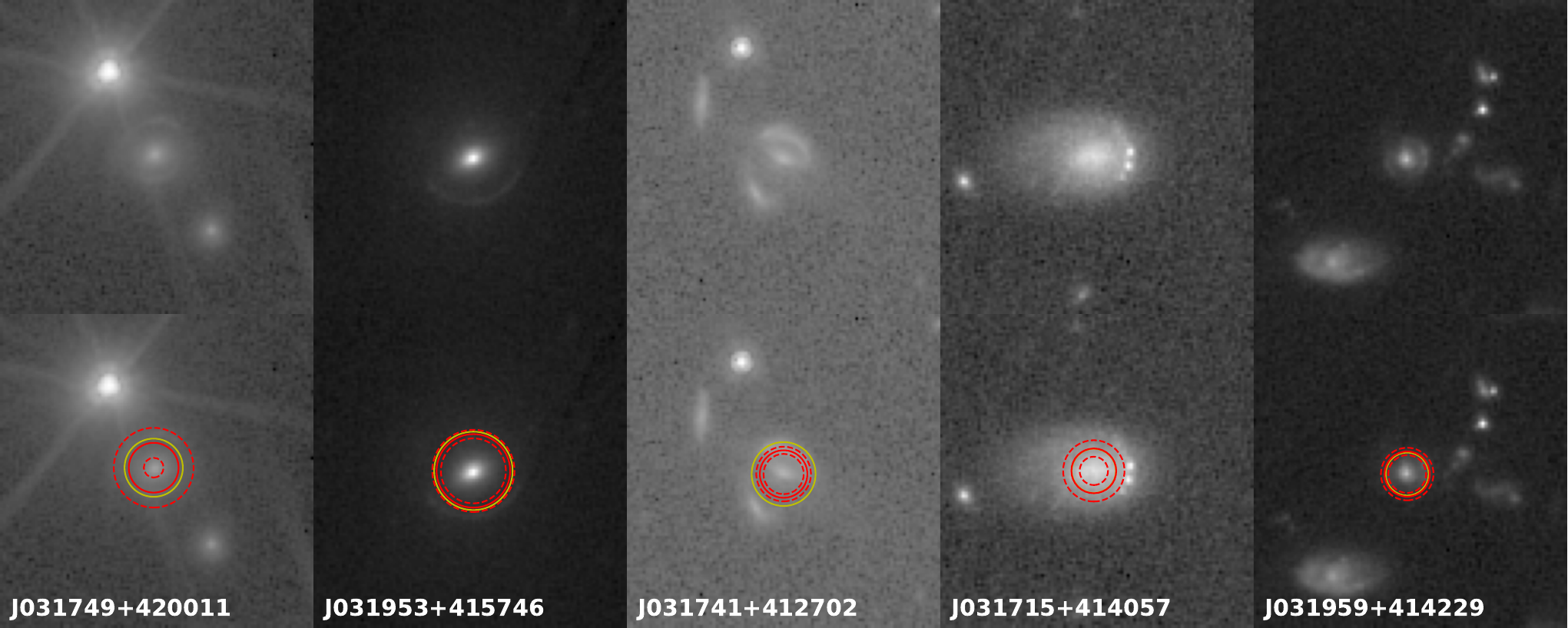}
    \caption{$10\arcsec \times 10\arcsec$ cut-outs of the five \Euclid\ gravitational lenses found in the Perseus ERO field and modelled in \protect\cite{AcevedoBarroso24}. \textit{Top row}: Unedited cut-outs centred on the lenses. \textit{Bottom row}: Same cut-outs, with the predicted Einstein radius (red circles, with dashed circles showing the uncertainty bands) and the value obtained from the classical modelling (yellow circle) superimposed on them.}
    \label{fig:Euclid_Perseus_ERO_lenses}
\end{figure*}

\begin{figure}
    \centering
    \includegraphics[width=1\linewidth]{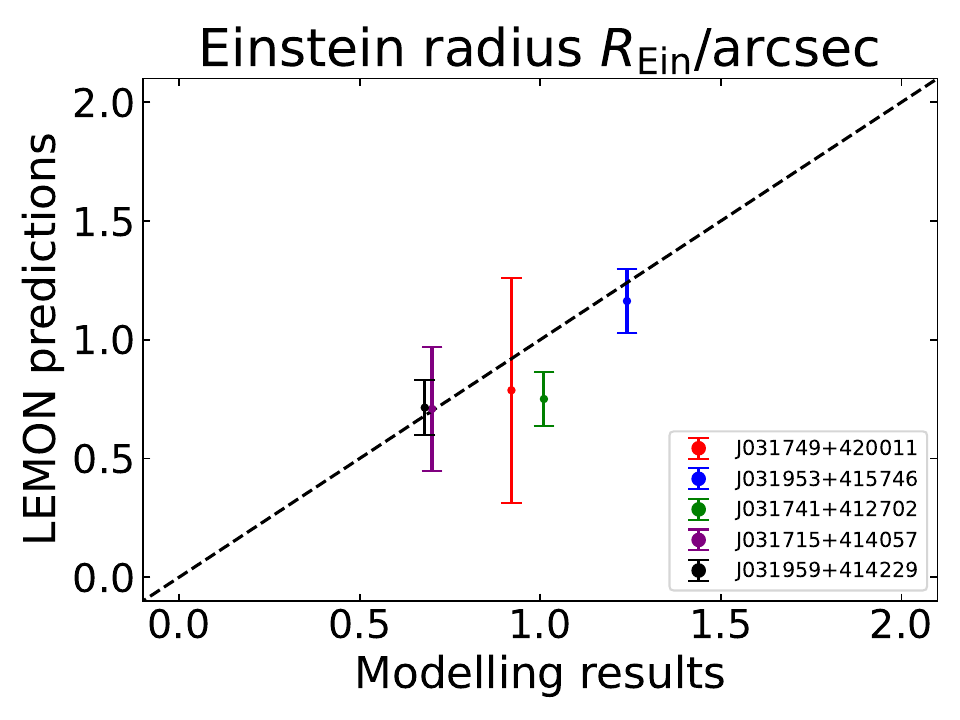}
    \caption{Comparison between Einstein radii values predicted by LEMON and those predicted by classical modelling for the \Euclid\ lenses found in the Perseus ERO field.}
    \label{fig:recovery_plot_Perseus_ERO}
\end{figure}

To analyse the behaviour of LEMON with respect to the Euclidised lenses, we first evaluated both the aleatoric and epistemic components of the uncertainties for these lenses, obtaining values of epistemic uncertainties much lower than the aleatoric ones (e.g. the maximum epistemic uncertainty for the Einstein radius is at most $\ang{;;0.25}$, while the corresponding maximum aleatoric uncertainty is $\ang{;;1.35}$), in agreement with the results for the mock \Euclid lenses' test set. These values, however, are much higher than the corresponding components of the uncertainties for the mock \Euclid lenses' test set (e.g. the maximum aleatoric uncertainty for the Einstein radius for the mock Euclid lenses' test set is $\ang{;;0.20}$, which is of the order of the maximum epistemic uncertainty for the Euclidised lenses set).

\begin{figure*}
    \centering
    \includegraphics[width=0.32\linewidth]{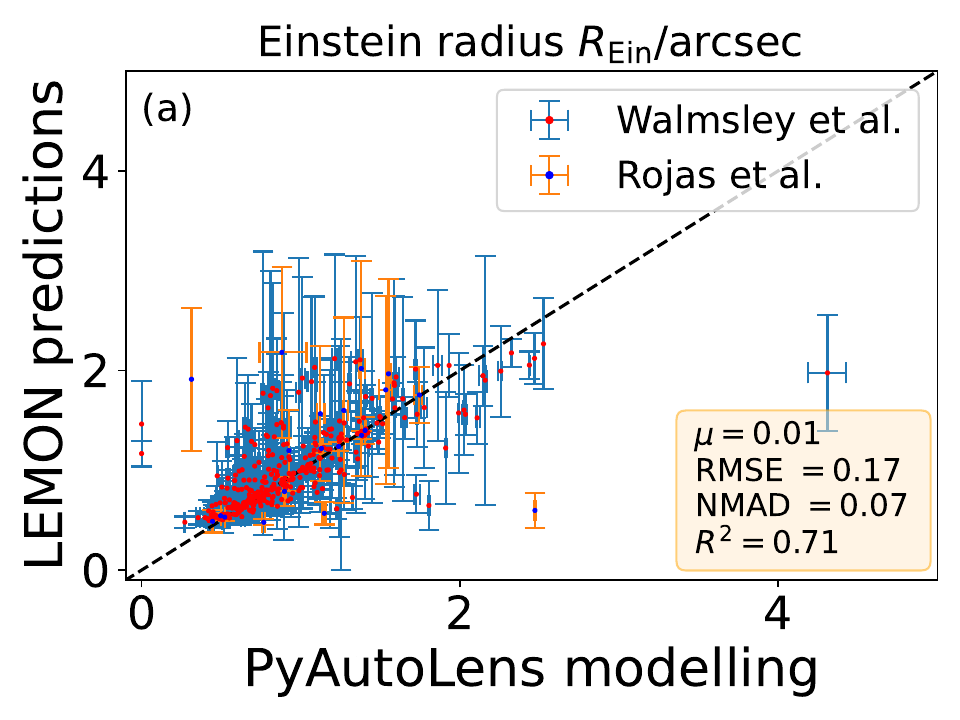}
    \includegraphics[width=0.32\linewidth]{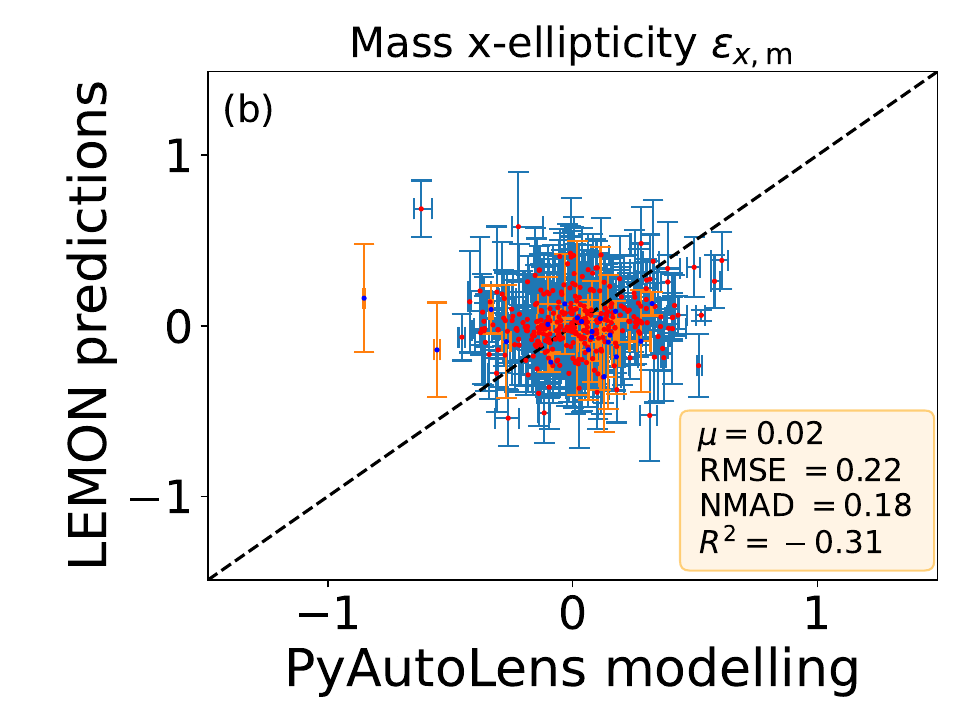}
    \includegraphics[width=0.32\linewidth]{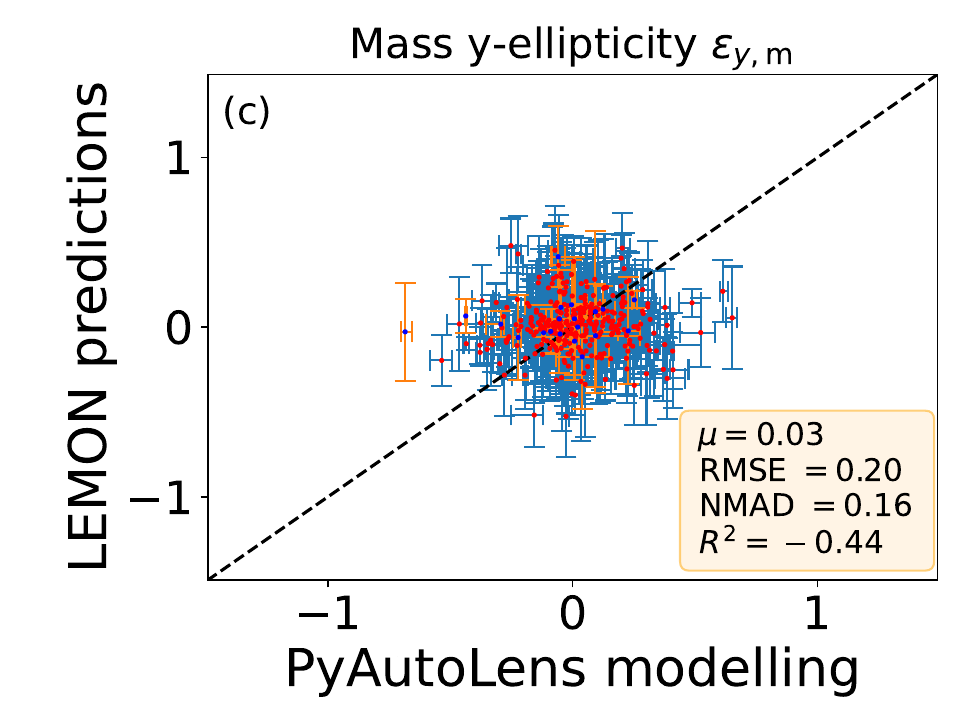}
\\
    \includegraphics[width=0.32\linewidth]{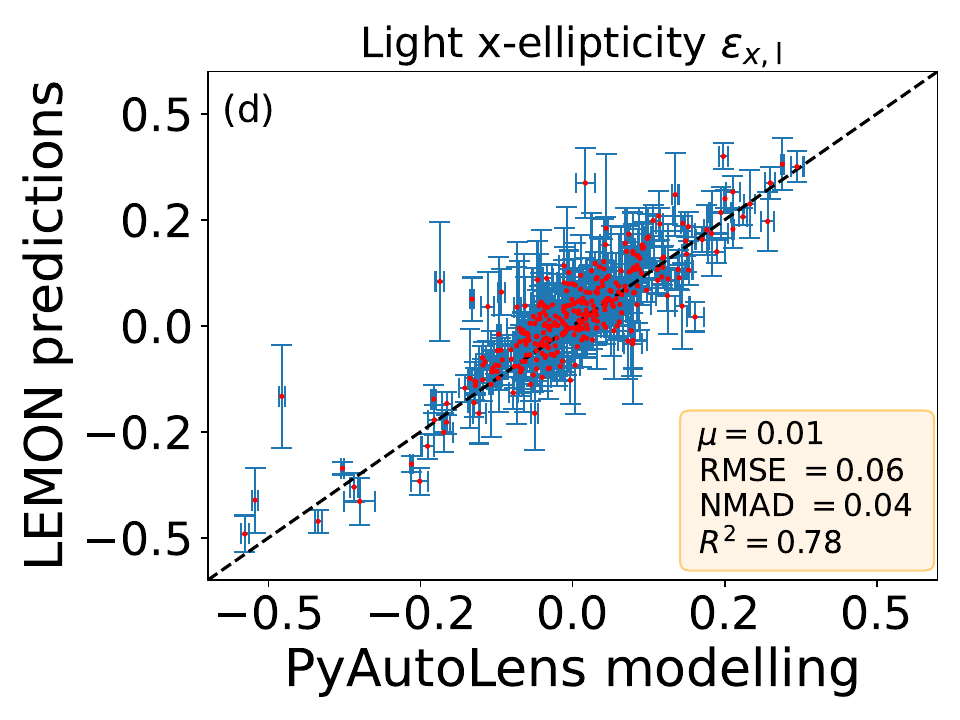}
    \includegraphics[width=0.32\linewidth]{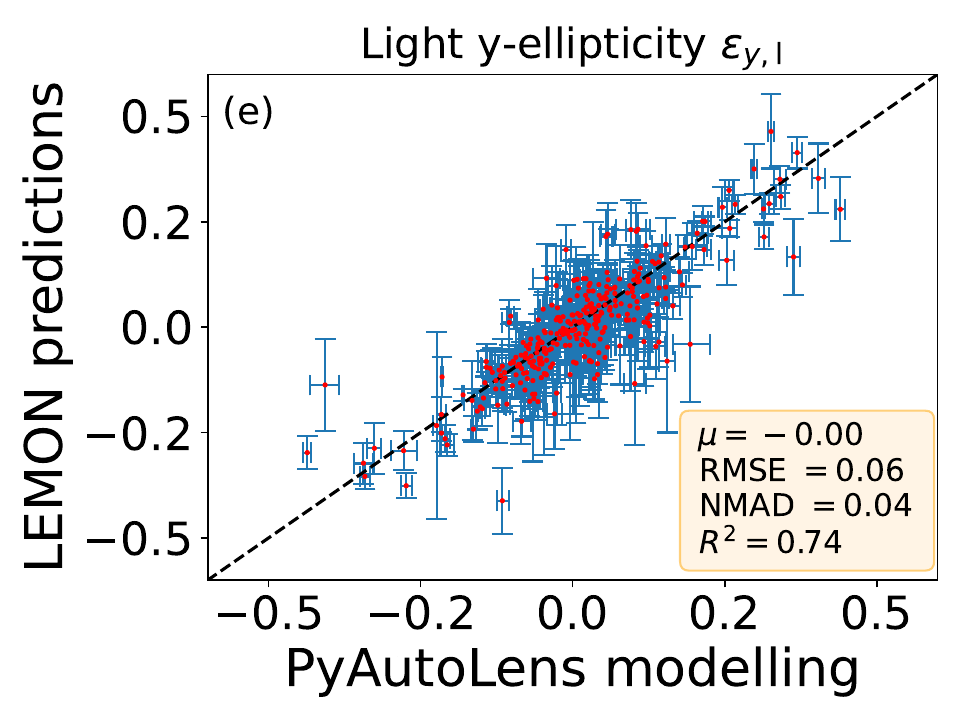}
    \includegraphics[width=0.32\linewidth]{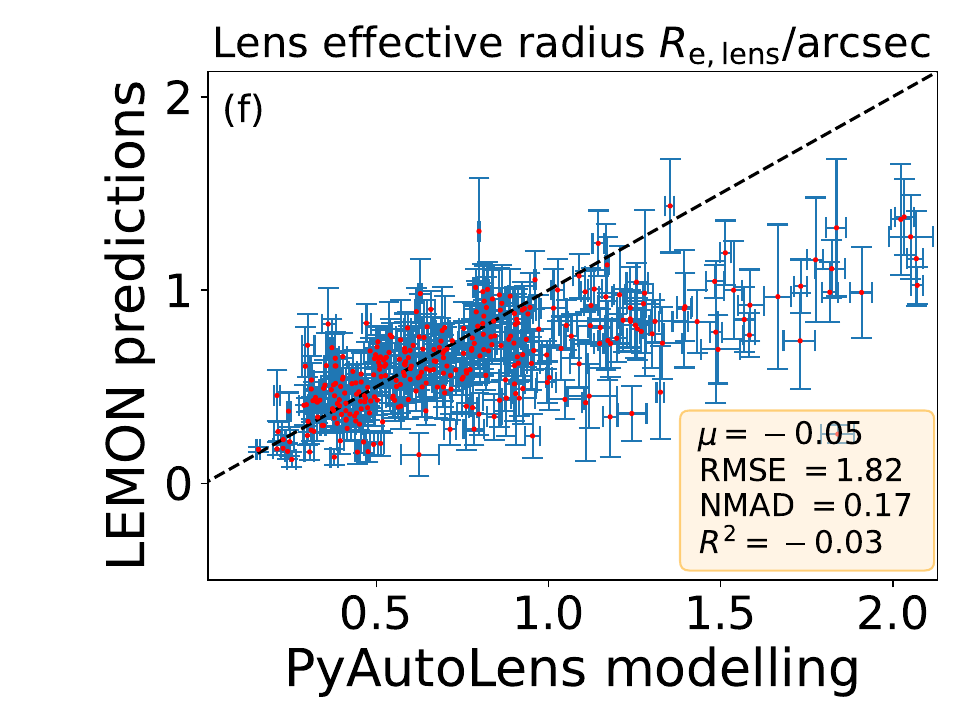}
\\
    \hspace{-5.5cm}
    \includegraphics[width=0.32\linewidth]{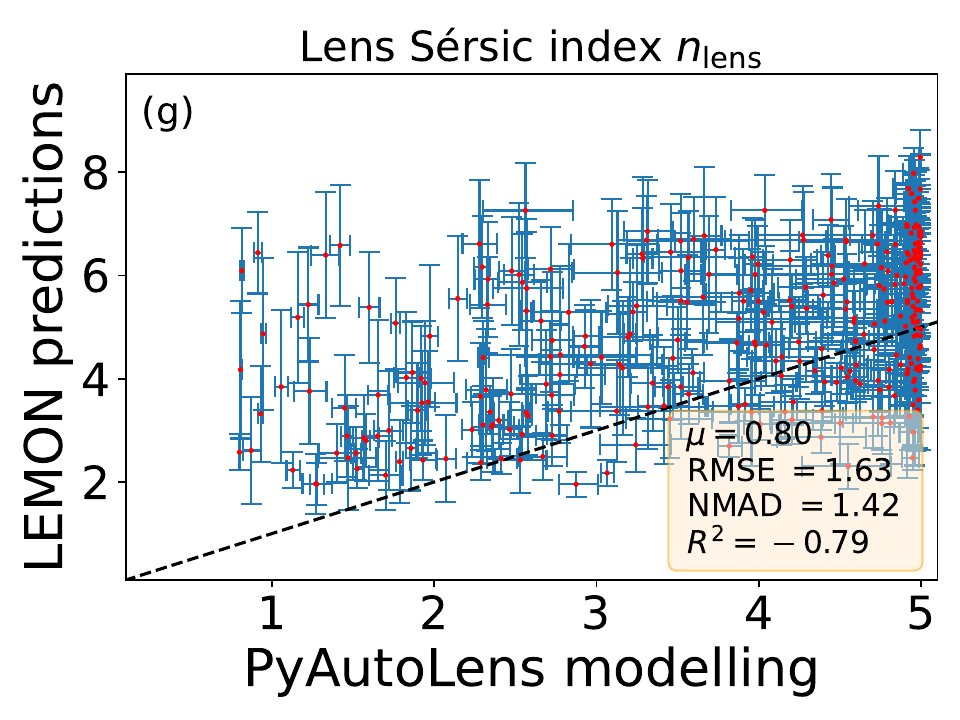}
    \includegraphics[width=0.32\linewidth]{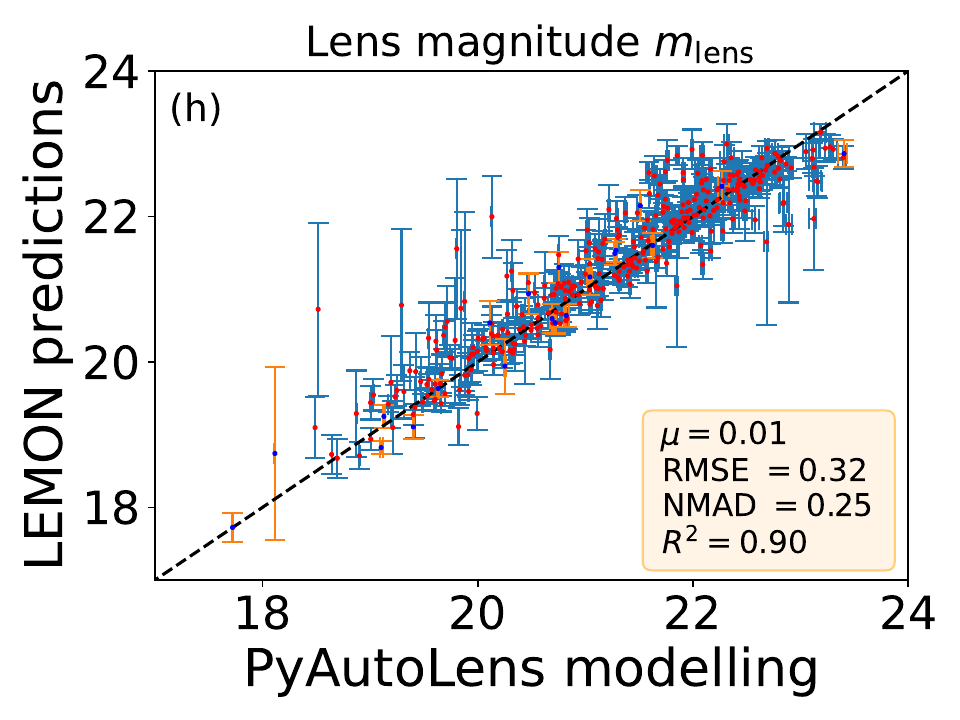}
    \caption{Parameter estimates of \Euclid lenses found in the Q1 fields given by LEMON against values obtained by traditional modelling with \texttt{PyAutoLens}. For each panel, we also report the respective cumulative metrics. Notice that predicted LEMON magnitudes have been subtracted by a factor of $0.22$ due to different zero-points between mock \Euclid images used for training and real Q1 images.}
    \label{fig:recovery_Q1}
\end{figure*}

A possible cause of these high uncertainty values is distributional shift, i.e. variations in the properties of the input image shift the parameters' distributions outside the training distribution space, which affects the network's output predictions (for more information, see \citealt{Filipp2025}). To check this, we analysed the distribution of the Einstein and effective radii as a function of the lens redshifts, as can be seen in Fig. \ref{fig:radii_vs_z}. From these, we can see that the parameters for the Euclidised lenses are often on the edge of the training-set distribution, if not completely outside it. This could influence negatively the recovery of these parameters, given that the network is not trained to recognise this region of the parameter space. To try and alleviate this issue, we preventively removed from our comparisons lenses with values of the Einstein radius and effective radius estimated from the literature smaller than $\ang{;;0.5}$ and larger than  $2\arcsec$, respectively, since they fall outside the training sample of LEMON, making their predicted values unreliable extrapolations. Another performance driver could be the fact that the estimates of the effective radii from classical modelling cannot always be compared homogeneously with LEMON's predictions, as SLACS and ACS lenses are modelled with a fixed De Vaucouleurs profile (i.e. $n_{\textrm{lens}} = 4$), while LEMON considers Sérsic profiles with variable values of $n_{\mathrm{lens}}$. Additionally, the fitting process should ideally remove the arc to obtain an accurate estimate; otherwise, the Sérsic fit may overestimate the radius due to bumps in the surface brightness profile. A final possible performance driver could be the fact that the Euclidisation procedure degrades the original image, and as such faint arcs or counterimages could become invisible after the procedure. This could produce lenses that would not be detectable with \Euclid resolution, and that would thus have lensing configurations that were not visible or that were outside the training space of LEMON. In future works, we shall expand the distribution of the training set to cover more of the parameter space, and check if the recovery of these parameters improves.

\subsection{LEMON applied to real \Euclid\ ERO lenses}
We also applied LEMON to real \Euclid lenses found in the Perseus ERO field and modelled in \cite{AcevedoBarroso24}. The modelling was performed with the \texttt{pronto} software \citep{Vegetti2009,Rybak2015a,Rybak2015b,Rizzo2018,Ritondale2019,Powell2021}, with the lens mass distribution modelled by a SIE and the light distribution modelled as a composite of three Sérsic profiles. The surface brightness distribution of the source was reconstructed using a pixelated, free-form approach. Regularisation was applied to penalise large gradients in the reconstructed source brightness. The parameters for the lens mass and light profiles were optimised non-linearly using the \texttt{MultiNest} algorithm \citep{Feroz2009}. The positions of assumed lens images were provided as inputs, with models constrained to ensure that these positions align in the source plane with a tolerance of $\ang{;;1}$. Each model was checked against three criteria: the presence of a SIE critical curve consistent with observed lens image; the alignment of the critical curve with the lens galaxy's light profile; and finally, the consistency of the reconstructed source with a compact object inside a caustic.

From all the lenses found in the field, we considered the ones that have a convincing model (identified as `valid' in the `modelling' column of table 2 in \citealt{AcevedoBarroso24}), and for which the value of the Einstein radius is available. An image of these lenses is shown in Fig. \ref{fig:Euclid_Perseus_ERO_lenses}. The comparison between LEMON predictions for the Einstein radius and those from classical modelling associated with these lenses is shown in Fig. \ref{fig:recovery_plot_Perseus_ERO}. From the figure, we can see that the recovery of the Einstein radius is within $1\sigma$ from the ideal equality line with respect to traditional modelling results for all lenses, except for the lens J031741+412702, for which the Einstein radius is slightly underestimated. It is remarkable that the lens J031749+420011 is recovered correctly, albeit with a large uncertainty, due to the fact that there is a very bright star present in the cut-out, which increases both the aleatoric uncertainty and the epistemic uncertainty, the former due to low visibility of the lens and the latter due to not having trained LEMON with stars as contaminants in the image.

\subsection{LEMON applied to real \Euclid\ Q1 lenses}\label{sec:res_Q1}
We finally applied LEMON to the sample of strong lens images found by the strong lensing discovery engine in the Euclid Quick Release 1 (Q1) data \citep{Q1-SP048}. Through an initial scan of the Q1 galaxy catalogues with machine-learning models, followed by citizen science inspection, expert vetting, and system-by-system modelling, \cite{Q1-SP048} has collected 500 strong lens candidates, divided in Grade A (246) and B (254) lens candidates after the expert vetting scoring. We complement these data with 38 Grade A and 40 Grade B candidates found after the expert visual classification performed on high-velocity dispersion galaxies in \cite{Q1-SP052}.

The strong lenses in \cite{Q1-SP048} and \cite{Q1-SP052} have been modelled via \texttt{PyAutoLens}, which is the standard algorithm for lens modelling in \Euclid\footnote{\url{https://github.com/Jammy2211/PyAutoLens}} \citep{Nightingale2015,Nightingale2018,Nightingale2021}. \texttt{PyAutoLens} is an open-source \texttt{Python} package for strong gravitational lensing, which includes both fully automated strong lens modelling of galaxies and galaxy clusters, and tools for simulating samples of strong lenses. In this work, \texttt{PyAutoLens} employs a SIE mass model, consistent with the model used to train LEMON. 

We predicted the mass and light parameters of 354 of these Q1 lenses, filtered such that the classical modelling is successful. Figure \ref{fig:recovery_Q1} shows the comparison between the parameters obtained through the two modelling techniques: from the figure, we can see that the Einstein radius recovery matches the traditional modelling well, with an $R^{2}$ value of $0.71$. The bias and NMAD are very low, respectively, $\ang{;;0.01}$ and $\ang{;;0.07}$. The mass ellipticity components do not correlate with \texttt{PyAutoLens}' modelling results, with all the predictions clustering around zero, while the light ellipticity components match well the classical modelling, with an $R^{2}$ of $0.74\textrm{--}0.78$ and very low values of bias and NMAD. The predicted effective radii from LEMON match those from \texttt{PyAutoLens} up until $\ang{;;1}$, after which we systematically underpredict with respect to the classical modelling results, similarly to the comparisons with the literature for the effective radius from Sect. \ref{sec:res_euclidezed}. The Sérsic indices predicted by LEMON do not correlate with \texttt{PyAutoLens}' results. It should first be noted that \texttt{PyAutoLens}' modelling of these lenses assumes a maximum value for the Sérsic index $n=5$. Moreover, we have verified that almost all of the points above $\ang{;;1}$ in Fig. \ref{fig:recovery_Q1}f correspond to points that lie on the $n=5$ strip in Fig. \ref{fig:recovery_Q1}g. This could imply that the assumption of a maximum value for the Sérsic index biases the classical modelling predictions, leading to differing values with respect to what LEMON predicts. Finally, the lens \IE magnitude has a very good match with classical modelling, having $R^{2} = 0.90$. Overall, these results show that LEMON matches well the results of classical modelling for real \Euclid lenses, at least for Einstein radius, light ellipticity, effective radius (up to $\ang{;;1}$) and \IE magnitude (after a zero-point correction).

\section{Enhancement of the \Euclid pipeline due to LEMON}\label{sec:classical_methods_speed_up}

One possible use of LEMON in the context of strong lensing applications within \Euclid, beside fast modelling of strong gravitational lenses by itself, is to both speed-up the modelling of strong gravitational lenses through the \Euclid pipeline by giving traditional lens modelling methods initial starting points from which to start the search, and to reduce the failure rate of the modelling pipeline by cross-checking LEMON results with those from the traditional methods.

The \Euclid modelling pipeline \citep{Q1-SP048} adopts the \texttt{Python} library \texttt{PyAutoFit}\footnote{\url{https://github.com/rhayes777/PyAutoFit}}\citep{Nightingale2021_PyAutoFit} to produce an initial lens model result, which is then used to initialise more complex lens models in the deeper parts of the pipeline (e.g. models that use a Delaunay mesh to reconstruct the source galaxy). \texttt{PyAutoFit} uses nested-sampling algorithms, such as \texttt{nautilus} \citep{Lange2023}, for Bayesian inference of the lens model. The posteriors obtained with this approach, however, are not used for any scientific result; rather, the point-wise results are used as initialisers for the more complex models discussed above.

To efficiently use the information from LEMON, \texttt{PyAutoLens} can also use a gradient ascent optimiser, which fully exploits the information from the initial starting point \citep{Amorisco2022}. The lens light and source galaxy during lens modelling are modelled using a multi-Gaussian expansion technique \citep[see][]{He2024}.

To assess the speed gain on the initial stages of the \Euclid lens modelling pipeline, we considered 20 random lenses from the test sample of \Euclid mock lenses, and first modelled them through \texttt{PyAutoLens} with the standard modelling pipeline. The modelling took approximately \num{25000} likelihood evaluations, which amount to around $7\,\textrm{h}$ on a single CPU. A total of 16 out of 20 fits successfully recover the correct lens model, demonstrating the reliability of the \texttt{PyAutoLens} standard modelling pipeline. In two of the four bad cases, the fit is biased due to additional line-of-sight emission from a galaxy not associated with the strong lens (a future iteration of the \texttt{PyAutoLens} pipeline will include modelling emission from line-of-sight galaxies).

We then used \texttt{PyAutoLens} with the gradient ascent optimiser, by modelling the lenses with LEMON and then using the obtained best values as starting points for the optimiser. The fits take about $1000$ likelihood evaluations, which amounts to around $0.27\,\textrm{h}$ on a single CPU, which is a substantial gain over the nested-sampling algorithm, with the hybrid method being $26$ times faster. The majority of the fits are good, except for four of the lenses, which return an incorrect lens model. This result aligns with the number of correctly recovered lenses from the standard pipeline alone. It should be noted that, except for one of the lenses having additional line-of-sight emission from a foreground galaxy, which is wrongly modelled by both techniques, the other three lenses not modelled correctly by the hybrid method are different lenses than the ones wrongly modelled by the standard modelling pipeline. This shows another important benefit of combining LEMON with \texttt{PyAutoLens}, namely the ability to complement each other's limitations: models wrongly inferred by the standard pipeline alone could be modelled correctly by this hybrid approach. Running independent fits with both the standard \texttt{PyAutoLens} pipeline and the combined LEMON-\texttt{PyAutoLens} approach for the same lens, and selecting the best-fit model from the two, will therefore provide a viable strategy to minimise failure rates in lens modelling.

Given that the initialisation stage takes the majority of the time of the pipeline computing time, the speed-up in the initial phase should provide a speed-up of the entire pipeline of up to $5$ times the speed associated with using the nested-sampling algorithm only as an initialiser. This speed-up of modelling is consistent with results from \cite{Pearson2021}, where a BNN was used to predict gravitational lens parameters, which are then given to \texttt{PyAutoLens} as priors. In their case, the combination of BNN and \texttt{PyAutoLens} increases the modelling speed by a mean factor of $1.73$. It should be noted, however, that in our case the standard modelling method is not a Markov chain Monte Carlo with priors set by a BNN like in \cite{Pearson2021}, and as such, the speed-up factor in our case can be bigger due to the use of a simpler algorithm. This is an optimistic estimate of the speed-up, however, given that on real images the predictions from LEMON are worse than on simulated lenses. In future iterations of LEMON, we plan to implement error estimation on the lens model parameters predicted by \texttt{PyAutoLens}, which are currently not computed with the gradient-ascent method. This can be achieved in different ways, for example by using Hessian matrix estimation or the Fisher information matrix. Investigating these approaches is, however, beyond the scope of this paper.

Finally, to assess whether the optimiser approach needs LEMON in order to correctly model gravitational lenses, we repeated the test by using random starting points for the optimiser. Results show that, with random initialisations, \texttt{PyAutoLens} is unable to successfully recover a correct lens model fit for any of the lenses. Thus, the robust initial starting points provided by LEMON prove to be required in order for \texttt{PyAutoLens} to be sped up by using the gradient ascent optimiser instead of the nested-sampling algorithm.

\section{Conclusions}\label{sec:Conclusions}
In this work, we expand the lens modelling machine-learning algorithm LEMON \citep{Gentile2023} by predicting both the mass and the light of the foreground deflector of \Euclid\ galaxy-galaxy gravitational lenses. The former is modelled as a SIE with external shear, while the latter is a Sérsic profile. We used mock \Euclid lenses with contaminants in the image for both training and testing of the architecture. The main results are as follows.

\begin{itemize}
    \item All parameters of the test set are recovered correctly, with very low values of bias and scatter. The parameter that is recovered worst is the Sérsic index, which shows a large scatter and a deviation from the linear trend for high values of the Sérsic index. This suggests that LEMON is able to recover the mass and light parameters for the lens even in the presence of companions contaminating the image.
    \item Applied to a selection of Euclidised HST images, LEMON manages to match classical modelling in recovering the Einstein radius, mass and light ellipticity components, lens effective radius of the lens up to $\ang{;;1}$, and lens magnitude for the SLACS sample, while we cannot reach any conclusion about the Sérsic index due to low statistics. In general, the results are worse than those obtained from the test set, given that Euclidised lenses may not be similar to the simulations used to train the network, as is shown in Fig. \ref{fig:radii_vs_z}.
    \item Applied to five real galaxy-galaxy \Euclid\ gravitational lenses observed in the Perseus ERO field \citep{AcevedoBarroso24}, for which the value of the Einstein radius is available thanks to classical modelling, and the 578 new candidates in the Q1 data \citep{Q1-SP048,Q1-SP052}, for which more parameters are available, we match \texttt{PyAutoLens} in recovering the Einstein radius, light ellipticity components, and lens \IE magnitude parameters. The effective radius predictions match \texttt{PyAutoLens} well up to $\ang{;;1}$, while the Sérsic index and mass ellipticities do not correlate as well with the predictions from \texttt{PyAutoLens}. On average, however, the inferred mass ellipticities are not strongly discrepant from the \texttt{PyAutoLens} values.
    \item By using the predictions of LEMON as starting points for a gradient-ascent version of the \texttt{PyAutoLens} modelling algorithm, it is possible to speed up the algorithm by up to 26 times compared to the standard \Euclid modelling pipeline. This would not be possible without the initial LEMON guesses, because the gradient-ascent optimiser with random initialisations does not produce valid model results for all lenses.
\end{itemize}

In future works, we plan to further advance the LEMON architecture, by predicting the parameters of more general lens mass models, such as the power-law model. We also plan on enhancing the prediction of certain parameters by deblending the gravitational lens image in lens-only and source-only images \citep{Fucheng2024}, and training LEMON separately on the two components.

Q1 data are now public, with DR1 data around the corner. We are therefore preparing to process the enormous amount of upcoming data, and have started applying LEMON to newly discovered strong lenses from the Q1 data release. Results confirm that LEMON is a robust and fast lens modelling tool for \Euclid. This will enable the generation of models for the approximately \num{100000} expected \Euclid lenses in a remarkably short time.

By applying LEMON and \texttt{PyAutoLens} to \Euclid gravitational lenses, we shall determine the total mass within the Einstein radius, or extrapolated to the effective radius, when lens and source redshifts are available. Combining this with stellar mass estimates from SED fitting (\citealt{Q1-SP031}), we shall derive dark matter fractions. We shall also provide constraints on the mass density slope and initial mass function. With such a vast statistical lens sample, we shall be able to investigate the evolution of these properties in unprecedented detail, as well as the scaling relations that relate them as a function of redshift \citep{Koopmans2006,Gavazzi2007,Bolton2008,Auger2010,Tortora2010,Tortora+14_DMslope,Tortora+18_KiDS_DMevol,Tortora+19_LTGs_DM_and_slopes,Sonnenfeld2013,Sonnenfeld+15_SL2SV,Sonnenfeld+17_IMF,ORiordan+25_NGC6505}. By comparing these scaling relations with predictions from cosmological simulations, we shall be able to constrain cosmological parameters and physical processes in galaxies \citep{Mukherjee2018,Mukherjee2021,Mukherjee2022,Busillo+23_CASCOI,Busillo+24_CASCO-II,Tortora+25_CASCOIII}.


%
%

\begin{acknowledgements}
VB and CT acknowledge the INAF grant 2022 LEMON. LL thanks the support from INAF theory Grant 2022: Illuminating Dark Matter using Weak Lensing by Cluster Satellites.
This research uses observations made with the NASA/ESA \textit{Hubble} Space Telescope obtained from the Space Telescope Science Institute, which is operated by the Association of Universities for Research in Astronomy, Inc., under NASA contract NAS 5–26555.
\AckERO
\AckQone
\AckEC
\end{acknowledgements}
%
%

\bibliography{bibliography}

%

\begin{appendix}
\section{Recovery of shear parameters}\label{sec:shear_recovery}
\begin{figure*}
    \centering
    \includegraphics[width=0.45\linewidth]{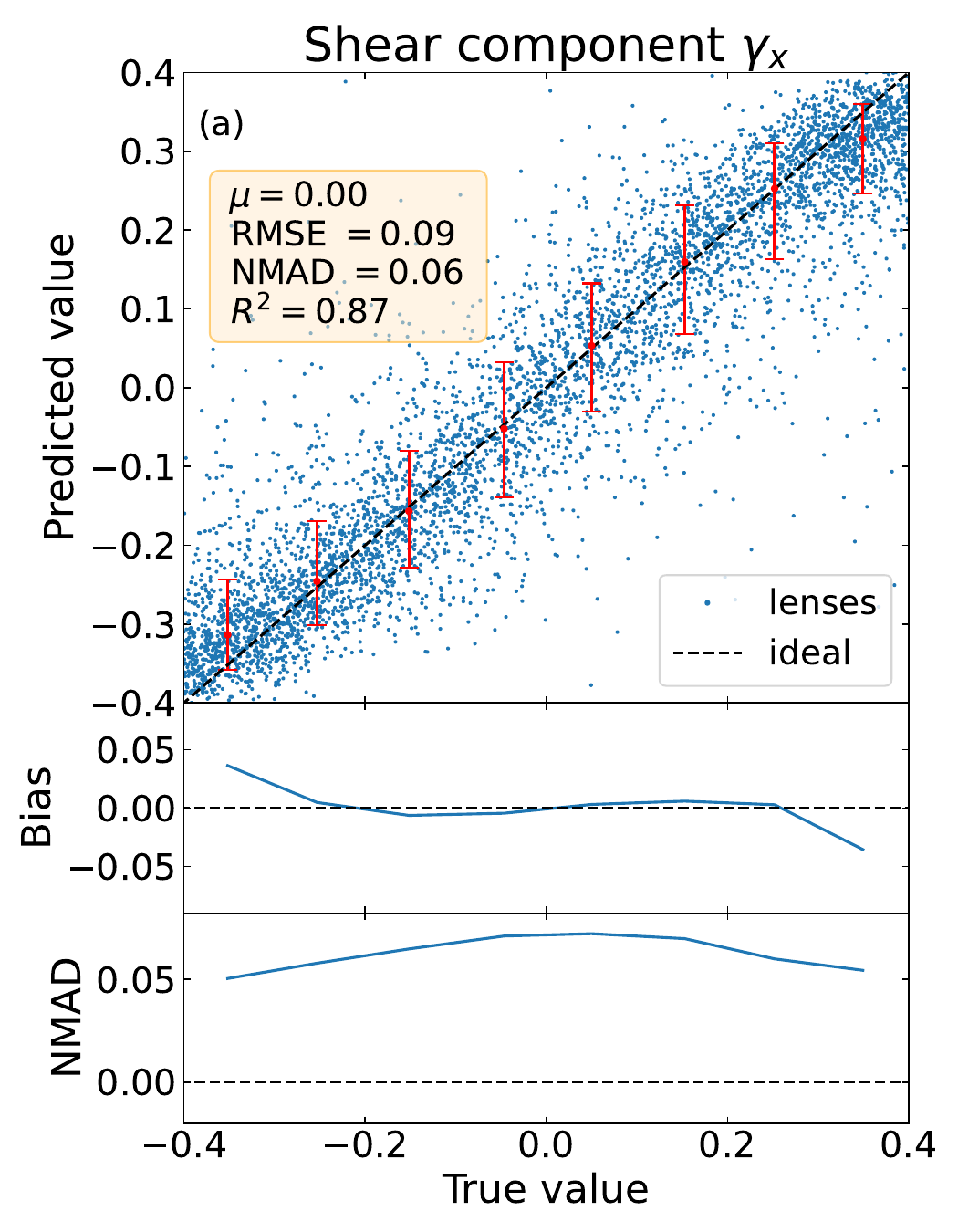}
    \includegraphics[width=0.45\linewidth]{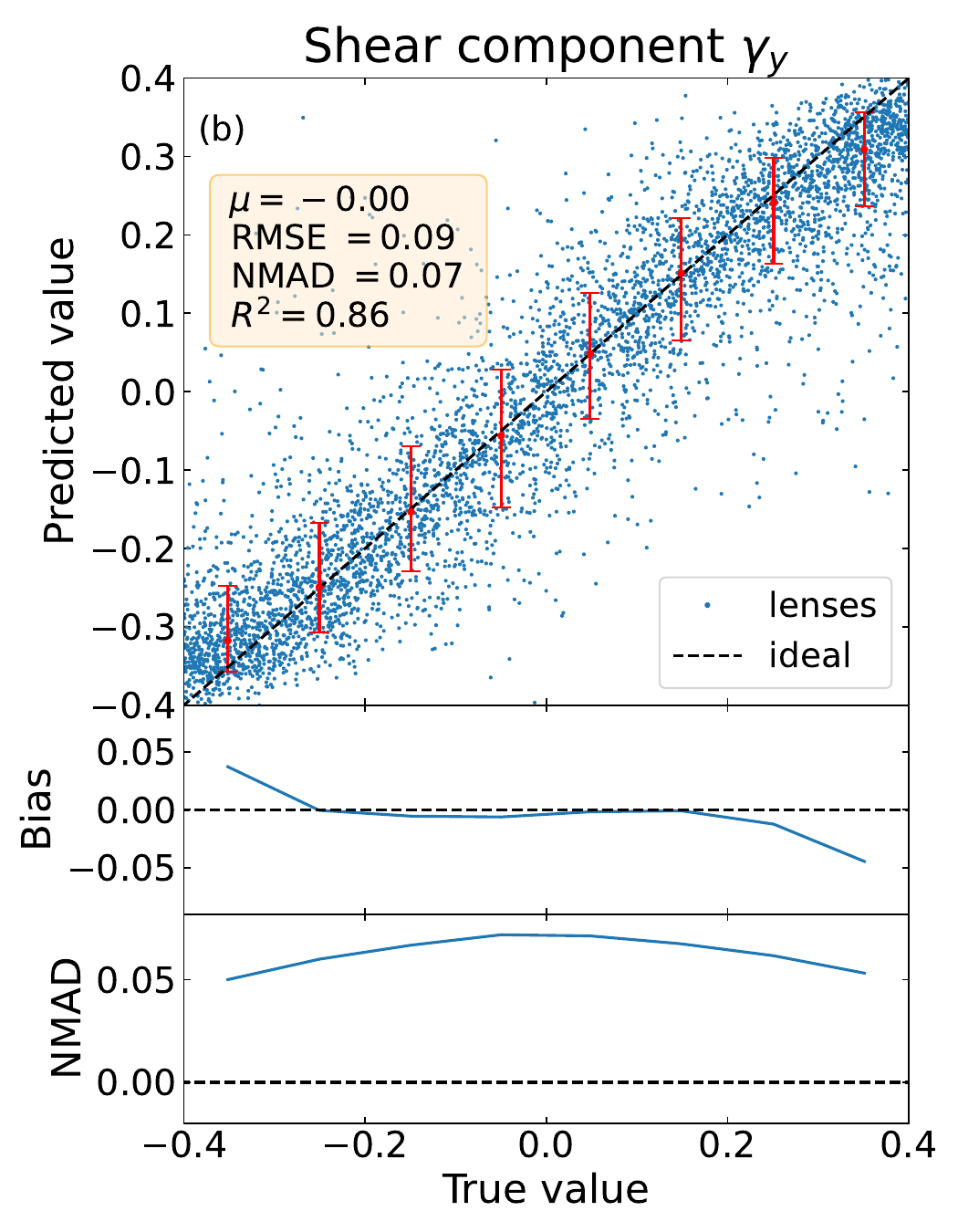}
    \caption{Same as \Fig\protect\ref{fig:recovery_plot}, but for the external shear components $\gamma_{x}$ and $\gamma_{y}$.}
    \label{fig:recovery_plot_shear_gamma}
\end{figure*}

\begin{figure*}
    \centering
    \includegraphics[width=0.85\linewidth]{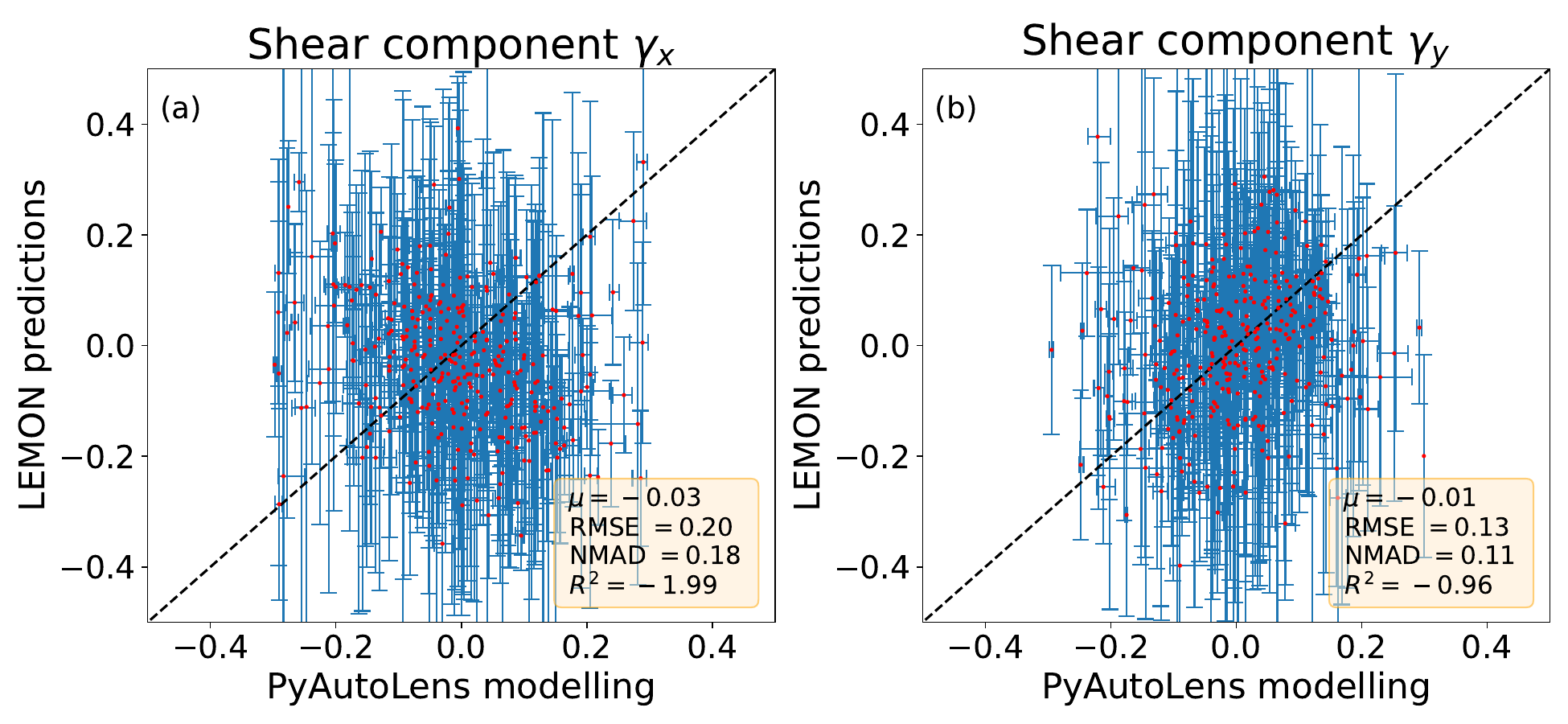}
    \caption{Same as \Fig\protect\ref{fig:recovery_Q1}, but for the external shear components $\gamma_{x}$ and $\gamma_{y}$.}
    \label{fig:James_shear_comparison}
\end{figure*}
As a side test, we also trained LEMON to specifically recover the shear components $\gamma_{x}$ and $\gamma_{y}$, using the second dataset of \Euclid mock lenses described in Sect. \ref{sec:Metcalf_lenses} for training, validation and testing. This dataset has the exact same properties as the main dataset, except for the extended interval of values for the shear components.

Figure \ref{fig:recovery_plot_shear_gamma} shows the recovery of the two shear components on the simulated test-set (constructed in an identical way as with the main dataset). This recovery is very good, showing no bias (except for a slight deviation on the edges of the parameter space), low scatter, and a very high value of the coefficient of determination ($R^{2}=0.86\textrm{--}0.87$). Overall, on simulated images LEMON manages to recover these parameters, under the condition that the parameters' intervals on which the inference is done is larger than the intrinsic scatter on the component. Indeed, when trying to recover the shear components of the main dataset, the inference fails, likely because the parameter values are too close to zero, and the shear effects on the lensing configurations are too small to be discerned. All the other parameters are recovered as well as with the network trained on the main dataset, with similar values for the metrics.

Figure \ref{fig:James_shear_comparison} shows a comparison between our estimates of the shear and the estimates from \texttt{PyAutoLens} on the same Q1 lenses used in Sect. \ref{sec:res_Q1}. We can see that the two modelling results strongly disagree, with negative coefficients of determination. This discrepancy is most likely due to the fact that our network finds ambiguity in the mapping from image to shear (hence the very large uncertainties in our predictions with respect to the classical modelling estimates). This ambiguity could be due to several reasons, such as intrinsic degeneracies with mass ellipticity or domain shift issues \citep[see e.g.][]{Filipp2025}.

It should be noted that recovery of external shear parameters with machine learning is notoriously difficult \citep{Gawade2025, Schuldt2023_shear}, and as such we did not expect a good match with literature results for real lenses. It is notable, however, that the network manages to recover the external shear values at least in the ideal case of simulated \Euclid lenses.

\section{Recovery of Einstein mass}\label{sec:Einstein_mass}
\begin{figure}
    \centering
    \includegraphics[width=1\linewidth]{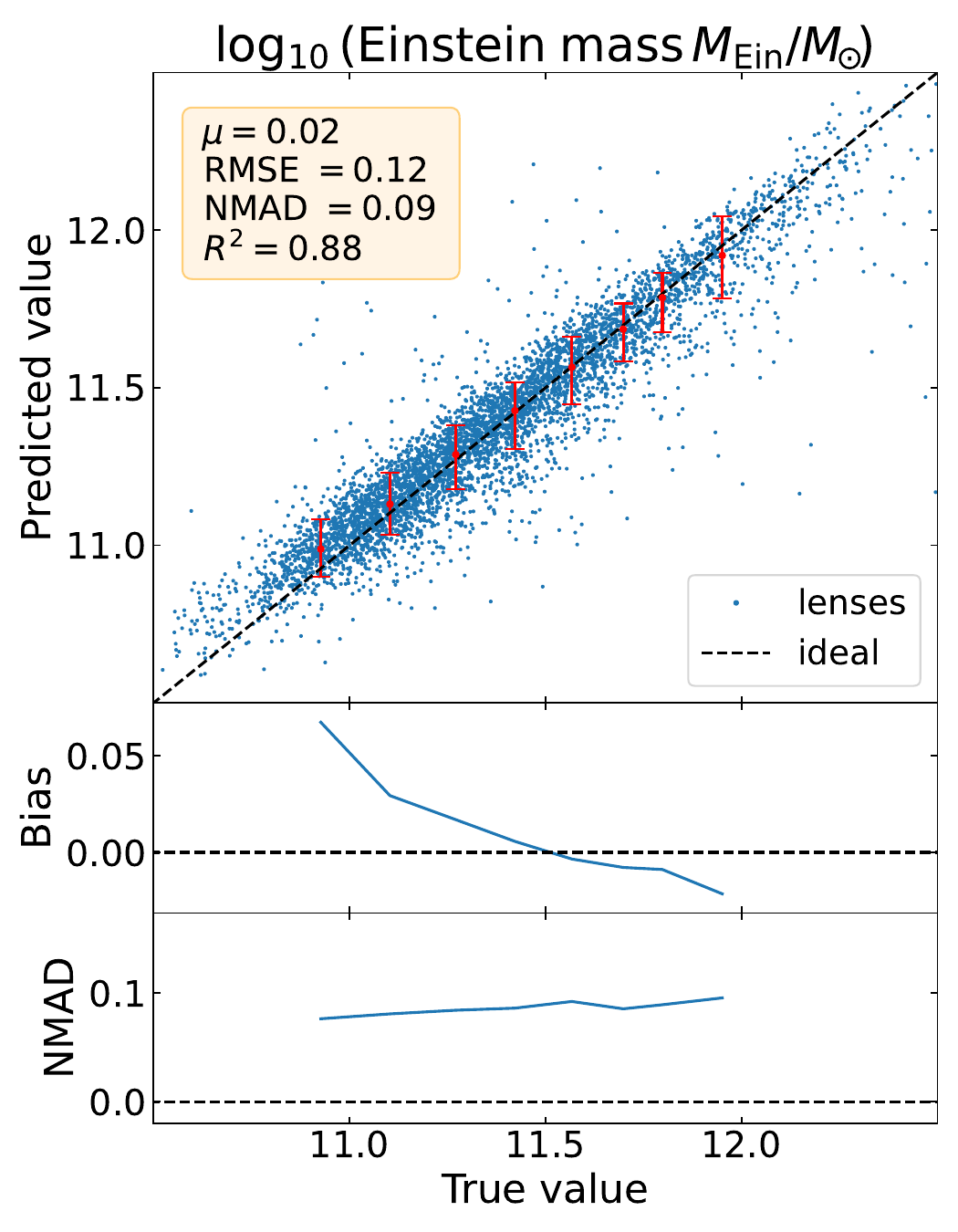}
    \caption{Same as Fig. \ref{fig:recovery_plot}, but for the Einstein mass parameter.}
    \label{fig:Einstein_mass_recovery_plot}
\end{figure}

\begin{figure}
    \centering
    \includegraphics[width=0.95\linewidth]{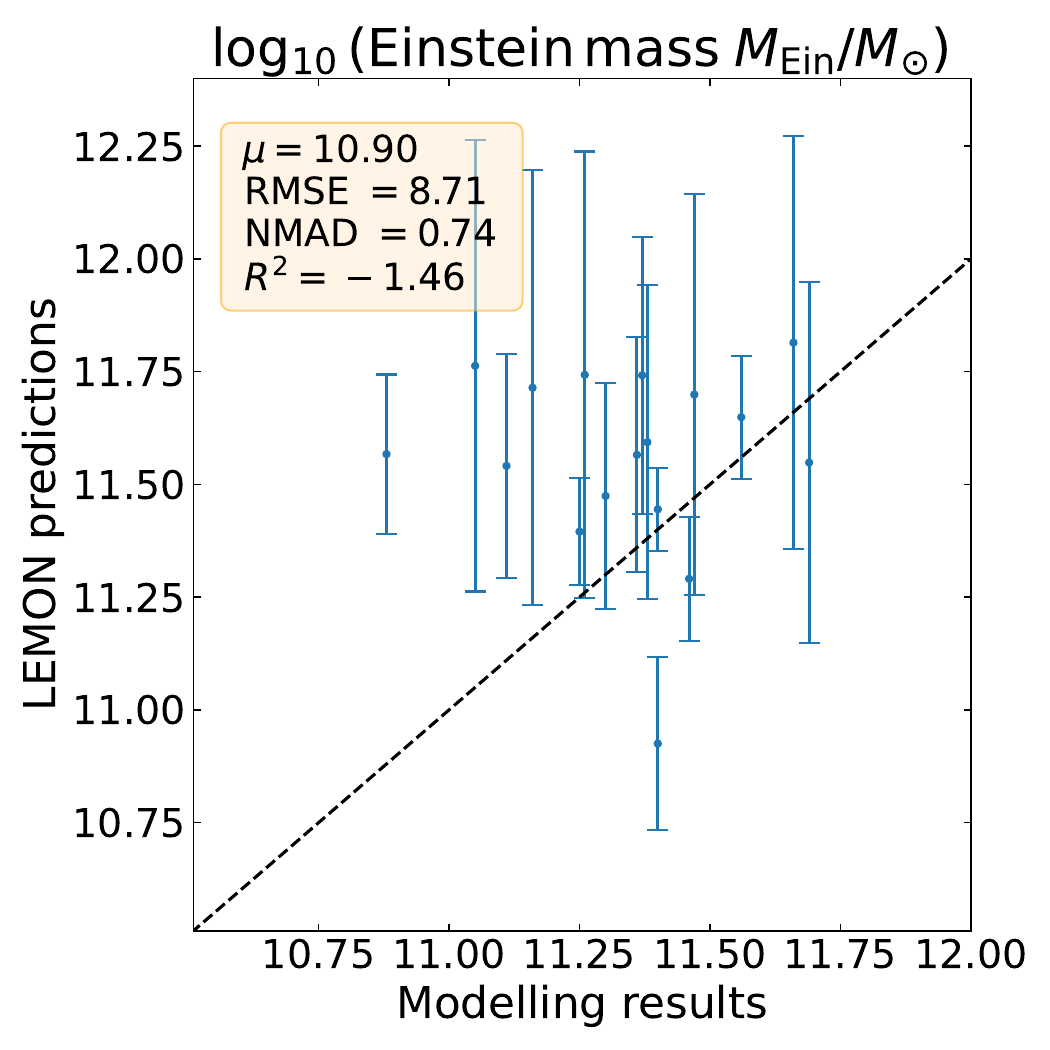}
    \caption{Comparison between predictions of Einstein mass obtained from LEMON and from classical modelling \protect\citep{Auger2009} for the Euclidised SLACS lenses.}
    \label{fig:Einstein_mass_recovery_plot_Euclidized_lenses}
\end{figure}

\begin{figure}
    \centering
    \includegraphics[width=1\linewidth]{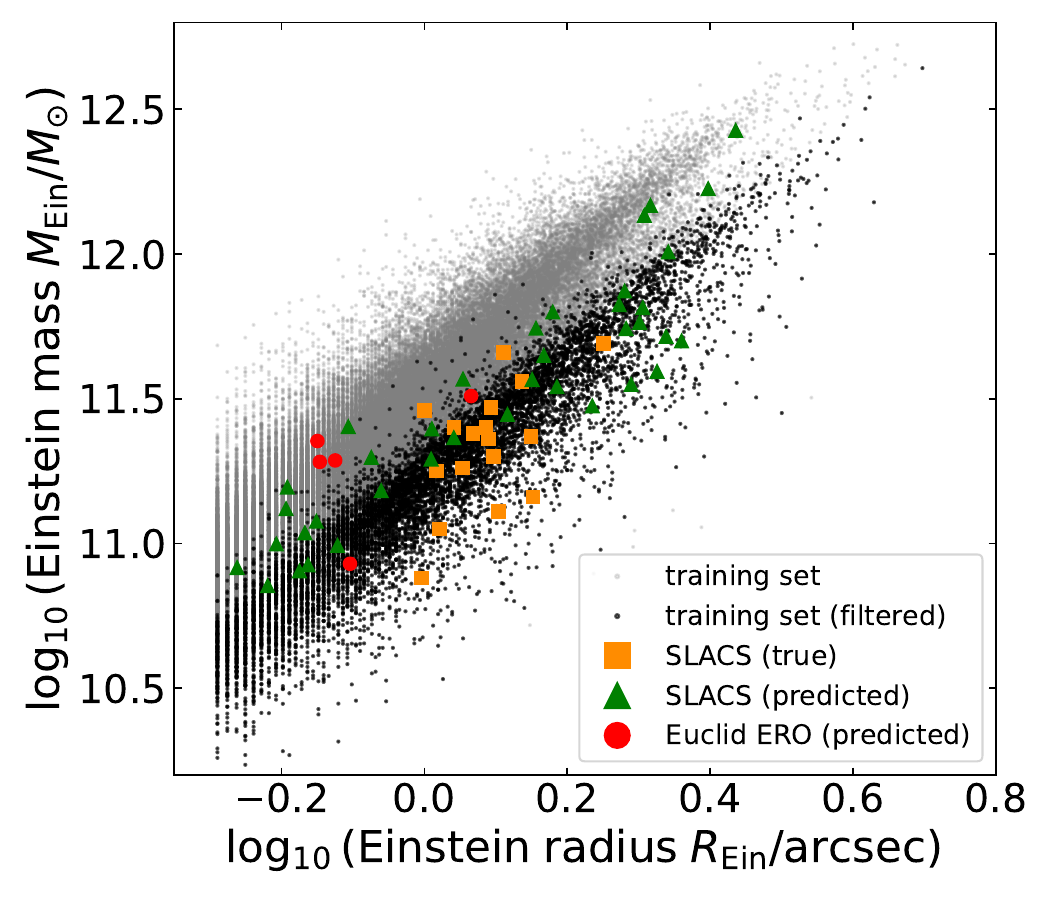}
    \caption{Einstein mass as a function of Einstein radius. Grey points are the values for the training set, with black points corresponding to those lenses of the training sample having values of lens redshift within the redshift range of the SLACS sample. Orange squares are the literature parameter values for SLACS lenses, while green triangles are the SLACS predicted values from LEMON. Red points correspond to the predicted values from LEMON for Euclid ERO lenses considered in Fig. \ref{fig:recovery_plot_Perseus_ERO}.}
    \label{fig:Einstein_mass_vs_Einstein_radius}
\end{figure}
As a side test, we trained LEMON to recover the Einstein mass parameter, defined as the total lens mass enclosed within the largest critical curve. This parameter is the effective total mass responsible for the observed lensing effect.

For any axisymmetric lens model, the Einstein mass is related to the (circularised) Einstein radius via the formula

\begin{equation}
    R_{\textrm{Ein}} = \sqrt{\frac{D_{\textrm{ds}}}{D_{\textrm{s}}D_{\textrm{d}}} \frac{4GM_{\textrm{Ein}}}{c^2}}\,,\label{eq:Einstein_mass_relation}
\end{equation}
where $D_{\textrm{ds}}$ is the source-lens distance, $D_{\textrm{d}}$ is the distance between the observer and the lens and $D_{\textrm{s}}$ is the distance between the observer and the source. Notice that these distances are angular diameter distances, and thus depend on the redshifts of the source and the lens. We thus expect $M_{\textrm{Ein}} \propto R_{\textrm{Ein}}^{2}$.

The recovery of the Einstein mass for the test set lenses is shown in Fig. \ref{fig:Einstein_mass_recovery_plot}, with the associated trends for bias and NMAD shown in the lower two panels. From the figure, we can see that the Einstein mass is well recovered over all the lens mass range. It should be noted that this is not expected: given the implicit dependence of \Eq\eqref{eq:Einstein_mass_relation} from the redshifts of source and lens, which we did not train the network to recover, the network should have difficulty in recovering the values of Einstein mass through the value of the Einstein radius.

We conjecture that the correct recovery could be due to the fact that the network is learning known or unknown strong correlations between some lensing features and the Einstein mass values, bypassing the need for direct inference of redshift values. Some of these features could be lensing distortion patterns (larger Einstein masses produce more pronounced arcs), the size and spread of the lensed images (larger Einstein masses produce wider separations of the lensed images), and the brightness of the arcs (larger Einstein masses focus light more intensely, which affects the intensity profiles of the arcs in the image). The network could also be learning implicit correlations between features related to the lens redshift, such as the relative brightness of source and lens, and the redshift of the lens itself. The network could then use these correlations to create a Bayesian mapping that reproduces the relation \eqref{eq:Einstein_mass_relation}, by learning a distribution of possible mappings between Einstein radius and Einstein mass. Thus, when making predictions, the network sees a particular arc size that often correlates with a higher Einstein mass, and gives a value that peaks around the higher value, with smaller or larger uncertainty associated with how often the value of $R_{\textrm{Ein}}$ is associated with a certain value of $M_{\textrm{Ein}}$ in the training phase.

Figure \ref{fig:Einstein_mass_recovery_plot_Euclidized_lenses} shows the recovery of the Einstein mass for the SLACS Euclidised lenses sample, which is the only one that has corresponding values from classical modelling of the Einstein mass reported \cite[][table 4]{Auger2009}. We preventively removed all lenses with effective radii higher than $\ang{;;2}$, as performed in Sect. \ref{sec:res_euclidezed}. The comparison between methods shows a systematic overestimation with respect to the classical modelling, with a mean bias of $\mu = 10.9$. The negative coefficient of determination also shows a complete mismatch with the results from \cite{Auger2009}.

To further investigate this mismatch, in Fig. \ref{fig:Einstein_mass_vs_Einstein_radius} we show the correlation between the Einstein mass and the Einstein radius. For a fixed redshift, we expect ideally a single power-law curve, corresponding to \Eq\eqref{eq:Einstein_mass_relation} with fixed values of $D_{\textrm{ds}}$, $D_{\textrm{s}}$, and $D_{\textrm{d}}$. For different redshifts, in log space we expect a family of curves with fixed slope and varying $y$ intercept, which is what we see in the figure with the training set region. The figure shows that the region of the training set corresponding to the minimum and maximum redshifts of the SLACS sample aligns with the lower boundary of the training set area, where most of the SLACS sample is concentrated. The ERO lenses from \Euclid, instead, lie fully within the training-set parameter space, which could explain the better recovery of the Einstein radius for these lenses. We thus conjecture that the mismatch with literature results for this parameter could be due to the fact that the sample lies on the edge of the training set, or that the parameter is intrinsically more difficult to recover with machine learning for real lens images, which are naturally more complex. Future estimates of Einstein mass for \Euclid lenses obtained from classical modelling could provide more substantial information on the recovery of this parameter.

\section{Robustness of Platt-scaling}\label{sec:robustness_scaling_test}
\begin{figure}
    \centering
    \includegraphics[width=1\linewidth]{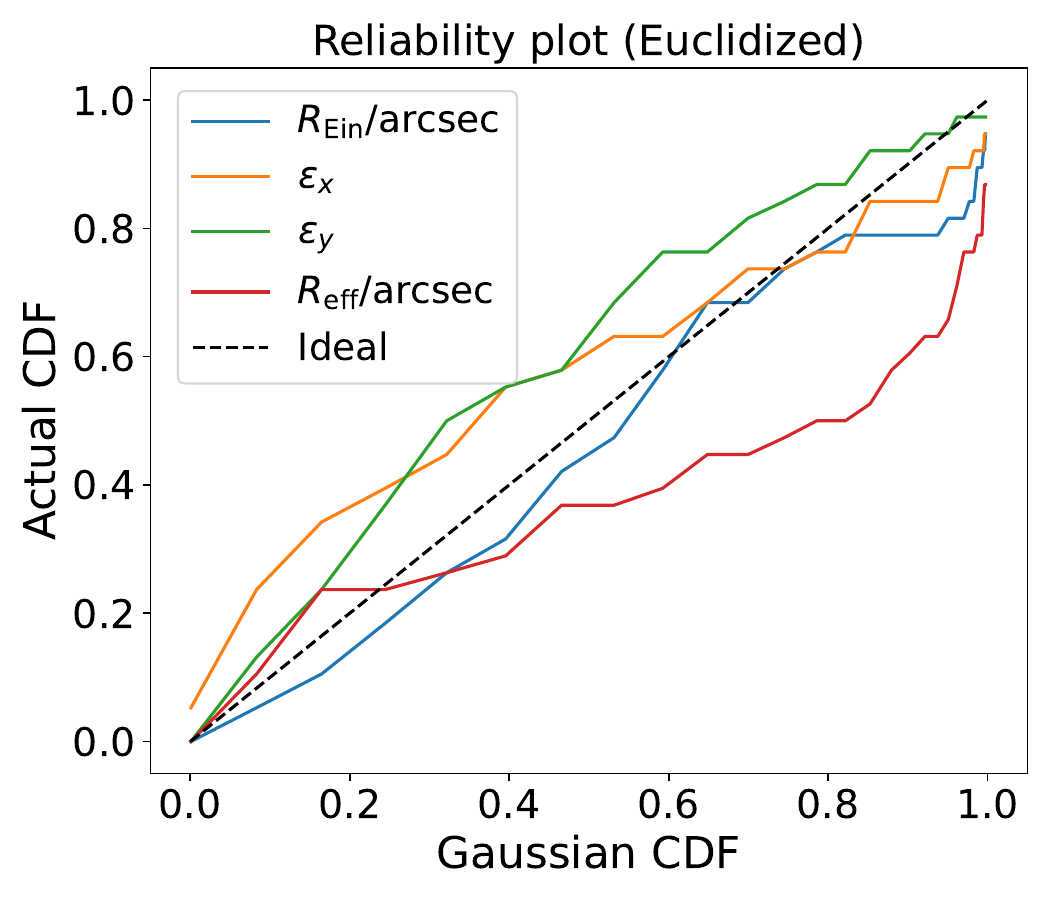}
    \caption{Empirical reliability plot generated from the Euclidised lenses' predictions of \Sec\protect\ref{sec:res_euclidezed}.}
    \label{fig:reliability_plot_Euclidized_lenses}
\end{figure}

\begin{figure}
    \centering
    \includegraphics[width=1\linewidth]{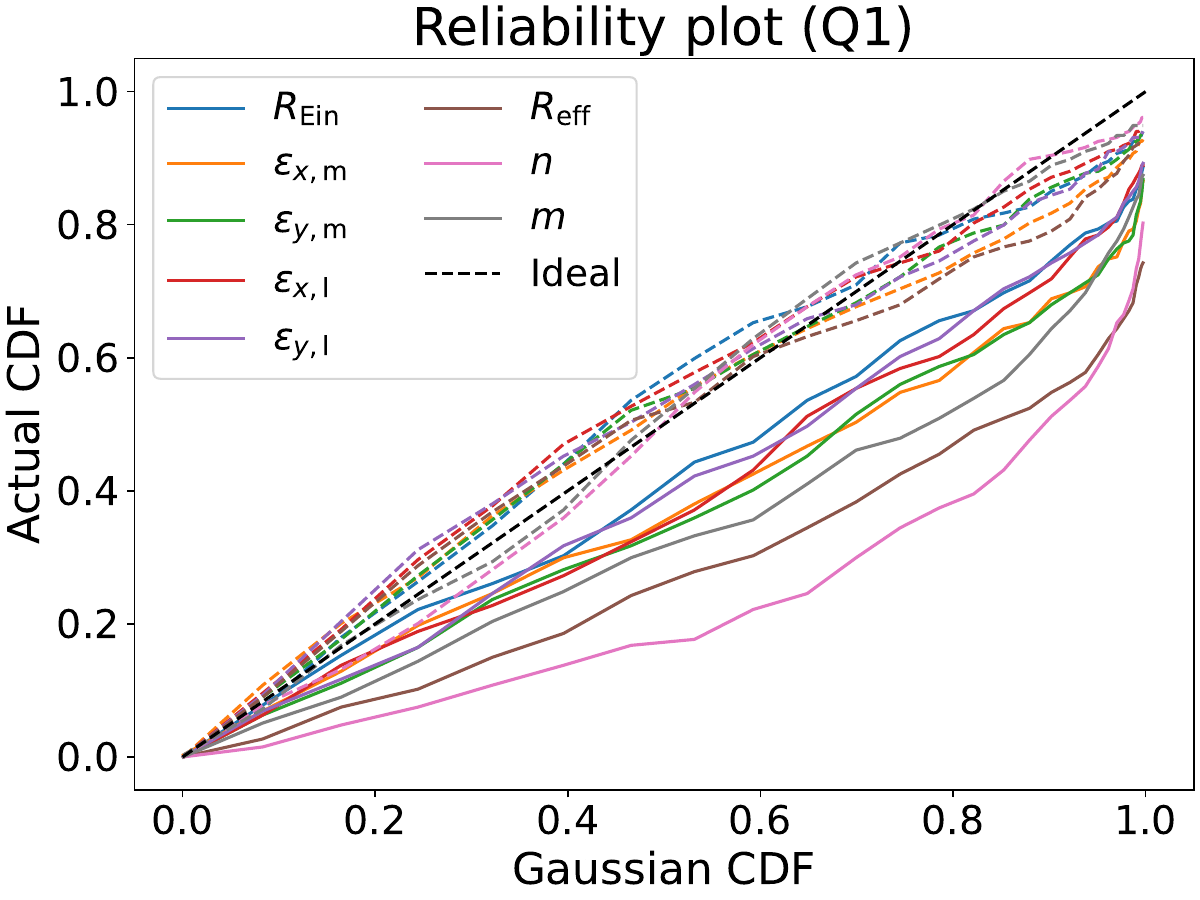}
    \caption{Empirical reliability plot generated from the Q1 lenses' predictions of \Sec\protect\ref{sec:res_Q1}. Continuous lines are associated with the default calibration obtained from the simulations, while dashed lines are the results after the manual correction with the values from \Tab\protect\ref{tab:Scaling_factors_Q1_corrected}.}
    \label{fig:reliability_plot_Q1}
\end{figure}

\begin{table}
    \centering
\caption{Scaling factors needed to produce Gaussian uncertainties for Q1 lenses' predictions.}
    \begin{tabular}{cc}
    \hline
    \noalign{\vskip 2pt}
    \hline
         Parameter& Scaling factor\\
    \hline
    \noalign{\vskip 2pt}
         Einstein radius $R_{\textrm{Ein}}/\textrm{arcsec}$& $1.2$\\
         Mass $x$-ellipticity $\epsilon_{x,\mathrm{m}}$& $1.5$\\
         Mass $y$-ellipticity $\epsilon_{y,\mathrm{m}}$& $1.5$\\
         Light $x$-ellipticity $\epsilon_{x,\mathrm{l}}$& $1.2$\\
         Light $y$-ellipticity $\epsilon_{y,\mathrm{l}}$& $1.2$\\
         Lens effective radius $R_{\textrm{e, lens}}/\textrm{arcsec}$& $2.0$\\
         Lens Sérsic index $n_{\textrm{lens}}$& $2.0$\\
         Lens magnitude $m_{\textrm{lens}}$& $1.5$\\
    \hline
    \end{tabular}
    \label{tab:Scaling_factors_Q1_corrected}
\end{table}
In this section we discuss the robustness of the Platt-scaling correction for the total uncertainty of our BNN. Due to distributional shift and other systematic effects that produce a difference between simulated and real images, it is possible that the total calibrated uncertainties predicted in output by LEMON, which have been made Gaussian for predictions on the mock \Euclid lenses with the Platt-scaling calibration procedure, may become unreliable on real lenses. To check this, we evaluated the empirical reliability plots for both the Euclidised and the Q1 sample of lenses, by using the results (predicted best values and calibrated uncertainties) reported in \Secs\ref{sec:res_euclidezed} and \ref{sec:res_Q1}. We show the one for the Euclidised lenses in Fig. \ref{fig:reliability_plot_Euclidized_lenses}, while the one for the Q1 lenses in Fig. \ref{fig:reliability_plot_Q1} (continuous lines).

For the Euclidised lenses, the Einstein radius and ellipticity uncertainties are still consistent with the assumption of Gaussianity, as can be seen by the fact that the reliability plot trends for these parameters roughly follow the identity line. The effective radius predictions, instead, show signs of overconfidence, with uncertainties lower than what they should be if they were calibrated correctly. This means that the output uncertainties of our network for this parameter, using the scaling factors that we assumed for the simulated lenses, underestimate the expected uncertainty under the assumption of Gaussianity. This systematic underestimation of the uncertainty appears also for all the predictions of the Q1 lenses.

It should be noted, however, that for the Q1 lenses it is possible to find new values of scaling such that the associated empirical reliability plot matches the identity line: such values are reported in \Tab\ref{tab:Scaling_factors_Q1_corrected}. These new values produce uncertainties that are systematically bigger than the ones we reported in Sect. \ref{sec:res_Q1}, with minimum median percent increases of around $41\%$ for the Einstein radii uncertainties, up to $136\%$ for the effective radii uncertainties.

\section{Uncalibrated aleatoric and epistemic uncertainty trends}\label{sec:epistemic_analysis}

\begin{figure*}
    \centering
    \includegraphics[width=1\linewidth]{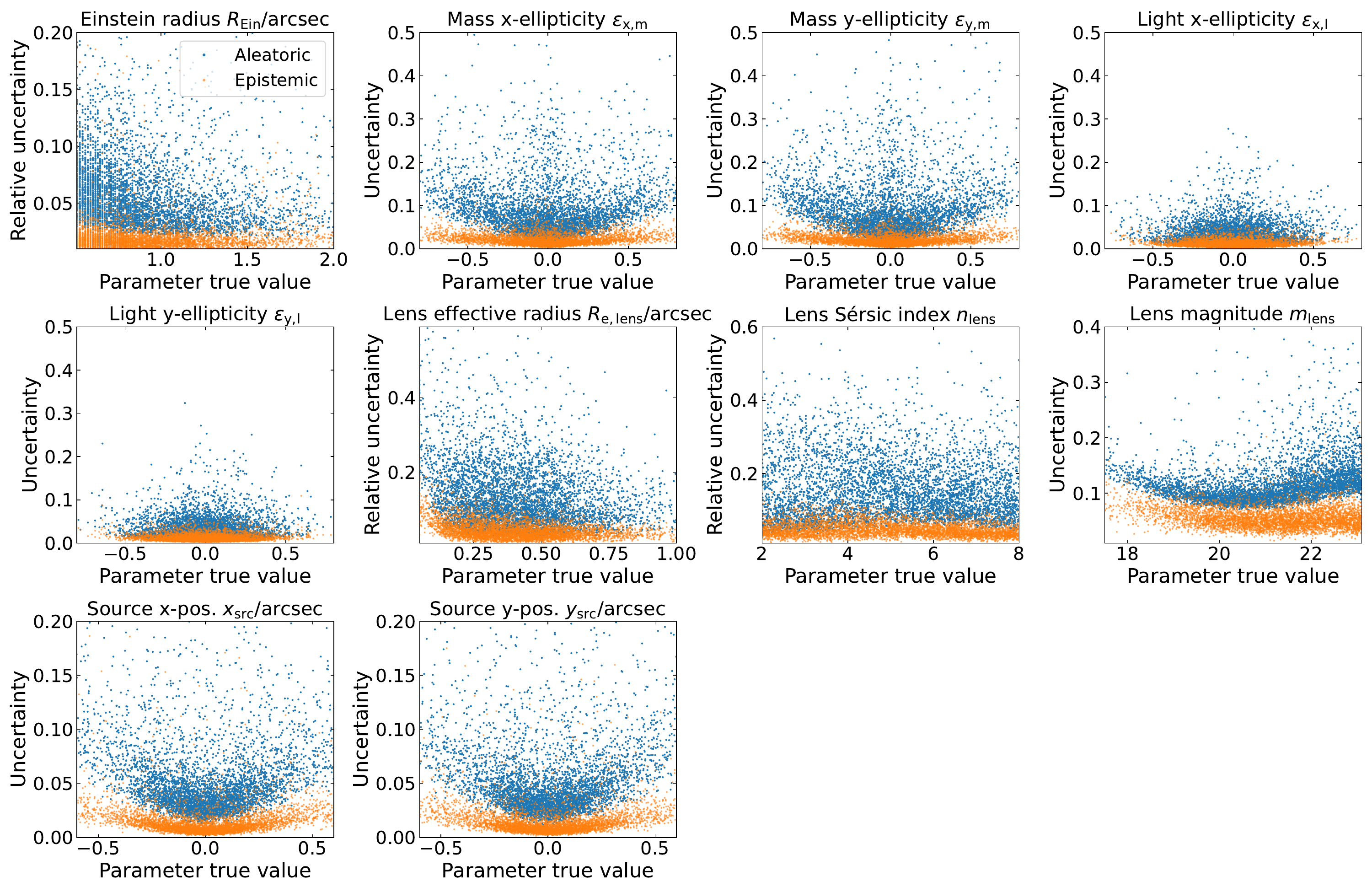}
    \caption{Trends of aleatoric (blue) and epistemic (orange) relative uncertainties as a function of the true value of the lens parameters. For the position angle and lens magnitude, we consider the absolute uncertainties. For clarity, we plotted only \num{5000} random points from the test set.}
    \label{fig:aleatoric_epistemic_uncertainties_comparison}
\end{figure*}
We report, in Fig. \ref{fig:aleatoric_epistemic_uncertainties_comparison}
, an alternative version of Fig. \ref{fig:calibrated_uncertainty_trend}, where the trends of the (uncalibrated) aleatoric and epistemic components of the total uncertainty are shown separately. The fact that the epistemic uncertainty is systematically lower than the aleatoric uncertainty is a good sign, because it indicates that, at least for the mock \Euclid-like images that compose the test-set, the network is confident that the learned mapping between images and parameters is well determined (i.e. the error associated with a lack of training is irreducible, and the dominant uncertainty is the aleatoric one). This means that more training data of the same type or a larger network is not necessary, because the model has seen enough similar examples to have consistent predictions in the range covered by the training set, and the model is considered expressive enough.

\section{Analysis of signal-to-noise ratio of mock \Euclid lenses}\label{sec:SNR_analysis}

\begin{figure*}
    \centering
    \includegraphics[width=1\linewidth]{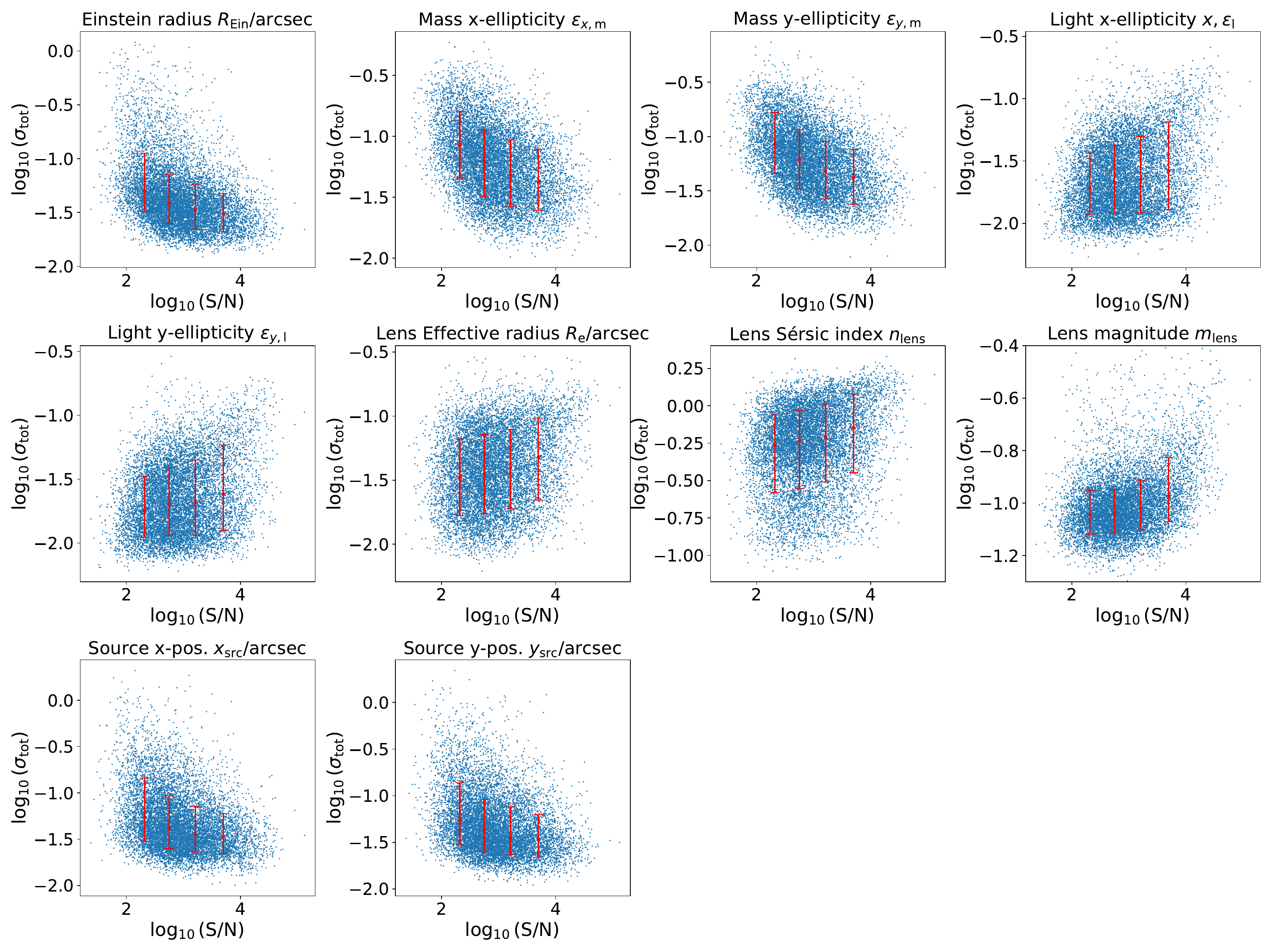}
    \caption{Total calibrated uncertainty for each lens against total S/N. The median trends of the scatter plots, along with the respective scatters associated with the 16th and 84th percentiles, are shown as red points and bars, respectively.}
    \label{fig:SNR_test_uncertainty}
\end{figure*}

\begin{figure*}
    \centering
    \includegraphics[width=1\linewidth]{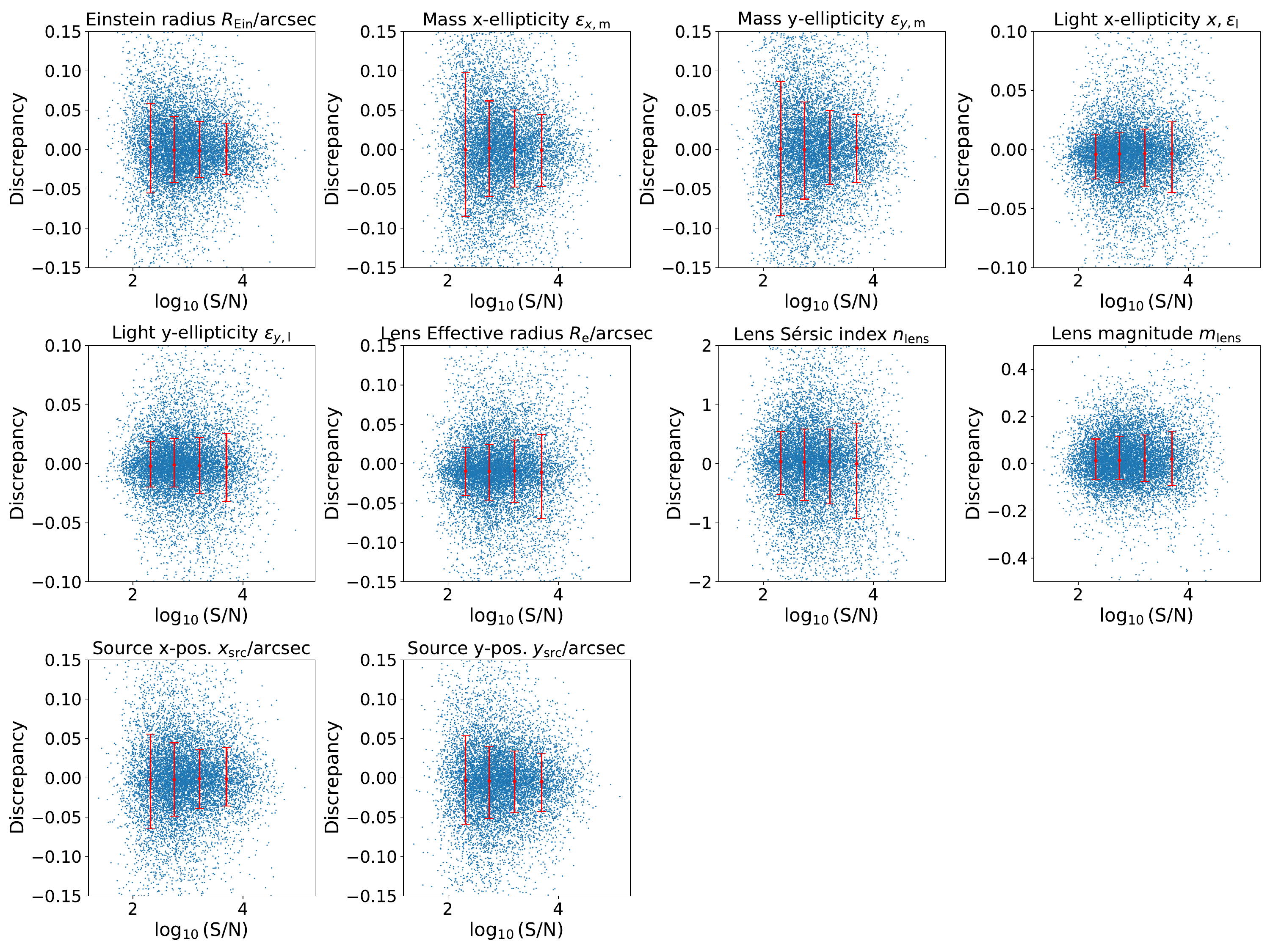}
    \caption{Same as \Fig\protect\ref{fig:SNR_test_uncertainty}, but for the discrepancy between predicted and true parameter value for each lens.}
    \label{fig:SNR_test_discrepancy}
\end{figure*}

In this appendix we check in regards to the mock \Euclid lenses whether there is any correlation between the total S/N in lensed source pixels above 4$\sigma$ (identified as `total S/N' from now on in this section) and two quantities: total predicted uncertainty and discrepancy between predictions and true values for each parameter.
Figure \ref{fig:SNR_test_uncertainty} shows the trend of total uncertainty against total S/N. The figure shows an interesting, but expected trend: quantities associated with the mass distribution, such as the Einstein radius, have a decreasing trend with an increasing total S/N, while the opposite is true for quantities associated with the light distribution, such as the effective radius. This seems to confirm that quantities associated with the mass are derived by the network from arc information (and thus the uncertainty decreases as the arc is more luminous with respect to the lens light), while quantities associated with the lens light distribution are derived from the lens light information (and thus the uncertainty increases as the lens light becomes weaker).

A similar trend is seen in Fig. \ref{fig:SNR_test_discrepancy}, but on the scatter of the discrepancy-$\mathrm{S/N}$ relation: quantities associated with the mass have a larger discrepancy scatter between predictions and true values at low values of S/N, due to the fact that the source configurations may be harder to detect if the source is not visible enough, while quantities associated with the light have larger discrepancy scatter at high values of S/N, due to the fact that light from the source interferes with the inference from the lens light.

\end{appendix}

\end{document}